\documentclass[11pt,english]{article}

\usepackage{times}
\usepackage{amsfonts}
\usepackage{amsmath,amssymb}
\usepackage{amsthm}
\usepackage{natbib}
\usepackage{bm}
\usepackage{graphicx}
\usepackage{subcaption}
\usepackage{appendix}
\usepackage{rotating}
\usepackage{float}
\usepackage{mathrsfs}
\usepackage{dsfont}
\usepackage{nicefrac}
\usepackage{bm}
\usepackage{xspace}
\usepackage{color}
\usepackage[colorlinks,linkcolor=blue,citecolor=blue,bookmarks=false,pagebackref]{hyperref}
\usepackage[top=1.0in, bottom=1.3in, left=1.15in, right=1.15in]{geometry}
\usepackage{setspace}
\usepackage{titling}
\allowdisplaybreaks

\theoremstyle{plain}

\theoremstyle{plain}

\theoremstyle{plain}
\newtheorem{assumption}{Assumption}[section]
\theoremstyle{plain}
\newtheorem{proposition}{Proposition}[section]
\theoremstyle{plain}
\newtheorem{corollary}{Corollary}[section]
\theoremstyle{remark}

\theoremstyle{definition}

\newcommand{\argmin}{\operatornamewithlimits{argmin\,}}

\newcommand{\was}{\mathcal{W}}
\newcommand{\pdist}{\mathcal{D}}
\mathchardef\mhyphen="2D

\floatstyle{ruled}
\newfloat{algorithm}{tbp}{loa}
\providecommand{\algorithmname}{Algorithm}
\floatname{algorithm}{\protect\algorithmname}

\title{Approximate Bayesian computation with the Wasserstein distance}
\author{Espen Bernton\thanks{Department of Statistics, Harvard University, USA.
Address correspondence to ebernton@g.harvard.edu.} , Pierre E. $\text{Jacob}^*$, Mathieu
Gerber\thanks{School of Mathematics, University of Bristol, UK.} , Christian P.
Robert\thanks{CEREMADE, Universit\'e Paris-Dauphine and Paris Sciences \& Lettres - PSL Research University, France, and
Department of Statistics, University of Warwick, UK.}}
\date{}

\begin{document}

\maketitle

\begin{abstract}
\noindent  {A growing number of generative statistical models do not permit the numerical evaluation of their likelihood functions.
Approximate Bayesian computation (ABC) has become a popular approach to overcome this
issue, in which one simulates synthetic data sets given parameters and compares summaries of
these data sets with the corresponding observed values. We propose to avoid the use of summaries and the ensuing loss of information by instead using the
Wasserstein distance between the empirical distributions of the observed and synthetic data. 
This generalizes the well-known approach of using order statistics within ABC to arbitrary dimensions.
We describe how recently developed approximations of the Wasserstein distance allow the method to scale to realistic data sizes, and propose
a new distance based on the Hilbert space-filling curve. We provide a theoretical study of the proposed method,
describing consistency as the threshold goes to zero while the observations are kept fixed, and concentration properties as the number of observations grows.
Various extensions to time series data are discussed.
The approach is illustrated on various examples,
including univariate and multivariate g-and-k distributions, a toggle switch model from systems biology, 
a queueing model, and a L\'evy-driven stochastic volatility model.}

\end{abstract}

% \tableofcontents

\section{Introduction \label{sec:introduction}}

The likelihood function plays a central role in modern statistics.  However, for many models of interest, the
likelihood cannot be numerically evaluated. It might still be possible to simulate
synthetic data sets from the model given parameters. A
popular approach to Bayesian inference in such generative models 
is approximate Bayesian computation
\citep[ABC,][]{beaumont2002ABC,marin2012approximate}. ABC constructs an approximation of the posterior distribution by simulating parameters and synthetic data sets, and retaining the parameters such that the associated data sets are similar enough to the observed data set. Measures of similarity between data sets
are often based on summary statistics, such as sample moments. In other words, data sets are 
considered close if some distance between their summaries is small.
The resulting ABC approximations have proven extremely useful, but can lead
to a systematic loss of information compared to the
original posterior distribution.

We propose here to instead view data sets as empirical distributions and to rely on the
Wasserstein distance between synthetic and observed data sets.  The
Wasserstein distance, also called the Gini, Mallows, or Kantorovich distance,
defines a metric on the space of probability distributions, and has become
increasingly popular in statistics and machine learning,
due to its appealing computational and statistical properties \citep[e.g.][]{cuturi2013sinkhorn,srivastava2015wasp,sommerfeld2016inference,panaretos2018statistical}. We will show 
that the resulting ABC posterior, which we term the Wasserstein ABC (WABC) posterior, can
approximate the posterior distribution arbitrarily well in the limit
of the threshold $\varepsilon$ going to zero, while bypassing the choice of
summaries. Furthermore, we derive asymptotic settings under which the WABC posterior behaves differently from the posterior, illustrating the potential impact of model misspecification and the effect of the dimension of the observation space, by providing upper bounds on
concentration rates as the number of observations goes to infinity.
 The WABC posterior is a particular case of coarsened posterior,
and our results are complementary to those of \citet{miller2015robust}.

We further develop two strategies to deal with the specific case of time series.  
The challenge is that the marginal empirical distributions of time series
might not contain enough information to identify all model parameters.
In the first approach, which we term curve matching,
each data point is augmented with the time at which it was observed. A new
ground metric is defined on this extended observation space, which in turn
allows for the definition of a Wasserstein distance between time series,
with connections to \citet{thorpe2017transportation}. A tuning parameter
$\lambda>0$ allows the proposed 
distance to approximate the Euclidean distance as $\lambda \to \infty$, 
and the Wasserstein distance between the marginal distributions as $\lambda \to 0$.
The second approach involves transforming the time series such that its
empirical distribution contains enough information for parameter estimation. 
We refer to such transformations as reconstructions and discuss 
delay reconstructions, as studied in dynamical systems \citep{stark2003},
and residual reconstructions, as already used in ABC settings \citep{mengersen2013bayesian}.

The calculation of Wasserstein distances is fast for empirical distributions in
one dimension, as the main computational task reduces to sorting. For multivariate data sets, we can leverage the rich literature
on the computation of Wasserstein distances and approximations thereof
\citep{peyre2018computational}. We also propose a new distance utilizing the idea of sorting, termed the
Hilbert distance, based on the Hilbert space-filling curve
\citep{sagan1994space,gerber2015sequential}. The proposed distance approximates the Wasserstein distance well in low dimensions, but can be
computed faster than the exact distance. We also shed light on some
theoretical properties of the resulting ABC posterior.

In the following subsections we set up the problem we consider in this work,
and  briefly  introduce ABC and   the Wasserstein distance; we refer to
\citet{marin2012approximate}  and to
\citet{villani2008} and \citet{santambrogio2015optimal} for more detailed presentations of
ABC and the Wasserstein distance, respectively.

\subsection{Set-up, notation and generative models \label{sec:generativemodels}}
Throughout this work we consider a probability space $(\Omega,\mathcal{F},\mathbb{P})$, with associated expectation operator $\mathbb{E}$, on which all the random variables are defined. The set of probability measures on a space $\mathcal{X}$ is denoted by $\mathcal{P}(\mathcal{X})$.   The data take values in $\mathcal{Y}$, a subset of $\mathbb{R}^{d_y}$ for $d_y\in\mathbb{N}$. We observe $n\in\mathbb{N}$ data points, $y_{1:n} = y_1,\ldots,y_n$, that are distributed according to $\mu_\star^{(n)}\in\mathcal{P}(\mathcal{Y}^{n})$. Let $\hat{\mu}_n = n^{-1}\sum_{i=1}^n \delta_{y_i}$, where $\delta_y$ is the Dirac distribution with mass on $y\in\mathcal{Y}$. With a slight abuse of language, we refer below to $\hat{\mu}_n$  as   the empirical distribution  of $y_{1:n}$, even in the presence of non i.i.d. observations.

Formally, a model refers to a collection of distributions on $\mathcal{Y}^n$,
denoted by $\mathcal{M}^{(n)}=\{\mu^{(n)}_\theta : \theta \in \mathcal{H}\} \subset
\mathcal{P}(\mathcal{Y}^n)$, where $\mathcal{H}\subset \mathbb{R}^{d_\theta}$ is the parameter space, endowed
with a distance $\rho_\mathcal{H}$ and of dimension $d_\theta\in \mathbb{N}$.  However, we will often assume that the sequence of models $(\mathcal{M}^{(n)})_{n\geq 1}$ is such that, for every $\theta\in\mathcal{H}$, the sequence $(\hat{\mu}_{\theta,n})_{n\geq 1}$ of random  probability measures on $\mathcal{Y}$ converges (in some sense) to a distribution $\mu_\theta\in\mathcal{P}(\mathcal{Y})$,  where $\hat{\mu}_{\theta,n}=n^{-1}\sum_{i=1}^n \delta_{z_i}$ with $z_{1:n}\sim \mu_\theta^{(n)}$. Similarly, we will often assume  that    $\hat{\mu}_n$ converges   to some distribution $\mu_\star\in  \mathcal{P}(\mathcal{Y})$ as $n\rightarrow \infty$. Whenever the notation $\mu_\star$ and $\mu_\theta$ is used, it is implicitly assumed that these objects exist. In such cases, we instead refer to $\mathcal{M} = \{\mu_\theta : \theta \in \mathcal{H}\} \subset \mathcal{P}(\mathcal{Y})$ as the model. 
We say that it is well-specified if there exists $\theta_\star\in\mathcal{H}$ such that $\mu_\star = \mu_{\theta_\star}$; otherwise it is misspecified. Parameters are identifiable if $\theta =\theta^\prime$ is implied by $\mu_\theta = \mu_{\theta^\prime}$.

We consider parameter inference for purely generative models: it is possible to generate
observations $z_{1:n}$ from $\mu_\theta^{(n)}$, for all
$\theta\in\mathcal{H}$, but it is not possible to numerically evaluate the
associated likelihood. In some cases, observations from the model are obtained as
$z_{1:n}=g_n(u, \theta)$, where $g_n$
is a known deterministic function and $u$ some known fixed-dimensional
random variable independent of $\theta$. Some methods
require access to $g_n$ and $u$ \citep{prangle2016rare,graham2017asymptotically};
by contrast, here we do not place assumptions on how data sets are generated from the model.

\subsection{Approximate Bayesian computation \label{sec:ABC}}

Let $\pi$ be a prior distribution  on
the parameter $\theta$. Consider the following algorithm, where $\varepsilon>0$ 
is referred to as the threshold, and $\pdist$ denotes
a discrepancy measure between two data sets $y_{1:n}$ and $z_{1:n}$, taking non-negative values.
\begin{enumerate}
    \item Draw a parameter $\theta$ from the prior distribution $\pi$, and a synthetic dataset $z_{1:n} \sim \mu_{\theta}^{(n)}$.
    \item If $\pdist(y_{1:n}, z_{1:n})\leq\varepsilon$, keep $\theta$, otherwise reject it.
\end{enumerate}
The accepted samples are drawn from the ABC posterior distribution 
\begin{equation}\label{eq:abcposterior}
    \pi^\varepsilon_{y_{1:n}}(d\theta)= \frac{\pi(d\theta)\int_{\mathcal{Y}^n}\mathds{1}\left(\pdist(y_{1:n},z_{1:n})\leq\varepsilon\right)\mu_\theta^{(n)}(dz_{1:n})}{\int_{\mathcal{H}}\pi(d\theta)\int_{\mathcal{Y}^n}\mathds{1}\left(\pdist(y_{1:n},z_{1:n})\leq\varepsilon\right)\mu_\theta^{(n)}(dz_{1:n})},
\end{equation}
where $\mathds{1}$ is the indicator function. A more sophisticated algorithm to approximate ABC posteriors, which we will apply in our numerical experiments, is described in Section \ref{sec:smcsamplers}.

Let $\rho$ be a distance on the observation space $\mathcal{Y}$, referred to as the ground distance.  Suppose that $\pdist$ is 
chosen as
\begin{equation} \label{eq:pdist_euclidean}
    \pdist(y_{1:n}, z_{1:n})^p = \frac{1}{n} \sum_{i=1}^n \rho(y_i, z_i)^p.
\end{equation}
Then, the resulting ABC posterior can be shown to have the desirable theoretical property of converging to the
standard posterior as $\varepsilon \to 0$  \citep[][see also Proposition \ref{prop:as_distn_fixedn}]{prangle2016rare}.
In the case where $p=2$, $\mathcal{Y} \subset \mathbb{R}$, and $\rho(y_i, z_i) = |y_i - z_i|$, $\pdist$ is a scaled version of the Euclidean distance between the vectors $y_{1:n}$ and $z_{1:n}$.

However, this approach is in most cases impractical due to the
large variation of $\pdist(y_{1:n},z_{1:n})$ over repeated samples from
$\mu_\theta^{(n)}$. A rare example of practical use of ABC with the Euclidean distance
is given in \citet{sousa2009approximate}.  A large proportion of the ABC literature is devoted to studying ABC posteriors
in the setting where $\pdist$ is the Euclidean distance between summaries, i.e. $\pdist(y_{1:n},z_{1:n}) = \|\eta(y_{1:n}) - \eta(z_{1:n})\|$,
where $\eta:\mathcal{Y}^n \to \mathbb{R}^{d_\eta}$ for some small $d_\eta$.
Using summaries can lead to a loss of information:
the resulting ABC posterior converges, at best, to the conditional distribution of $\theta$ given $\eta(y_{1:n})$,
as $\varepsilon \to 0$.
A trade-off ensues, where using more summaries reduces the information loss, 
but increases the variation in the distance over repeated model simulations \citep{fearnhead:prangle:2012}.

\subsection{Wasserstein distance \label{sec:wassersteindistance}}
A natural approach to reducing the variance of the distance defined in
\eqref{eq:pdist_euclidean}, while hoping to avoid the loss of information incurred
by the use of summary statistics, is to instead consider the distance
\begin{equation} \label{eq:wass_def_assignment}
    \was_p(y_{1:n}, z_{1:n})^p = \inf_{\sigma \in \mathcal{S}_n} \frac{1}{n} \sum_{i=1}^n \rho(y_i, z_{\sigma(i)})^p, 
\end{equation}
where $\mathcal{S}_n$ is the set of permutations of $\{1, \dots,n\}$.
Indeed, when the observations are univariate and $\rho(y_i,z_j) = |y_i - z_j|$, the above infimum is achieved by sorting 
$y_{1:n}$ and $z_{1:n}$ in increasing order and matching the order statistics.
Using order statistics as a choice of
summary within ABC has been suggested multiple times in the literature, see
e.g. \citet{sousa2009approximate,fearnhead:prangle:2012}. 
It turns out that $\was_p(y_{1:n}, z_{1:n})$ is the $p$-Wasserstein distance between the empirical
distributions supported on the data sets $y_{1:n}$ and $z_{1:n}$. 
From this perspective, our proposal of using the Wasserstein distance between empirical distributions
can be thought of as generalizing the use of order statistics within ABC to arbitrary dimensions.

More formally, let $\mathcal{P}_p(\mathcal{Y})$ with $p\geq1$ (e.g. $p=1$ or $2$) be the set of
distributions $\mu\in\mathcal{P}(\mathcal{Y})$
with finite $p$-th moment:
there exists $y_0\in\mathcal{Y}$ such that $\int_\mathcal{Y} \rho(y,y_0)^p d\mu(y) < \infty$.
The space $\mathcal{P}_p(\mathcal{Y})$ is referred to as the $p$-Wasserstein space of distributions on $\mathcal{Y}$ \citep{villani2008}. 
The $p$-Wasserstein distance is a finite metric on $\mathcal{P}_p(\mathcal{Y})$, defined by the transport problem
\begin{equation} \label{eq:wass_def} 
\was_p(\mu,\nu)^p = \inf_{\gamma \in \Gamma(\mu,\nu)} \int_{\mathcal{Y}\times \mathcal{Y}} \rho(x,y)^p d\gamma(x,y),
\end{equation}
where $\Gamma(\mu,\nu)$ is the set of probability measures on $\mathcal{Y}\times \mathcal{Y}$ with marginals $\mu$ and $\nu$ respectively;
see the notes in Chapter 6 of \citet{villani2008} for a brief history of this distance and its central role in optimal transport.

As in  \eqref{eq:wass_def_assignment}, we also write $\was_p(y_{1:n},z_{1:m})$ for $\was_p(\hat\mu_n,\hat\nu_m)$,
where $\hat\mu_n$ and $\hat\nu_m$ stand for the empirical distributions
$n^{-1}\sum_{i=1}^n \delta_{y_i}$ and $m^{-1}\sum_{i=1}^m \delta_{z_i}$.
In particular, the Wasserstein distance between two empirical distributions with 
unweighted atoms takes the form 
\begin{equation} \label{eq:wass_def_discrete}
    \was_p(y_{1:n}, z_{1:m})^p = \inf_{\gamma\in \Gamma_{n,m}} \sum_{i=1}^n \sum_{j=1}^m \rho(y_i, z_j)^p \gamma_{ij}
\end{equation}
where $\Gamma_{n,m}$ is the set of $n\times m$ matrices with non-negative
entries, columns summing to $m^{-1}$, and rows summing to $n^{-1}$. We focus on
the case $n=m$, for which it is known that the solution to the optimization problem,
$\gamma^\star$, corresponds to an assignment matrix with only one non-zero
entry per row and column, equal to $n^{-1}$ \citep[see e.g. the introductory chapter
in][]{villani2003topics}.  In this special case, the Wasserstein distance can thus
be represented as in  \eqref{eq:wass_def_assignment}.
Computing the Wasserstein distance between two samples of the same size can therefore also be
thought of as a matching problem; see Section
\ref{sec:distancecalculations}. 

\subsection{Related works and plan\label{subsect:related_works}}

The minimum Wasserstein estimator (MWE), first studied in
\citet{bassetti2006minimum}, is an example of a minimum distance estimator
\citep{basu2011statistical} and is defined as $\hat{\theta}_n = \argmin_{\theta
\in \mathcal{H}} \was_p(\hat{\mu}_n, \mu_\theta)$.  To extend this approach
to generative models,  \citet{bernton2017inference} introduce the minimum
expected Wasserstein estimator (MEWE), defined as $\hat{\theta}_{n,m} =
\argmin_{\theta \in \mathcal{H}} \mathbb{E}\left[\was_p(\hat{\mu}_n,
\hat\mu_{\theta,m})\right]$, where the expectation refers to  the
distribution of $z_{1:m}\sim \mu_\theta^{(m)}$.  General results on both
the MWE and MEWE are  obtained in the technical report of
\citet{bernton2017inference}.  Another method related to our approach was proposed by \citet{park2015k2}, who bypass the choice of
summary statistics in the definition of the ABC posterior    in
\eqref{eq:abcposterior} by using a discrepancy measure  $\pdist$   such that
$\pdist(y_{1:n},z_{1:n})$ is an estimate of  the maximum mean discrepancy
(MMD)  between $\hat{\mu}_n$ and  $\mu_\theta^{(n)}$.

Our contributions are structured as follows: the proposed approach to Bayesian inference 
in generative models using the Wasserstein distance is described in Section \ref{sec:inference}, 
some theoretical properties of the Wasserstein ABC posterior is detailed in Section \ref{sec:asymptoticproperties},
methods to handle time series are proposed in Section \ref{sec:timeseries},
and numerical illustrations in Section \ref{sec:numerics}, 
where in each example we make comparisons to existing methods, such as semi-automatic ABC \citep{fearnhead:prangle:2012}.
The supplementary material is available on \href{https://espenbernton.github.io}{espenbernton.github.io} and includes additional theoretical results and details on computational aspects, as referenced in the present article. Code to reproduce the numerical results is available at
\href{https://github.com/pierrejacob/winference}{github.com/pierrejacob/winference}.

\section{Wasserstein ABC  \label{sec:inference}}

The distribution $\pi^\varepsilon_{y_{1:n}}(d\theta)$ of
  \eqref{eq:abcposterior}, with  $\pdist$ replaced by $\was_p$ for some choice of $p\geq 1$,
is referred to as the Wasserstein ABC (WABC) posterior; that is, the  WABC  posterior is defined by
\begin{equation} \label{eq:Wabcposterior}
  \pi^\varepsilon_{y_{1:n}}(d\theta)= \frac{\pi(d\theta)\int_{\mathcal{Y}^n}\mathds{1}\left(\was_p(y_{1:n},z_{1:n})\leq\varepsilon\right)\mu_\theta^{(n)}(dz_{1:n})}{\int_{\mathcal{H}}\pi(d\theta)\int_{\mathcal{Y}^n}\mathds{1}\left(\was_p(y_{1:n},z_{1:n})\leq\varepsilon\right)\mu_\theta^{(n)}(dz_{1:n})}
\end{equation}
with $\was_p(y_{1:n},z_{1:n})$ defined in  \eqref{eq:wass_def_assignment}. Throughout the experiments of this article we set $p=1$,
which makes minimal assumptions on the existence of moments of the data-generating process.

As   mentioned in the introductory section, the motivation for choosing
$\pdist$ to be the Wasserstein distance is to have a discrepancy measure
$\pdist(y_{1:n},z_{1:n})$ that has both  a small variance and results in an ABC
posterior that has satisfactory theoretical properties. In particular, we show in
Section \ref{sec:asymptoticproperties} that,  as per ABC based on the Euclidean
distance, the WABC posterior converges to the true posterior distribution as
$\varepsilon\rightarrow 0$. In that section, we also provide a result showing
that, as $n\to\infty$ and the threshold $\varepsilon$ converges slowly enough
to some minimal value $\varepsilon_\star\geq 0$, the WABC posterior
concentrates around $\theta_\star:= \argmin_{\theta\in \mathcal{H}}
\was_p(\mu_\star,\mu_\theta)$.  In the well-specified case, $\theta_\star$
coincides with the data-generating parameter.  In the misspecified case,
$\theta_\star$ is typically different from where the actual posterior
concentrates, which is around the minimizer of $\theta\mapsto
\text{KL}(\mu_\star|\mu_\theta)$, where KL refers to the Kullback--Leibler
divergence. The experiments in Section \ref{sec:numerics}
contains examples where the WABC posterior provides a practical and accurate
approximation of the standard posterior, and examples where it does not, partly because
of the computational difficulty of sampling from the WABC posterior when $\varepsilon$ is small.

\subsection{Sampling sequentially from the WABC posterior\label{sec:smcsamplers}}

Instead of the rejection sampler of Section \ref{sec:ABC}, we will target the
WABC and other ABC posteriors using a sequential Monte Carlo
(SMC) approach, with $N$ particles exploring the parameter space
\citep{del2012adaptive}. The algorithm starts with a threshold $\varepsilon_0 = +\infty$,
for which the WABC posterior is the prior. 
Given the Monte Carlo approximation of the WABC posterior for $\varepsilon_{t-1}$,
the next value $\varepsilon_t$ is chosen so as to maintain a number
of unique particles of at least $\alpha N$, with $\alpha \in (0,1]$. Upon choosing $\varepsilon_t$, resampling and rejuvenation steps
are triggered and the algorithm proceeds. In the experiments, we will run
the algorithm until a fixed budget of model simulations is reached.
At the end of the run, the algorithm provides $N$ parameter samples 
and synthetic data sets, associated with a threshold $\varepsilon_T$.

The algorithm is parallelizable over the $N$ particles, and thus over 
equally many model simulations and distance calculations. Any choice of
MCMC kernel can be used within the rejuvenation steps. In particular, we use
the r-hit kernel of \citet{lee2012rhit}, shown to be advantageous compared to
standard ABC-MCMC kernels in \citet{lee2014variance}. We choose the number of
hits to be $2$ by default. For the proposals of the MCMC steps, we use a
mixture of multivariate Normal distributions, with $5$ components by default. 
We set $N$ to be $2,048$ and $\alpha$ to be $50\%$. These default tuning parameters are used throughout all the numerical experiments of Section \ref{sec:numerics}, unless otherwise specified.
Full details on the SMC algorithm are given in the supplementary material.

\subsection{Illustration on a Normal location model\label{sec:firstexample}}

Consider $100$ i.i.d. observations generated from a bivariate Normal distribution. The mean components are drawn from a standard
Normal distribution, and the generated values are approximately $-0.71$ and $0.09$. The covariance is equal to $1$ on the
diagonal and $0.5$ off the diagonal. The parameter $\theta$ is the mean vector, and is
assigned a centered Normal prior with variance $25$ on each component. 

\begin{figure}[h]
        \begin{subfigure}[t]{0.32\textwidth}
            \centering
            \includegraphics[width=\textwidth]{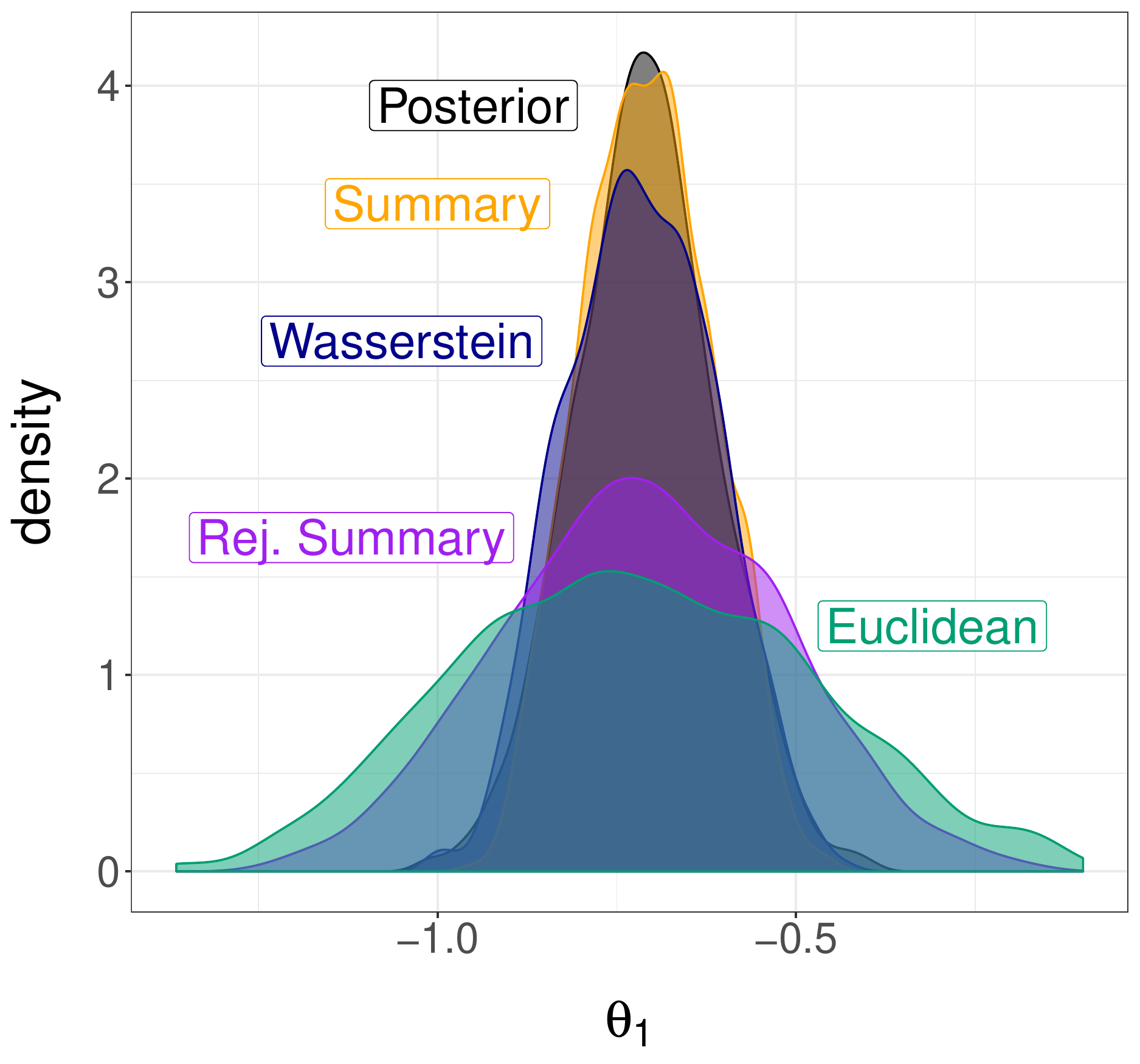}
            \caption{{\small Posteriors of $\theta_1$. }}  
            \label{fig:mvnorm:post1}
        \end{subfigure}
        %\hspace*{1cm}
        \begin{subfigure}[t]{0.32\textwidth}
            \centering
            \includegraphics[width=\textwidth]{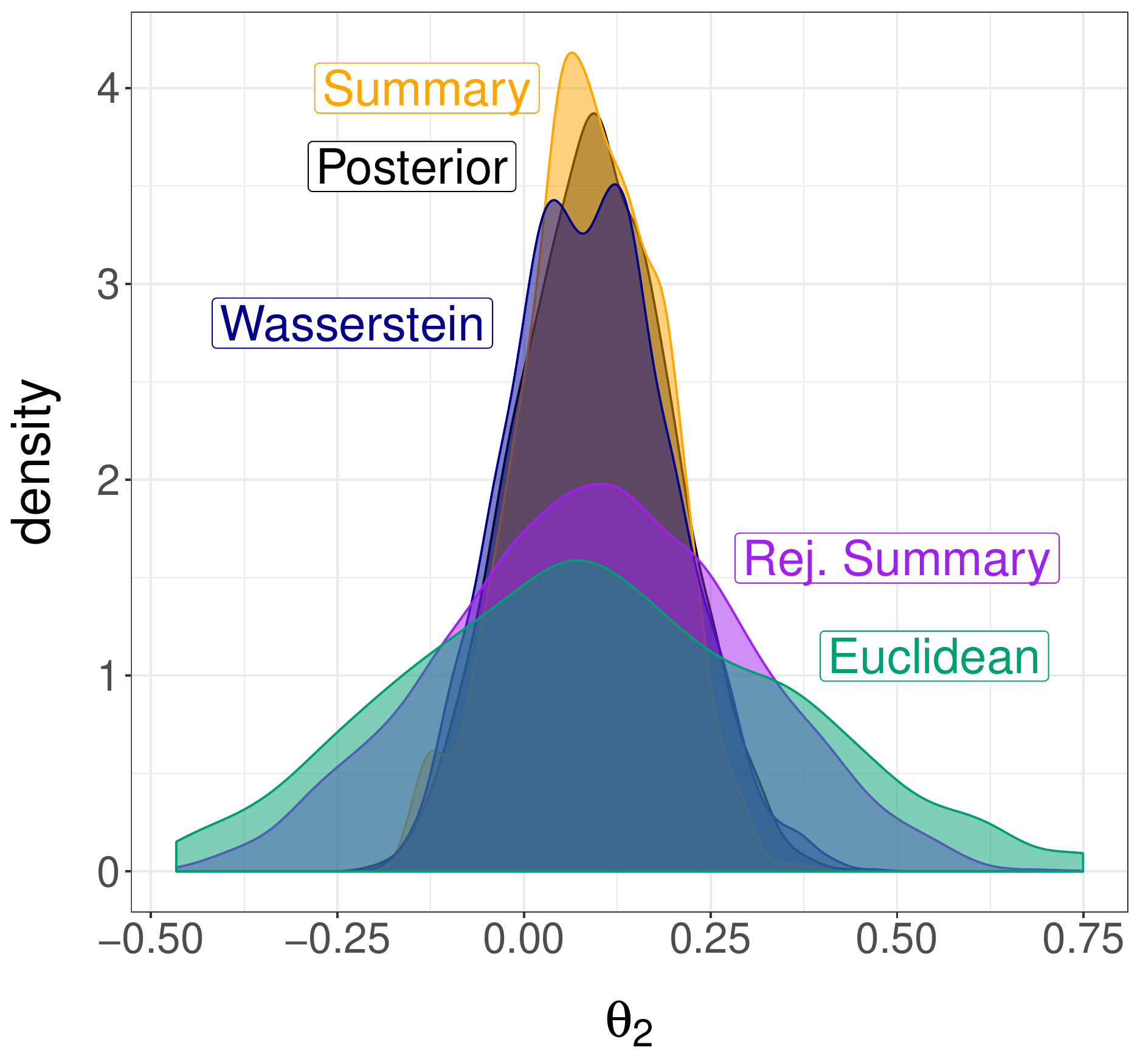}
            \caption{{\small Posteriors of $\theta_2$.}}  
            \label{fig:mvnorm:post2}
        \end{subfigure}
        %\hspace*{1cm}
        \begin{subfigure}[t]{0.32\textwidth}
            \centering
            \includegraphics[width=\textwidth]{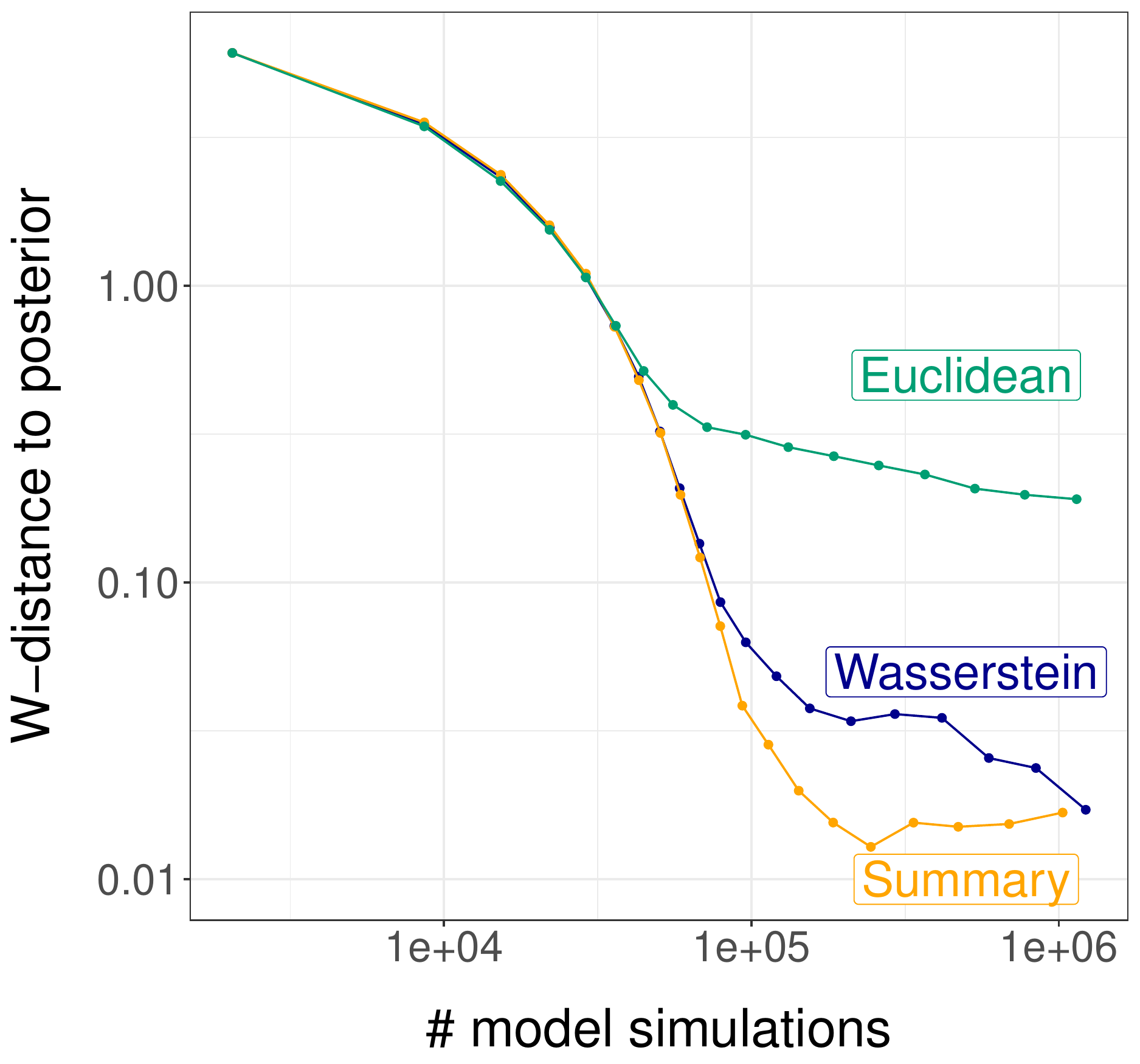}
            \caption{{\small $\was_1$-distance to posterior, versus number of model simulations (log-log scale).}}    
            \label{fig:mvnorm:distances}
        \end{subfigure}
        \caption{\small ABC in the bivariate Normal location model of Section \ref{sec:firstexample}. ABC approximations of the posterior after $10^6$ model simulations
        (left and middle), overlaid the actual posterior. On the right, the Wasserstein distance between ABC posterior samples and exact posterior samples is plotted
        against the number of model simulations (in log-log scale). In principle, these ABC approximations converge to the posterior as $\varepsilon\to 0$. Yet, for a given number of model simulations, the quality  
    of the ABC approximation is sensitive to the choice of distance and sampling algorithm.} 
        \label{fig:mvnorm}
\end{figure}

We compare WABC with two other methods: ABC using the
Euclidean distance between the data sets, and ABC using the Euclidean
distance between sample means, which for this model are sufficient summary statistics. 
All three ABC posteriors are approximated using the SMC sampler described in Section \ref{sec:smcsamplers}.
The summary-based ABC posterior is also approximated using the simple
rejection sampler given in Section \ref{sec:ABC} to illustrate the benefit of the SMC approach.
All methods are run for a budget of $10^6$ model simulations, using $N = 2,048$ particles in the SMC sampler. 
The rejection sampler accepted only the $2,048$ draws yielding the smallest distances. 
Approximations of the marginal posterior distributions of the parameters are given in Figures \ref{fig:mvnorm:post1} and \ref{fig:mvnorm:post2},
illustrating that the SMC-based ABC methods with the Wasserstein distance and with sufficient statistics both approximate the posterior accurately.

To quantify the difference between the obtained ABC samples and the posterior, we again use the Wasserstein distance. Specifically, we independently draw $2,048$ samples
from the posterior distribution, and compute the
Wasserstein distance between these samples and the $N = 2,048$ ABC samples produced by the SMC algorithm.  We plot the resulting distances against the number of model
simulations in Figure \ref{fig:mvnorm:distances}, in log-log scale.  As expected, ABC with
sufficient statistics converges fastest to the posterior. It should be noted that sufficient statistics are almost never available in realistic applications of ABC.  The proposed WABC
approach performs almost as well, but requires more model simulations to yield comparable results.
In contrast, the ABC approach with the Euclidean distance struggles to
approximate the posterior accurately. Extrapolating from the plot, it would
seemingly take billions of model simulations for the latter ABC approach to approximate the posterior as accurately as the other two
methods. Similarly, despite being based on the sufficient statistic, the rejection sampler does not adequately estimate the posterior distribution for the given sample budged. The estimated 1-Wasserstein distance between the $2,048$ accepted samples and the posterior was $0.63$.

In terms of computing time, based on our R implementation on an Intel Core i5
(2.5GHz), simulating a data set took on average $4.0\times 10^{-5}s$.
Computing the discrepancy between data sets took on average $6.4\times
10^{-5}s$ for the summary-based distance, $3.8\times 10^{-4}s$ for the
Euclidean distance, and $1.2\times 10^{-2}s$ for the Wasserstein distance; see
Section \ref{sec:distancecalculations} for fast approximations of the
Wasserstein distance. The SMC sampler is algorithmically more involved than the
rejection sampler, and one could ask whether the added computational
effort is justified.
In this example, the total time required by the SMC algorithm using
the summary statistic was $169s$, whereas the analogous rejection sampler took
$141s$. This illustrates that even when one can very cheaply simulate data and
compute distances, the added costs associated with an SMC sampler are relatively small;
see \citet{del2012adaptive,filippi2013optimality,sisson2018abc} for more details on SMC samplers for ABC purposes.

\subsection{Computing and approximating the Wasserstein distance \label{sec:distancecalculations}}

Computing the Wasserstein distance between the empirical distributions $\hat\mu_n =
n^{-1}\sum_{i=1}^n \delta_{y_i}$ and $\hat\nu_n = n^{-1}\sum_{i=1}^n
\delta_{z_i}$ reduces to a linear sum assignment problem, as in 
\eqref{eq:wass_def_assignment}.  In the univariate case, finding the optimal
permutation can be done by sorting the vectors $y_{1:n}$ and $z_{1:n}$ in
increasing order, obtaining the orders $\sigma_{y}(i)$ and $\sigma_{z}(i)$ for
$i\in\{1,\ldots,n\}$.  Then, one associates each $y_i$ with $z_{\sigma(i)}$
where $\sigma(i)= \sigma_{z}\circ \sigma^{-1}_{y}(i)$.  The cost of the
Wasserstein distance computation is thus of order $n\log n$ for distributions
on one-dimensional spaces.

In multivariate settings,  \eqref{eq:wass_def_assignment} can be solved by
the Hungarian algorithm for a cost of order $n^3$. Other algorithms have a cost
of order $n^{2.5}\log(n\, C_n)$, with $C_n=\max_{1\leq i,j\leq
n}\rho(y_i,z_j)$, and can therefore be more efficient when $C_n$ is small
\citep[Section 4.1.3]{assignmentproblems}.  In our numerical experiments, we
use the short-list method presented in \citet{gottschlich2014shortlist} and
implemented in \citet{transportpackage}. This simplex algorithm-derived method
comes without guarantees of polynomial running times, but
\citet{gottschlich2014shortlist} show empirically that their method tends to
have sub-cubic cost.

The cubic cost of computing Wasserstein distances in the multivariate setting
can be prohibitive for large data sets. However, many applications of ABC
involve relatively small numbers of observations from complex models which are
expensive to simulate. In these settings, the cost of simulating synthetic data sets might
dominate the model-free cost of computing distances.  
Note also that the dimension $d_y$ of the observation space only enters
the ground distance $\rho$, and thus the cost of computing the Wasserstein
distance under a Euclidean ground metric is linear in $d_y$.

\subsubsection{Fast approximations \label{sec:otherdistances}}

In conjunction with its increasing popularity as a tool for inference in
statistics and machine learning, there has been a rapid growth in the number of
algorithms that approximate the Wasserstein distance at reduced computational
costs; see \citet{peyre2018computational}.  In particular, they provide an in-depth
discussion of the popular method proposed by \citet{cuturi2013sinkhorn}, in
which the optimization problem in  \eqref{eq:wass_def_discrete} is regularized
using an entropic constraint on the joint distribution $\gamma$.  Consider 
$ \gamma^\zeta = \argmin_{\gamma \in \Gamma_n} \sum_{i,j=1}^n \rho(y_i, z_j)^p \gamma_{ij} 
    + \zeta \sum_{i,j=1}^n \gamma_{ij} \log \gamma_{ij},$
which includes a negative penalty on the entropy of $\gamma$,
and define the dual-Sinkhorn divergence 
$S^\zeta_p(y_{1:n},z_{1:n})^p =  \sum_{i,j=1}^n \rho(y_i, z_j)^p \gamma^\zeta_{ij}$.
The regularized problem can be solved iteratively by Sinkhorn's algorithm, which involves
matrix-vector multiplications resulting in a cost of order $n^2$ per iteration.
If $\zeta\to0$, the dual-Sinkhorn divergence converges to the Wasserstein distance, whereas if $\zeta \to \infty$ it converges to the maximum mean discrepancy \citep{ramdas2017wasserstein}.
It can therefore be seen as an interpolation between optimal transport and kernel-based distances.
Further properties of the dual-Sinkhorn divergence and other algorithms to approximate it are discussed in \citet{peyre2018computational}. 

Unlike the optimal coupling that yields the exact Wasserstein distance, the coupling obtained in the regularized problem is typically not an assignment matrix. In the following subsections, we discuss two simple approaches with different computational complexities that yield couplings that are assignments. This has the benefit of aiding the theoretical analysis in Section \ref{sec:asymptoticproperties}.

\subsubsection{Hilbert distance \label{sec:hilbert}} 

The assignment problem in  \eqref{eq:wass_def_assignment}
can be solved in $n\log n$ in the univariate case by sorting the samples. We propose
a new distance generalizing this idea when $d_y>1$, by sorting samples according to their
projection via 
the Hilbert space-filling curve.
As shown in \citet{gerber2015sequential} and \cite{Schretter2016}, transformations
through the Hilbert space-filling curve and its inverse  preserve a notion of
distance between probability measures.  The Hilbert curve $H:[0,1]\to[0,1]^{d_y}$ is a H\"older continuous mapping
from $[0,1]$ into $[0,1]^{d_y}$. One can define a measurable
pseudo-inverse $h:[0,1]^{d_y}\to [0,1]$ verifying $h(H(x))=x$ for all $x\in
[0,1]$ \citep{Resampling}.  We assume in this subsection that
$\mathcal{Y}\subset\mathbb{R}^{d_y}$ is such that there exists a mapping
$\psi:\mathcal{Y}\rightarrow (0,1)^{d_y}$ verifying, for
$y=(y_1,\dots,y_{d_y})\in\mathcal{Y}$,
$
\psi(y)=\left(\psi_1(y_1),...,\psi_{d_y}(y_{d_y})\right)
$
 where the $\psi_i$'s are continuous and strictly monotone. For instance, if
$\mathcal{Y}=\mathbb{R}^{d_y}$,  one can take $\psi$ to be the component-wise logistic transformation; see \citet{gerber2015sequential}
for more details. By construction, the mapping
$h_\mathcal{Y}:=h\circ\psi:\mathcal{Y}\rightarrow (0,1)$ is one-to-one.
For two vectors $y_{1:n}$ and $z_{1:n}$, denote 
by $\sigma_y$ and $\sigma_z$ the permutations obtained
by mapping the vectors through $h_\mathcal{Y}$ and sorting the resulting univariate vectors
in increasing order.
We define the Hilbert distance $\mathfrak{H}_p$
between the empirical distributions of $y_{1:n}$ and $z_{1:n}$ by
\begin{equation}
    \label{eq:hilbertdistance}
    \mathfrak{H}_p(y_{1:n}, z_{1:n})^p = \frac{1}{n}\sum_{i=1}^n \rho\big(y_{i}, z_{\sigma(i)} \big)^p,
\end{equation}
where $\sigma(i)= \sigma_{z} \circ \sigma^{-1}_{y}(i)$ for all $i\in \{1,\ldots,n\}$.

\begin{proposition} \label{prop:hilbert}
For any integer $n\geq 1$ and real
number $p\geq 1$, $\mathfrak{H}_p$ defines a distance on the space of empirical
distributions of size $n$.
\end{proposition}

The Hilbert distance can be computed at a cost in the order of $n \log n$   and
an implementation is provided by the function \texttt{hilbert\_sort} in
\citet{cgal:eb-16a}. From a practical point of view, this implementation has the
attractive property of not having to map the samples to $(0,1)^{d_y}$ and
hence having to choose a specific mapping $\psi$. Instead,  this function directly
constructs the Hilbert curve around the input point set. 

Despite not being defined in terms of a transport problem, the Hilbert distance
yields approximations of the Wasserstein distance that are accurate for small
$d_y$, as illustrated in the supplementary material.  More  importantly for
its use within ABC, the level sets of the (random) map $\theta \mapsto
\mathfrak{H}_p(\hat{\mu}_n, \hat{\mu}_{\theta,n})$ appear to be close to those
of the analogous Wasserstein distance. The two distances therefore discriminate
between parameters in similar fashions.  However, this behavior tends to
deteriorate as the dimension $d_y$ grows.
 
The coupling produced by Hilbert sorting is feasible for the assignment problem
in  \eqref{eq:wass_def_assignment}. Therefore, it is always greater than the
Wasserstein distance, which minimizes the objective therein. This property
plays an important role in showing that the ABC posterior based on the Hilbert
distance concentrates on $\theta_\star$ as $n\rightarrow\infty$ and the
threshold $\varepsilon$ decreases sufficiently slowly. In the supplementary
materials, we provide such a result under the assumption that the model is
well-specified, but leave further theoretical analysis under milder conditions
for future research.  Other one-dimensional projections of
multivariate samples, followed by Wasserstein distance computation using the projected samples,
have been proposed in the computational optimal transport literature \citep{rabin2011wasserstein,bonneel2015sliced}, 
also leading to computational costs in $n\log n$.

\subsubsection{Swapping distance \label{sec:swapping}}

Viewing the Wasserstein distance calculation as the assignment problem 
in  \eqref{eq:wass_def_assignment}, 
\citet{puccetti2017algorithm}  proposed a greedy swapping algorithm
to approximate the optimal assignment. 
Consider an arbitrary permutation $\sigma$ of $\{1,\ldots,n\}$,
and the associated transport cost 
$\sum_{i=1}^n \rho(y_i, z_{\sigma(i)})^p$. The swapping algorithm consists in 
checking, for all $1\leq i < j \leq n$, whether 
$\rho(y_i, z_{\sigma(i)})^p + \rho(y_j, z_{\sigma(j)})^p$
is less or greater than 
$\rho(y_i, z_{\sigma(j)})^p + \rho(y_j, z_{\sigma(i)})^p$. 
If it is greater, then one swaps $\sigma(i)$ and $\sigma(j)$,
resulting in a decrease of the transport cost. One can repeat these sweeps
over $1\leq i < j \leq n$, until the assignment is left unchanged,
and denote it by $\tilde{\sigma}$.
Each sweep has a cost of order $n^2$ operations.
There is no guarantee that the resulting assignment $\tilde{\sigma}$ corresponds to the optimal one.
Note that we initialize the algorithm with the assignment obtained
by Hilbert sorting for a negligible cost of $n \log n$. We refer to the resulting
distance $(n^{-1}\sum_{i=1}^n \rho(y_i, z_{\tilde{\sigma}(i)})^p)^{1/p}$ as
the swapping distance. 

The swapping distance between $y_{1:n}$ and $z_{1:n}$ takes values that are, by construction, between the Wasserstein distance $\was_p(y_{1:n}, z_{1:n})$ 
and the Hilbert distance $\mathfrak{H}_p(y_{1:n}, z_{1:n})$.  Thanks to this property, 
we show in the supplementary material that the associated ABC posterior  concentrates on $\theta_\star$ as $n\rightarrow\infty$ and the threshold $\varepsilon$ decreases sufficiently slowly. As with the Hilbert distance, this result is obtained under
the assumption that the model is well-specified and leave further theoretical analysis 
under milder conditions for future research. In the supplementary material, we also observe that the swapping
distance can approximate the Wasserstein distance more accurately than the Hilbert distance as the dimension $d_y$ grows.

\subsubsection{Sub-sampling} \label{sec:subsampling}
Any of the aforementioned distances can be computed faster by 
first sub-sampling $m < n$ points from $y_{1:n}$ and $z_{1:n}$, and then computing the distance between the 
resulting distributions.  This increases the variance of the calculated distances, introducing a trade-off with computation time. 
In the case of the Wasserstein distance, this approach could be studied formally using the results of \citet{sommerfeld2016inference}.
Other multiscale approaches can also be used to accelerate computation \citep{merigot2011multiscale}. We remark that computing the distance between vectors  containing subsets of order statistics \citep{fearnhead:prangle:2012} can be viewed as an example of a multiscale approach to approximating the Wasserstein distance.

\subsubsection{Combining distances}
It might be useful to combine distances. For instance, 
one might want to start exploring the parameter space with a cheap approximation,
and switch to the exact Wasserstein distance in a region of interest;
or use the cheap approximation to save computations in a delayed acceptance scheme. One might also
combine a transport distance with a distance between summaries.
We can combine distances in the ABC framework
by introducing a threshold for each distance, and define the ABC 
posterior as in  \eqref{eq:abcposterior},  with a product of indicators corresponding to each distance.
We explore the combination of distances in the numerical experiments of Section \ref{sec:levydriven}.

\section{Theoretical properties \label{sec:asymptoticproperties}}

We study the behavior of the Wasserstein ABC posterior under different asymptotic regimes.
First, we give conditions on a discrepancy measure for the associated ABC
posterior to converge to the posterior as the threshold $\varepsilon$ goes
to zero, while keeping the observed data fixed. We then
discuss the behavior of the WABC posterior as $n\to \infty$ for fixed
$\varepsilon>0$.  Finally, we establish bounds on the rates of concentration of
the WABC posterior as the data size $n$ grows and the threshold
$\varepsilon$ shrinks sufficiently slowly at a rate dependent on $n$, similar to \citet{frazier2016} in the case of summary-based ABC. Proofs are deferred to the appendix. 

We remark that the assumptions underlying our results are typically hard to check in practice,
due to the complexity and intractable likelihoods of the models to which ABC methods are applied.
This is also true for state-of-the-art asymptotic results about 
summary-based ABC methods, which, for example, require injectivity and growth conditions on the ``binding function'' to which 
the summary statistics converge \citep{frazier2016}. Nonetheless, we believe that our results provide insight
into the statistical properties of the WABC posterior. For instance, in Corollary \ref{cor:concentration} we give conditions
under which the WABC posterior concentrates around $\theta_\star = \argmin_{\theta \in \mathcal{H}}
\was_p(\mu_\theta,\mu_\star)$ as $n$ grows. When the model is misspecified, this is in contrast with the posterior, which is known to 
concentrate around $\argmin_{\theta \in \mathcal{H}}\text{KL}(\mu_\theta,\mu_\star)$ \citep[see e.g.][]{muller2013risk}.

\subsection{Behavior as $\varepsilon \to  0$ for fixed observations \label{sec:asympotics:threshold}}

The following result establishes conditions under which a non-negative
measure of discrepancy between data sets $\mathfrak{D}$ yields an ABC posterior
that converges to the true posterior as $\varepsilon \to  0$, while the
observations are kept fixed.
\begin{proposition}\label{prop:as_distn_fixedn}
    Suppose that $\mu_\theta^{(n)}$ has a continuous density $f_\theta^{(n)}$ and that 
    \[\sup_{\theta\in \mathcal{H}\setminus\mathcal{N}_\mathcal{H}} f_\theta^{(n)}(y_{1:n})<\infty,\]
   where $\mathcal{N}_{\mathcal{H}}$ is a set such that $\pi(\mathcal{N}_{\mathcal{H}}) = 0$. Suppose that 
    there exists $\bar{\varepsilon} > 0$ such that 
$$\sup_{\theta\in\mathcal{H}\setminus \mathcal{N}_{\mathcal{H}}}\sup_{z_{1:n}\in \mathcal{A}^{\bar{\varepsilon}}} f_\theta^{(n)}(z_{1:n}) < \infty,$$
where $\mathcal{A}^{\bar{\varepsilon}} = \{z_{1:n} : \pdist(y_{1:n},z_{1:n}) \leq \bar{\varepsilon}\}$. Suppose also that $\pdist$ is continuous in the sense that $\pdist(y_{1:n},z_{1:n})\to \pdist(y_{1:n},x_{1:n})$ whenever $z_{1:n} \to x_{1:n}$ component-wise in the metric $\rho$. If either
\begin{enumerate}
    \item $f_\theta^{(n)}$ is $n$-exchangeable, such that $f_\theta^{(n)}(y_{1:n}) = f_\theta^{(n)}(y_{\sigma(1:n)})$ for any $\sigma \in \mathcal{S}_n$, and $\mathfrak{D}(y_{1:n},z_{1:n}) = 0$ if and only if $z_{1:n} = y_{\sigma(1:n)}$ for some $\sigma\in \mathcal{S}_n$, or \label{cond:exch}
\item $\mathfrak{D}(y_{1:n},z_{1:n}) = 0$ if and only if $z_{1:n} = y_{1:n}$, \label{cond:exact}
\end{enumerate}
then, keeping $y_{1:n}$ fixed, the ABC posterior converges strongly to the posterior as $\varepsilon\to 0$.
\end{proposition}

The Wasserstein distance applied to unmodified data satisfies
$\mathfrak{W}(y_{1:n},z_{1:n}) = 0$ if and only if $z_{1:n} = y_{\sigma(1:n)}$
for some $\sigma\in\mathcal{S}_n$, making condition \eqref{cond:exch} of Proposition
\ref{prop:as_distn_fixedn} applicable. In Section \ref{sec:timeseries}, we will discuss two methods applicable to time series that lead to discrepancies 
for which condition \eqref{cond:exact} holds.
Note that this result does not guarantee that the Monte Carlo algorithm employed 
to sample the ABC posterior distribution, with an adaptive mechanism to decrease the threshold, 
will be successful at reaching low thresholds in a reasonable time.

\subsection{Behavior as $n \to \infty$ for fixed $\varepsilon$ \label{sec:cposterior}} 
Under weak conditions, the WABC posterior  $\pi^\varepsilon_{y_{1:n}}(d\theta)$
in  \eqref{eq:abcposterior} converges to $\pi(d\theta| \was_p(\mu_\theta, \mu_\star) < \varepsilon)$
as $n\to \infty$ for a fixed threshold $\varepsilon$, following the reasoning in \citet{miller2015robust}
for general weakly-continuous distances, which include the Wasserstein distance.
Therefore, the WABC distribution with a fixed $\varepsilon$ does not converge to a Dirac mass, contrarily to the standard posterior. As argued
in \citet{miller2015robust}, this can have some benefit in case of model misspecification: the WABC posterior is less sensitive
to perturbations of the data-generating process than the standard posterior.

\subsection{Concentration as $n$ increases and $\varepsilon$ decreases \label{sec:asymptotics:double}}

A sequence of distributions $\pi_{y_{1:n}}$ on $\mathcal{H}$, depending on the data $y_{1:n}$,
is consistent at $\theta_\star$ if, for any $\delta > 0$,
$\mathbb{E}[\pi_{y_{1:n}}\left(\{\theta\in\mathcal{H}
:\rho_\mathcal{H}(\theta,\theta_\star) > \delta\}\right)]  \to  0$,
where the expectation is taken with respect to $\mu_\star^{(n)}$. Finding rates
of concentration for $\pi_{y_{1:n}}$ involves finding the fastest decaying
sequence $\delta_n>0$ such that the limit above holds. More precisely, we say
that the rate of concentration of $\pi_{y_{1:n}}$ is bounded above by the sequence $\delta_n$ if $\mathbb{E}[\pi_{y_{1:n}}\left(\{\theta\in\mathcal{H}
:\rho_\mathcal{H}(\theta,\theta_\star) > \delta_n\}\right)] \to  0$.

We establish upper bounds on the rates of concentration of
the sequence of WABC posteriors around $\theta_\star = \argmin_{\theta \in
\mathcal{H}} \was_p(\mu_\theta,\mu_\star)$, as the data size $n$ grows and
the threshold shrinks slowly  towards $\varepsilon_\star =
\was_p(\mu_{\theta_\star},\mu_\star)$ at a rate dependent on $n$. Although we
focus on the Wasserstein distance in this section, the reasoning also holds for other
metrics on $\mathcal{P}(\mathcal{Y})$; see Section \ref{sec:distancecalculations}
and the supplementary material.

Our first assumption is on the convergence of the empirical distribution of the data.
\begin{assumption}\label{as:cvgwas}
   The data-generating process is such that $\was_p(\hat{\mu}_n, \mu_\star) \to 0$, in $\mathbb{P}$-probability, as $n\to \infty$.
\end{assumption}
In the supplementary material, we derive a few different conditions under which Assumption 1 holds for i.i.d. data and certain classes of dependent processes.
Additionally, the moment and concentration inequalities of \citet{fournier_guillin2015, weed2017sharp} can also be used to verify both this and the next assumption.
\begin{assumption}\label{as:concentration}
For any $\varepsilon>0$,
$\mu^{(n)}_\theta(\was_p(\mu_\theta,\hat{\mu}_{\theta,n}) > \varepsilon) \leq
c(\theta)f_n(\varepsilon)$, where $f_n(\varepsilon)$ is a sequence of functions
that are strictly decreasing in $\varepsilon$ for fixed $n$ and $f_n(\varepsilon)
\to 0$ for fixed $\varepsilon$ as $n\to \infty$. The function $c:\mathcal{H}\to\mathbb{R}^+$ is
$\pi$-integrable, and satisfies $c(\theta) \leq c_0$ for some $c_0>0$, for all $\theta$ such that, 
for some $\delta_0>0$, 
$\was_p(\mu_{\star},\mu_\theta) \leq \delta_0 + \varepsilon_\star$.
\end{assumption}
For well-specified models, note that Assumption \ref{as:concentration} implies Assumption \ref{as:cvgwas}.
The next assumption states that the prior distribution puts enough mass on the sets of parameters $\theta$ that yield distributions $\mu_\theta$ close to $\mu_\star$ in the Wasserstein distance.
\begin{assumption}\label{as:prior}
There exist $L>0$ and $c_\pi >0$ such that, for all $\varepsilon$ small enough,
$$\pi\left(\{\theta\in\mathcal{H} : \mathfrak{W}_p(\mu_\star, \mu_{\theta})  \leq \varepsilon +\varepsilon_\star\}\right) \geq c_\pi \varepsilon^L.$$
\end{assumption}

The main result of this subsection is on the concentration of the WABC posteriors on the aforementioned sets.

\begin{proposition}\label{theorem:concentration}
    Under Assumptions \ref{as:cvgwas}-\ref{as:prior}, consider a sequence $(\varepsilon_n)_{n\geq 0}$ 
such that, as $n\to \infty$, $\varepsilon_n \to 0$, $f_n(\varepsilon_n) \to 0$, and $\mathbb{P}(\was_p(\hat{\mu}_n,\mu_\star) \leq \varepsilon_n)
\to 1$.
Then, the WABC posterior with threshold $\varepsilon_n + \varepsilon_\star$ satisfies, for some $0 < C < \infty$ and any $0 < R < \infty$,
$$\pi^{\varepsilon_n+\varepsilon_\star}_{y_{1:n}}\left(\{\theta\in\mathcal{H}: \mathfrak{W}_p(\mu_\star, \mu_{\theta}) > \varepsilon_\star + 4\varepsilon_n/3 + f_n^{-1}(\varepsilon_n^L/R)\}\right) \leq \frac{C}{R},$$
with $\mathbb{P}$-probability going to $1$ as $n\to \infty$.
\end{proposition}

The assumptions that $f_n(\varepsilon_n) \to 0$ and that $\mathbb{P}(\was_p(\hat{\mu}_n,\mu_\star) \leq \varepsilon_n) \to 1$ imply that $\varepsilon_n$
has to be the slowest of the two convergence rates: that of $\hat\mu_n$ to $\mu_\star$ and that of $\hat\mu_{\theta,n}$ to $\mu_\theta$.  
We can further relate concentration on the sets $\{\theta:
\was_p(\mu_\theta,\mu_\star) < \delta' + \varepsilon_\star\}$, for some $\delta' >0$,
to concentration on the sets $\{\theta: \rho_{\mathcal{H}}(\theta,\theta_\star) < \delta\}$,
for some $\delta>0$, assuming the parameter ${\theta_\star} = \argmin_{\theta\in\mathcal{H}}
\was_p(\mu_\star,\mu_\theta)$ is well-defined.
In turn, this leads to concentration rates of the WABC posteriors.
To that end, consider the following assumptions.

\begin{assumption}\label{as:wellsep}
The parameter ${\theta_\star} = \argmin_{\theta\in\mathcal{H}} \was_p(\mu_\star,\mu_\theta)$ exists, and is well-separated in the sense that, for all $\delta >0$, there exists $\delta'>0$ such that
$$\inf_{\{\theta\in\mathcal{H}: \rho_{\mathcal{H}}(\theta,\theta_\star) > \delta\}} \was_p(\mu_\theta,\mu_\star) > \was_p(\mu_{\theta_\star},\mu_\star) + \delta'.$$
\end{assumption}

This assumption is akin to those made in the
study of the asymptotic properties of the maximum likelihood estimator under
misspecification, where $\theta_\star$ is defined in terms of the Kullback--Leibler
divergence. In the supplementary material, we give a proposition establishing conditions under which Assumption 4 holds.

Under Assumption \ref{as:wellsep}, note that the last part of Assumption
\ref{as:concentration} is implied by $c(\theta) \leq c_0$ for all $\theta$
with $\was_p(\mu_{\theta_\star},\mu_\theta) \leq \delta_0$, for some
$\delta_0>0$. Indeed, $\was_p(\mu_{\theta_\star},\mu_\theta) \leq \delta_0$
implies that $\was_p(\mu_\theta,\mu_{\star}) -  \was_p(\mu_\star,
\mu_{\theta_\star}) \leq \delta_0$. Since $\varepsilon_\star =
\was_p(\mu_\star, \mu_{\theta_\star})$, the argument follows. By the same
reasoning, Assumption \ref{as:prior} is implied by
$\pi\left(\{\theta\in\mathcal{H}  :\mathfrak{W}_p(\mu_{\theta_\star},
\mu_{\theta})  \leq \varepsilon\}\right) \geq c_\pi \varepsilon^L$,
for some $c_\pi >0$ and $L>0$.

\begin{assumption} \label{as:functional}
The parameters are identifiable, and there exist $K>0$, $\alpha > 0$ and an open neighborhood $U\subset \mathcal{H}$ of $\theta_\star$, such that, for all $\theta\in U$, 
\[\rho_\mathcal{H}(\theta,\theta_\star) \leq K(\was_p(\mu_\theta,\mu_\star) - \varepsilon_\star)^\alpha.\]
\end{assumption}

\begin{corollary}\label{cor:concentration}
Under Assumptions \ref{as:cvgwas}-\ref{as:functional},
consider a sequence $(\varepsilon_n)_{n\geq 0}$ 
such that, as $n\to \infty$, $\varepsilon_n \to 0$, $f_n(\varepsilon_n) \to 0$, $f_n^{-1}(\varepsilon_n^{L}) \to 0$  and $\mathbb{P}(\was_p(\hat{\mu}_n,\mu_\star) \leq \varepsilon_n )
\to 1$.
Then the WABC posterior with threshold $\varepsilon_n + \varepsilon_\star$ satisfies, for some $0 < C < \infty$ and any $0 < R < \infty$,
$$\pi^{\varepsilon_n+\varepsilon_\star}_{y_{1:n}}\left(\{\theta\in\mathcal{H}: \rho_\mathcal{H}(\theta,\theta_\star) > K(4\varepsilon_n/3 + f_n^{-1}(\varepsilon_n^L/R))^{\alpha}\}\right) \leq \frac{C}{R},$$
with $\mathbb{P}$-probability going to $1$.
\end{corollary}

This result bounds the concentration rate from above through the expression
$\delta_n = K(4\varepsilon_n/3 + f_n^{-1}(\varepsilon_n^L/R))^{\alpha}$, but we
remark that it is not clear whether this bound is optimal in any sense. Explicit
upper bounds for certain classes of models and data-generating
processes, such as location-scale models and the AR(1) model in Section
\ref{example:ar}, are given in the supplementary material. 
Important aspects of the method that appear in these bounds 
include the dimension of the observation space $\mathcal{Y}$, 
the order $p$ of the Wasserstein distance, and model misspecification, through the exponent $\alpha$ in Assumption \ref{as:functional}.

The result provides some insight into the behavior of the method when $\varepsilon_n$ converges \textit{slowly} to $\varepsilon_\star$.
However, it is unclear what happens when $\varepsilon_n$ decays to a value
smaller than $\varepsilon_\star$ at a rate faster than that prescribed by
Corollary \ref{cor:concentration}. As shown in Proposition
\ref{prop:as_distn_fixedn}, the WABC posterior converges to
the true posterior when $\varepsilon \to 0$ for fixed observations. The
posterior itself is known to concentrate around the point in $\mathcal{H}$
minimizing the KL divergence between $\mu_\star$ and $\mu_\theta$ when $n \to \infty$ \citep[see e.g.][]{muller2013risk}, and it might be that the WABC posterior inherits similar properties for faster decaying thresholds. 

In high dimensions, the rate of convergence of the Wasserstein distance between empirical measures is known to be slow \citep{talagrand1994transportation}. On the other hand,  recent results establish that it concentrates quickly around its expectation: For instance, \citet{del2017central} show that regardless of dimension, $\was_2^2(\hat{\mu}_n,\hat{\mu}_{\theta,n}) - \mathbb{E}\was_2^2(\hat{\mu}_n,\hat{\mu}_{\theta,n})$ converges weakly at the $\sqrt{n}$ rate to a centered Gaussian random variable with known (finite) variance $\sigma^2(\mu_\star,\mu_\theta)$. If the map $\theta \mapsto \mathbb{E}\was_2^2(\hat{\mu}_n,\hat{\mu}_{\theta,n})$ offers discrimination between the parameters that is similar to $\theta \mapsto \was_2^2(\mu_\star,\mu_\theta)$, it is not clear how the Wasserstein distance's convergence rate would impact the WABC posterior. Detailed analysis of WABC's dependence on dimension is an interesting avenue of future research.

\section{Time series \label{sec:timeseries}}

Viewing data sets as empirical distributions requires
some additional care in the case of dependent data, 
which are common in settings where ABC methods are applied.
A na{\"\i}ve approach consists in ignoring dependencies,
which  might be enough to estimate all parameters
in some cases, as illustrated in Section \ref{sec:queue}.
However, in general, ignoring dependencies might prevent some parameters from being identifiable, 
as illustrated in the examples of this section. We propose two main approaches to extend the WABC methodology to time series.

\subsection{Curve matching \label{sec:curvematching}}

Visually, we might consider two time series to be similar if their curves are similar, 
in a trace plot of the series in the vertical axis against the time indices on the horizontal axis.
The Euclidean vector distance between curves sums the vertical differences between pairs of points with identical
time indices. We can instead introduce the points $\tilde{y}_t = (t,y_t)$ and $\tilde{z}_t = (t,z_t)$
for all $t \in 1:n$, viewing the trace plot as a scatter plot.
The distance between two points, $(t,y_t)$ and $(s,z_s)$, can be measured by 
a weighted distance $\rho_\lambda \left( (t,y_t), (s,z_s) \right) =  \|y_t - z_s\| + \lambda |t-s|$, where $\lambda$
is a non-negative weight, and $\|y - z\|$ refers to the Euclidean distance between $y$ and $z$.
Intuitively, the distance $\rho_\lambda$ takes into account both vertical and horizontal differences between points
of the curves, $\lambda$ tuning the importance of horizontal differences relative to vertical differences.
We can then define the Wasserstein distance between two empirical measures supported by $\tilde{y}_{1:n}$ and $\tilde{z}_{1:n}$,
with $\rho_\lambda$ as a ground distance on the observation space $\{1,\ldots,n\}\times\mathcal{Y}$.
Since computing the Wasserstein distance can be thought of as solving
an assignment problem, a large value of $\lambda$ implies that $y_t$
will be assigned to $z_t$, for all $t$.  The transport cost
will then be $n^{-1}\sum_{t=1}^n \|y_t - z_t\|$, corresponding to the Euclidean
distance (up to a scaling factor). If $\lambda$ is smaller, $(t,y_t)$ is
assigned to some $(s,z_{s})$, for some $s$ possibly different than $t$. If
$\lambda$ goes to zero, the distance coincides with the Wasserstein distance between
the marginal empirical distributions of $y_{1:n}$ and $z_{1:n}$, where the time
element is entirely ignored. Thus curve matching provides a compromise 
between the Euclidean distance between the series seen as vectors,
and the Wasserstein distance between marginal empirical distributions.

For any $\lambda > 0$, the curve matching distance satisfies condition \eqref{cond:exact} of Proposition \ref{prop:as_distn_fixedn}, implying that
the resulting WABC posterior converges to the standard posterior distribution as $\varepsilon \to 0$. To estimate the WABC posterior, we can utilize any of the methods for computing and approximating the Wasserstein distance discussed in Section \ref{sec:distancecalculations} in combination with the SMC algorithm of Section \ref{sec:smcsamplers}. In Section \ref{example:cosine}, we use the exact Wasserstein curve matching distance to infer parameters in a cosine model.
The choice of $\lambda$ is open, but a simple heuristic for univariate time series goes as follows. Consider the aspect ratio of the trace plot of the time series $(y_t)$,
with horizontal axis spanning from $1$ to $n$, and vertical axis from $\min_{t \in 1:n} y_t $ to $\max_{t\in 1:n} y_t $. For an aspect
ratio of $H:V$, one can choose $\lambda$ as $((\max_{t \in 1:n} y_t - \min_{t \in 1:n} y_t) / V)\times (H/n)$. 
For this choice $\rho_\lambda$ corresponds to the Euclidean
distance in a rectangular plot with the given aspect ratio. 

Generalizations of the curve matching distance have been proposed independently
by \citet{thorpe2017transportation} under the name  ``transportation $L_p$
distances''. In that paper, the properties of the curve matching distance are
studied in detail, and compared to and combined with the related notion of
dynamic time warping \citep{berndt1994using}. Other related distances between
time series include the Skorokhod distance between curves
\citep{majumdar2015computing} and the Fr{\'e}chet distance between polygons
\citep{buchin2008computing}, in which $y_t$ would be compared to $z_{r(t)}$,
where $r$ is a retiming function to be optimized.

\subsubsection{Example: Cosine model}\label{example:cosine}
\begin{figure}[h]
        \begin{subfigure}[t]{0.42\textwidth}
            \centering
            \includegraphics[width=\textwidth]{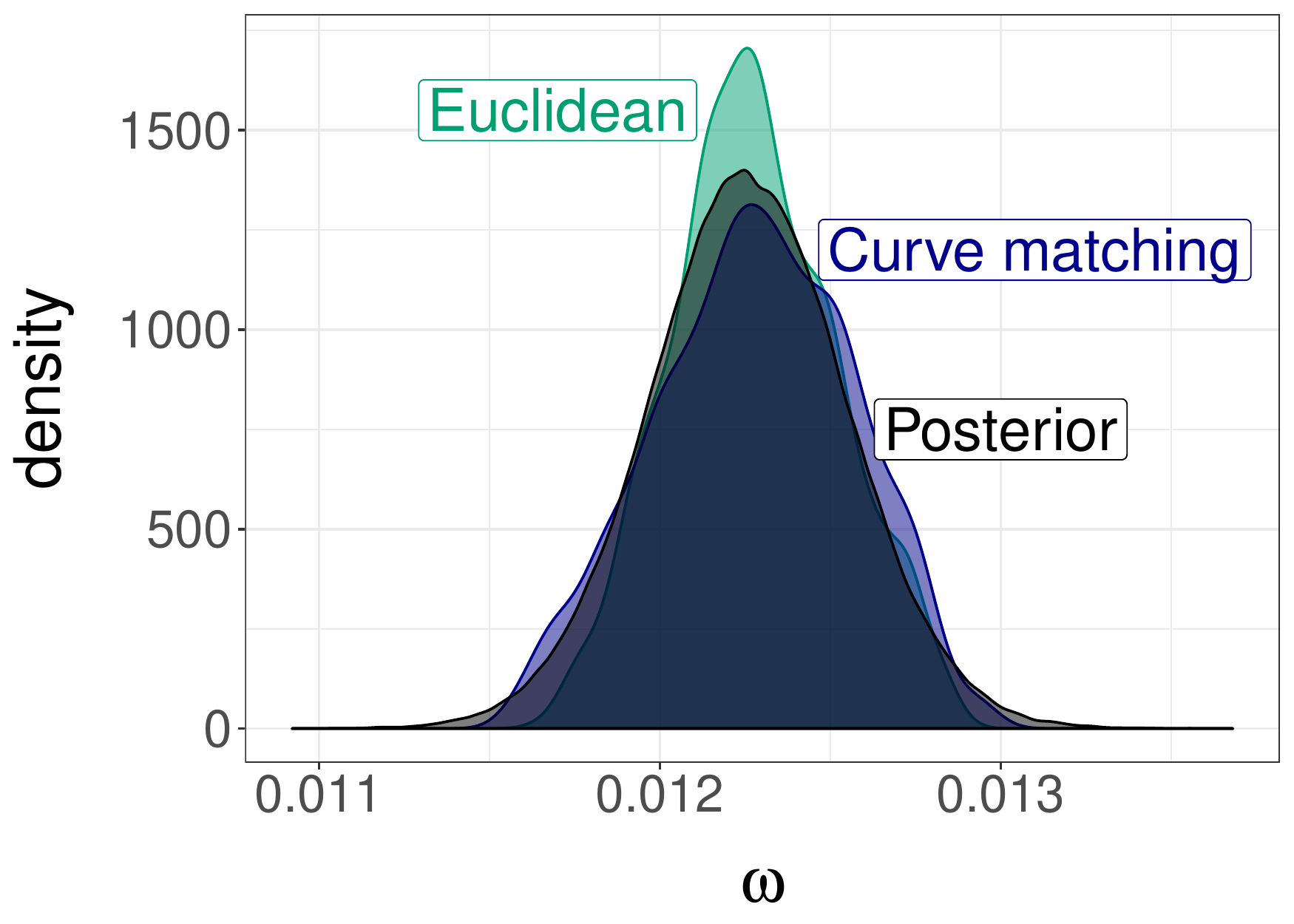}
            \caption{{\small Posteriors of $\omega$.}}    
            \label{fig:cosine:post1}
        \end{subfigure}
        \hspace*{1cm}
        \begin{subfigure}[t]{0.42\textwidth}
            \centering
            \includegraphics[width=\textwidth]{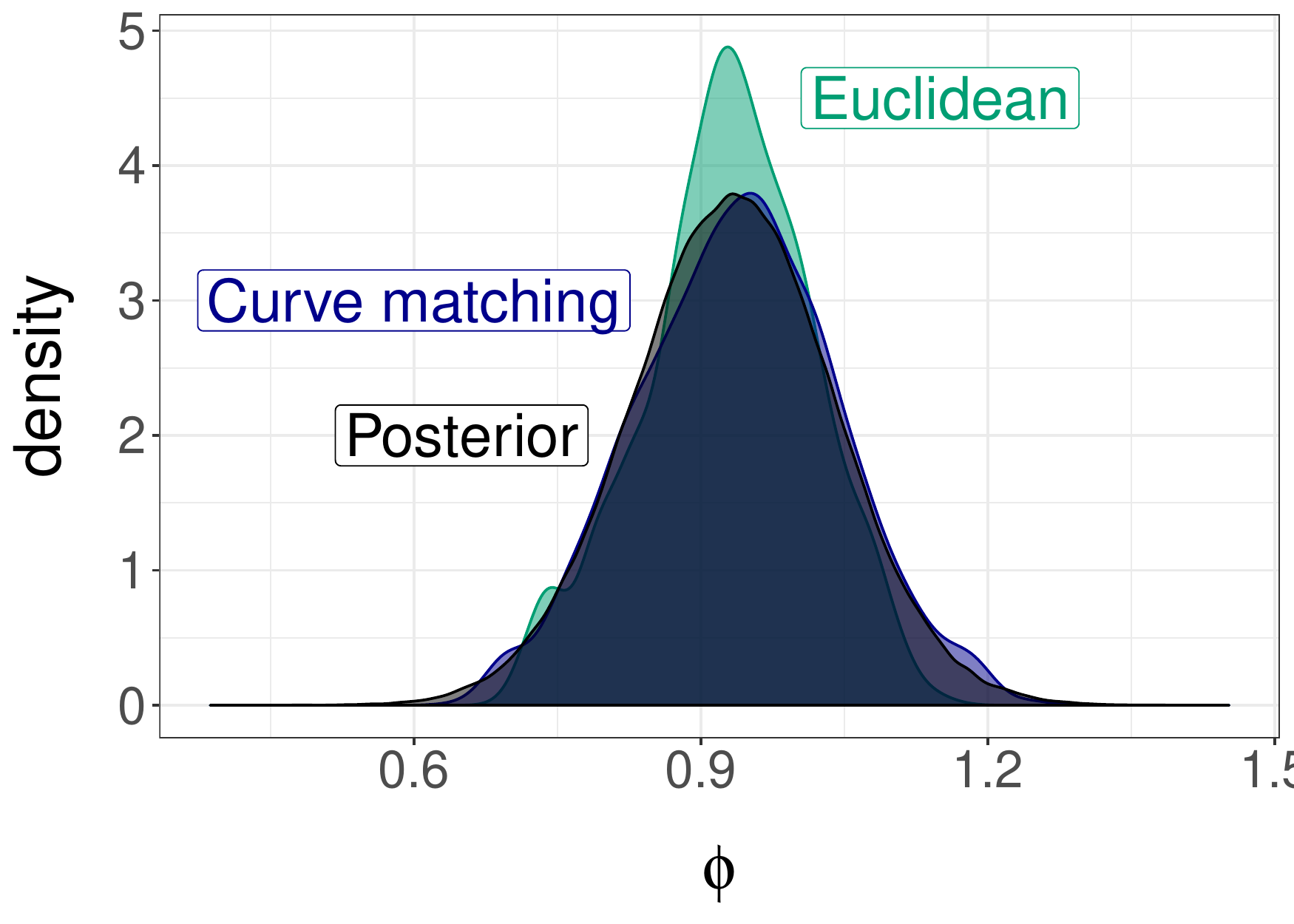}
            \caption{{\small Posteriors of $\phi$.}}    
            \label{fig:cosine:post2}
        \end{subfigure}

        \begin{subfigure}[t]{0.42\textwidth}
            \centering
            \includegraphics[width=\textwidth]{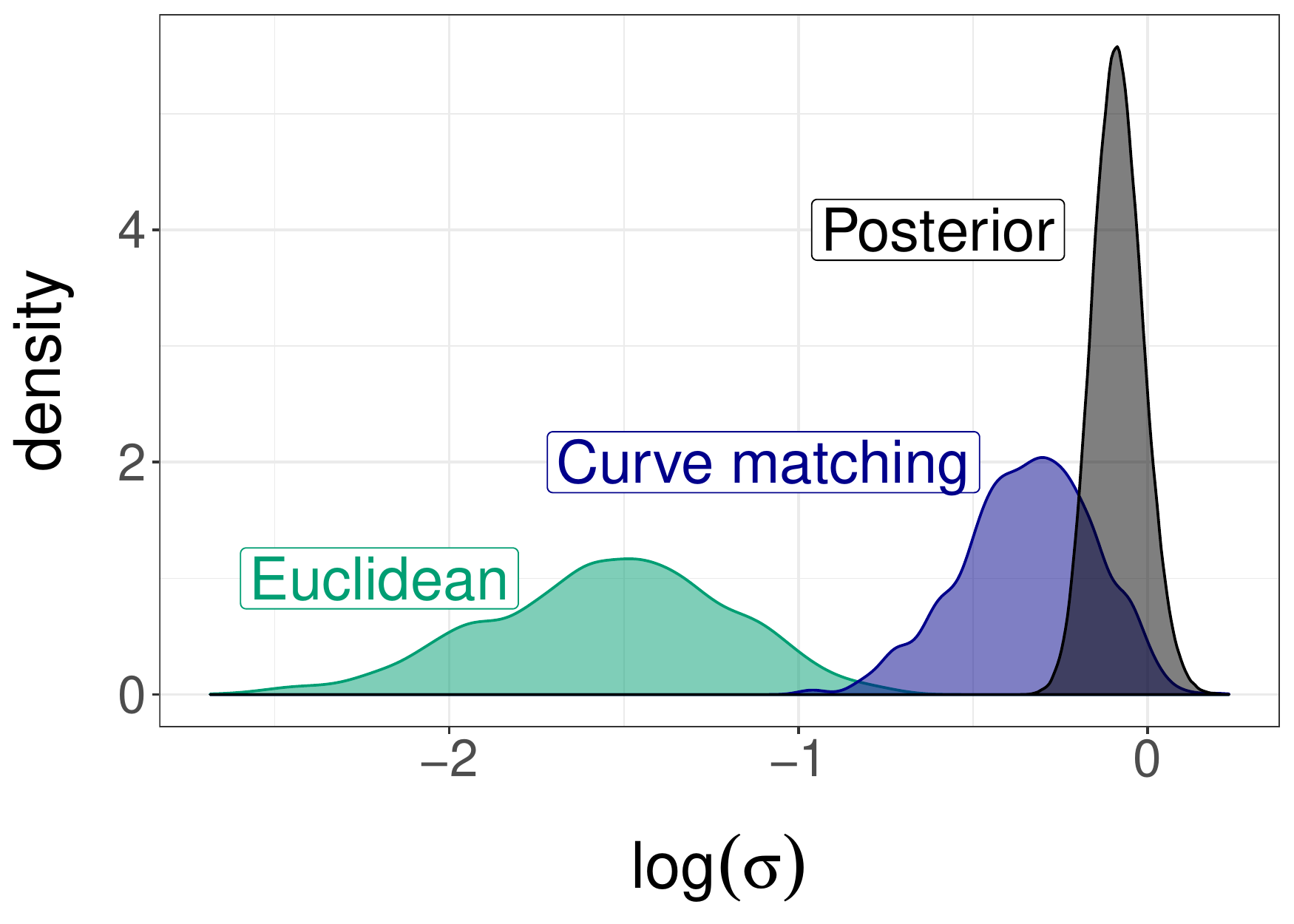}
            \caption{{\small Posteriors of $\log(\sigma)$.}}    
            \label{fig:cosine:post3}
        \end{subfigure}
        \hspace*{1cm}
        \begin{subfigure}[t]{0.42\textwidth}
            \centering
            \includegraphics[width=\textwidth]{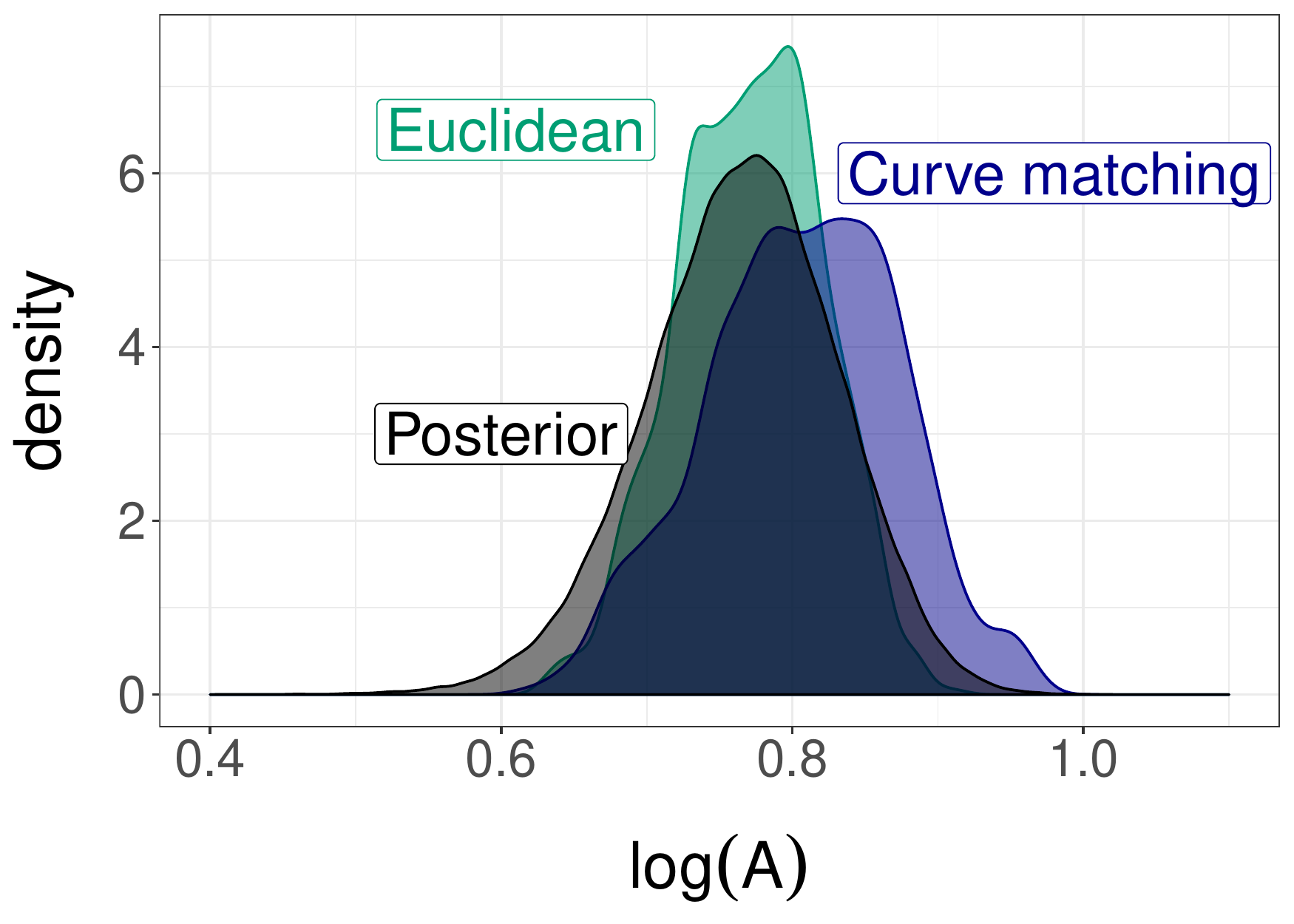}
            \caption{{\small Posteriors of $\log(A)$.}}    
            \label{fig:cosine:post4}
        \end{subfigure}
        %\hspace*{1cm}
        \caption{\small ABC posterior samples in the cosine model of Section \ref{example:cosine}, using either the Euclidean distance
            or curve matching with the exact Wasserstein distance and $\lambda = 1$, after $10^6$ model simulations. We compare to the posterior distribution,
            obtained using the 50,000 last samples in a Metropolis--Hastings chain of length 100,000. The standard deviation
    of the noise $\sigma$ is better estimated with curve matching than with the Euclidean distance between time series.} 
        \label{fig:cosine:curvematching}
\end{figure}

    Consider a cosine model where $y_t = A \cos(2\pi\omega t + \phi) + \sigma
    w_t$,  where $w_t \sim \mathcal{N}(0,1)$, for all $t\geq 1$, are independent.
    Information about $\omega$ and $\phi$ is
    mostly lost when considering the marginal empirical distribution of $y_{1:n}$.
    In Figure \ref{fig:cosine:curvematching}, we compare the ABC posteriors obtained either with the Euclidean distance
    between the series, or with curve matching, with an aspect ratio of one; in both cases the algorithm
    is run for $10^6$ model simulations. The figure also shows an approximation of the exact posterior distribution,
    obtained via Metropolis--Hastings.  The prior distributions are uniform on $[0,1/10]$ and $[0,2\pi]$ for $\omega$
    and $\phi$ respectively, and standard Normal on $\log(\sigma)$ and
    $\log(A)$.  The data are generated using $\omega = 1/80$, $\phi = \pi/4$,
    $\log(\sigma) = 0$ and $\log(A) = \log(2)$, with $n=100$. We see that curve matching
    yields a more satisfactory estimation of $\sigma$ in Figure \ref{fig:cosine:post3},
    and a similar approximation for the other parameters. By contrast, 
    an ABC approach based on the marginal distribution of $y_{1:n}$ 
    would fail to identify $\phi$.

\subsection{Reconstructions \label{sec:reconstructions}}

Our second approach consists in transforming the time series to define an empirical distribution
$\tilde{\mu}_n$ from which parameters can be estimated.

\subsubsection{Delay reconstruction \label{sec:delayreconstruction}}
In time series analysis, the lag-plot is a scatter plot of the pairs
$(y_t,y_{t-k})_{t=k+1}^n$, for some lag $k\in \mathbb{N}$, from which one can
inspect the dependencies between lagged values of the series. In ABC applied to time series models,
lag-$k$ autocovariances, defined as the sample covariance of $(y_t,y_{t-k})_{t=k+1}^n$, 
are commonly used statistics to summarize these dependencies \citep{marin2012approximate,mengersen2013bayesian,li2015asymptotic}.
Here, we also propose to use the joint samples $(y_t,y_{t-k})_{t=k+1}^n$, but bypass their summarization into sample covariances.
In particular, we define delay
reconstructions as $\tilde{y}_t =
(y_t,y_{t-\tau_1},\ldots,y_{t-\tau_k})$ for some integers
$\tau_1,\ldots,\tau_k$.  The sequence, denoted $\tilde{y}_{1:n}$ after
relabelling and redefining $n$, inherits many properties from the original
series, such as stationarity.  Therefore, the empirical distribution of
$\tilde{y}_{1:n}$, denoted by $\tilde{\mu}_n$, might converge to a limit
$\tilde{\mu}_{\star}$. In turn,
$\tilde{\mu}_\star$ is likely to capture more features of the dependency structure 
than $\mu_\star$, and the resulting 
procedure might provide more accurate inference on the model parameters than 
if we were to compare the lag-$k$ autocovariances alone.

Delay reconstructions (or embeddings) play a central role in dynamical systems
\citep{kantz2004nonlinear}, for instance in Takens' theorem 
and variants thereof \citep{stark2003}.  The Wasserstein distance between the
empirical distributions of delay reconstructions has previously been proposed
as a way of measuring distance between time series
\citep{moeckel1997,muskulus2011}, but not as a device for parameter inference.
In the ABC setting, we propose to construct the delay reconstructions of each synthetic time series,
and to compute the Wasserstein distance between their empirical distribution 
and the empirical distribution of $\tilde{y}_{1:n}$. We refer to this approach
as WABC with delay reconstruction.

Denote by $\tilde{\mu}_{\theta,n}$ the empirical distribution of the delay reconstructed series $\tilde{z}_{1:n}$, formed from $z_{1:n} \sim \mu^{(n)}_{\theta}$. Then,
provided that $\tilde{\mu}_{\theta,n}$ converges to an identifiable distribution $\tilde{\mu}_\theta$ as $n\to\infty$,
we are back in a setting where we can study the concentration behavior of the WABC posterior around $\tilde{\theta}_\star = \argmin_{\theta\in\mathcal{H}}\was_p(\tilde{\mu}_\star,\tilde{\mu}_\theta)$, assuming existence and uniqueness (see Section \ref{sec:asymptotics:double}). In well-specified settings, $\tilde{\theta}_\star$ must correspond to the data-generating parameters.

When the entries of the vectors $y_{1:n}$ and $z_{1:n}$ are all unique, which happens with probability
one when $\mu_\star^{(n)}$ and $\mu_\theta^{(n)}$  are continuous
distributions, then $\was_p(\tilde{y}_{1:n}, \tilde{z}_{1:n}) = 0$ if and only
if $y_{1:n} = z_{1:n}$. To see this, consider the setting where $\tilde{y}_t =
(y_t,y_{t-1})$, and $\tilde{z}_t = (z_t,z_{t-1})$. For the empirical distributions
of $\tilde{y}_{1:n}$ and $\tilde{z}_{1:n}$ to be equal, we require that for every $t$ there
exists a unique $s$ such that $\tilde{y}_t = \tilde{z}_s$. However, since the
values in $y_{1:n}$ and $z_{1:n}$ are unique, the values $y_1$ and $z_1$ appear
only as the second coordinates of $\tilde{y}_2$ and $\tilde{z}_2$ respectively.
It therefore has to be that $y_1 = z_1$ and $\tilde{y}_2 = \tilde{z}_2$. In
turn, this implies that $y_2 = z_2$, and inductively, $y_{t} =
z_{t}$ for all $t \in 1:n$. A similar reasoning can be done for any $k \geq 2$
and $1\leq \tau_1<\ldots <\tau_k$.
This property can be used to establish the convergence of 
the WABC posterior based on delay reconstruction to the posterior, as
$\varepsilon \to 0$, which can be deduced using condition \eqref{cond:exact} of Proposition \ref{prop:as_distn_fixedn}.

In practice, for a non-zero value of $\varepsilon$, the obtained ABC posteriors
might be different from the posterior, but still identify the parameters with a reasonable accuracy, as illustrated in 
Section \ref{example:ar}. The quality of the approximation will depend on the choice of lags
$\tau_1,\dots,\tau_k$. Data-driven ways of making such choices are discussed by \citet{kantz2004nonlinear}.
Still, since the order of the original data is only partly reflected in delay reconstructions,
some model parameters might be difficult to estimate with delay reconstruction, such as 
the phase shift $\phi$ in Section \ref{example:cosine}.

\subsubsection{Example: AR(1) model}\label{example:ar}
\begin{figure}
    \centering
        \begin{subfigure}[t]{0.42\textwidth}
            \centering
            \includegraphics[width=\textwidth]{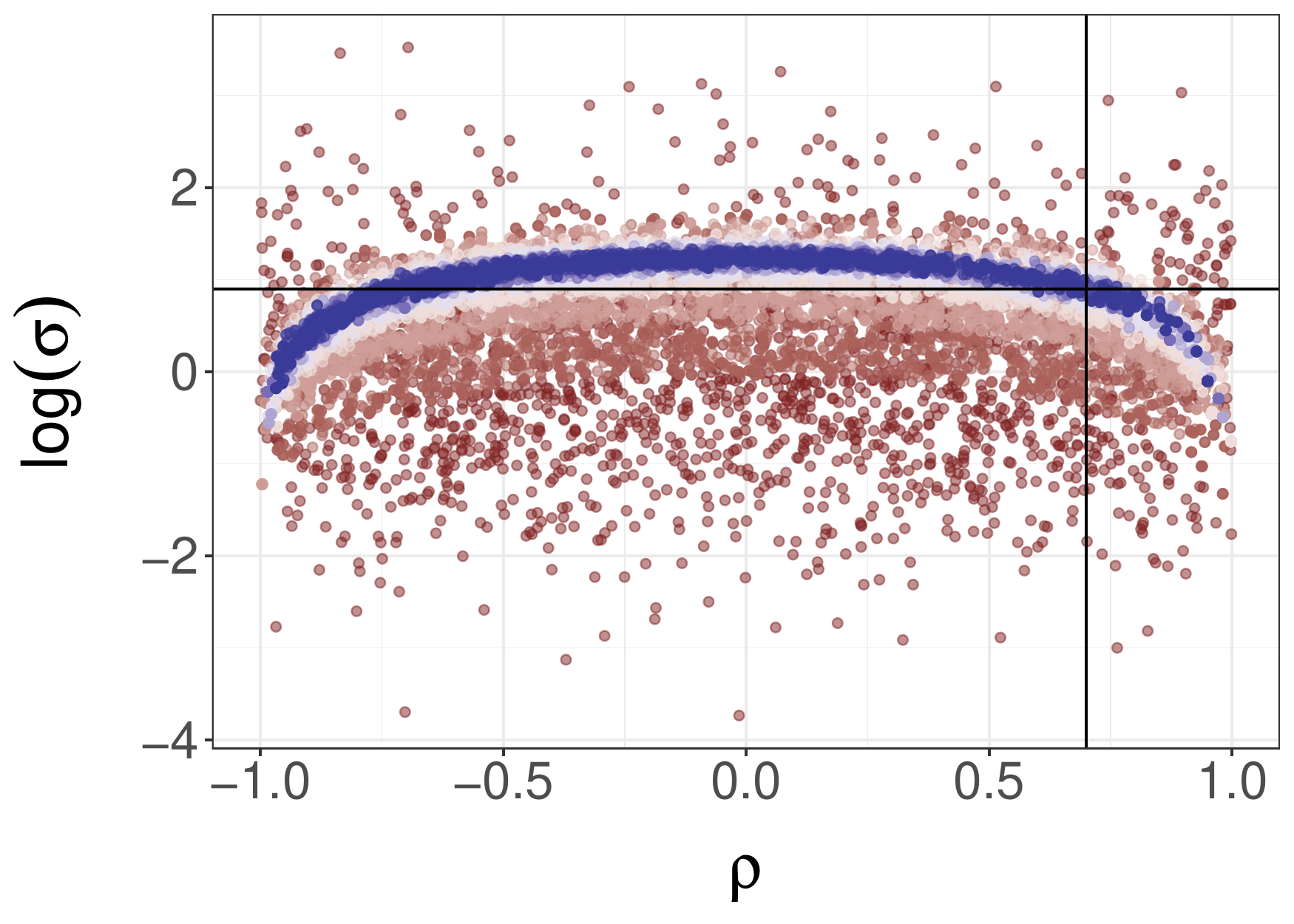}
            \caption{{\small WABC using the Wasserstein distance between marginal distributions.}}    
            \label{fig:ar1:withoutlag}
        \end{subfigure}
        \hspace*{1cm}
        \begin{subfigure}[t]{0.42\textwidth}
            \centering
            \includegraphics[width=\textwidth]{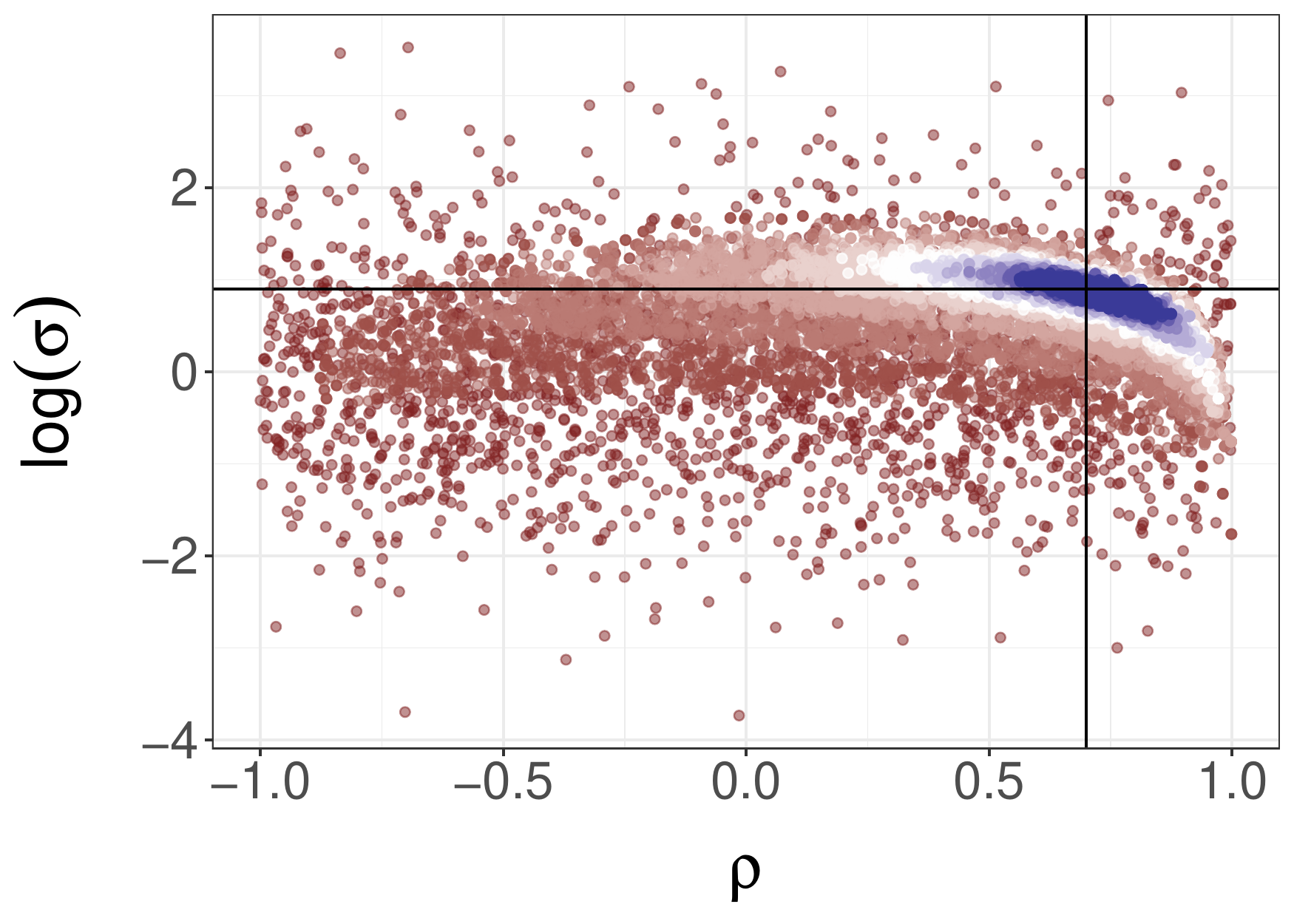}
            \caption{{\small 
            WABC using the Wasserstein distance between empirical distributions of delay reconstructions.}}    
            \label{fig:ar1:withlag1}
        \end{subfigure}
        \caption{{\small Samples from the WABC posteriors of $(\phi, \log(\sigma))$ in the
            AR(1) model of Section \ref{example:ar}, as $\varepsilon$ decreases over the steps of the SMC sampler (colors from red to white to blue). 
            On the left, using the marginal empirical distribution of 
            the series, the WABC posteriors concentrate around a
            ridge of values such that $\sigma^2/(1-\phi^2)$ is constant. On the
            right, using delay reconstruction with lag $k=1$, the
            WABC posteriors concentrate around the data-generating parameters, $\phi = 0.7, \log(\sigma) = 0.9$,
        indicated by full lines. Both methods had a total budget of $10^5$ model simulations.}}
        \label{fig:ar1:identification}
\end{figure}

Consider an autoregressive process of order 1, written AR(1), where
    $y_1\sim \mathcal{N}(0,\sigma^{2}/(1-\phi^{2}))$, for some $\sigma>0$ and
    $\phi\in(-1,1)$. For each $t\geq 2$, let $y_t = \phi y_{t-1} + \sigma w_t$, where $w_t
    \sim \mathcal{N}(0,1)$ are independent.  The marginal distribution of
    each $y_t$ is $\mathcal{N}(0,\sigma^{2}/(1-\phi^{2}))$.  Furthermore, by an ergodic theorem,
    the empirical distribution $\hat{\mu}_n$ of the time series converges to this marginal
    distribution. The two parameters $(\phi, \sigma^2)$ are not
    identifiable from the limit $\mathcal{N}(0,\sigma^{2}/(1-\phi^{2}))$. Figure \ref{fig:ar1:withoutlag} shows
    WABC posterior samples derived while ignoring time dependence, obtained for decreasing values of $\varepsilon$, using a budget of $10^5$ model simulations.
    The prior is uniform on $[-1,1]$ for $\phi$, and standard Normal on $\log(\sigma)$.
    The data are generated using $\phi = 0.7,\log(\sigma)=0.9$ and $n = 1,000$.
    The WABC posteriors concentrate on a ridge of values with
    constant $\sigma^2/(1-\phi^2)$.

    Using $k=1$, we consider $\tilde{y}_t=(y_t,y_{t-1})$ for $t\geq 2$. 
    The reconstructions are then sub-sampled to $500$ values, $\tilde{y}_2 = (y_2,y_1), \tilde{y}_4 = (y_4,y_3), \ldots, \tilde{y}_{1000} = (y_{1000},y_{999})$;
    similar results were obtained with the $999$ reconstructed values, but sub-sampling 
    leads to computational gains in the exact Wasserstein distance calculations; see Section \ref{sec:subsampling}.
    The stationary distribution of
    $\tilde{y}_{t}$ is given by
    \begin{equation}\label{eq:arstationarydistribution}
        \mathcal{N}\left(\begin{pmatrix} 0 \\ 0 \end{pmatrix},
        \frac{\sigma^{2}}{1-\phi^{2}} \begin{pmatrix} 1 & \phi \\ \phi & 1
        \end{pmatrix}\right).  \end{equation}
    Both parameters $\sigma^2$ and
    $\phi$ can be identified from a sample approximating the above
    distribution.  Figure \ref{fig:ar1:withlag1} shows the WABC posteriors 
    obtained with delay reconstruction and a budged of $10^5$ model simulations concentrating around the data-generating values as $\varepsilon$
    decreases. 

\subsubsection{Residual reconstruction \label{sec:residualreconstruction}}

Another approach to handle dependent data is advocated in \citet{mengersen2013bayesian}, in the context
of ABC via empirical likelihood.  In various time series models, the observations
are modeled as transformations of some parameter $\theta$ and residual
variables $w_1, \ldots,w_n$. Then, given a parameter $\theta$, one might
be able to reconstruct the residuals corresponding to the observations.  In Section
\ref{example:cosine}, one can define $w_t = (y_t - A \cos(2\pi \omega t +
\phi))/\sigma$.  In 
Section \ref{example:ar}, one can define $w_t = (y_t - \phi y_{t-1})/\sigma$; other 
examples are given in \citet{mengersen2013bayesian}.
Once the residuals have been reconstructed, their empirical distribution
can be compared to the distribution that they would follow under the model,
e.g. a standard Normal in Sections \ref{example:ar} and \ref{example:cosine}.

\section{Numerical experiments\label{sec:numerics}}
We illustrate the proposed approach and make comparisons to existing methods in various models taken from the literature. In each example, we approximate the WABC posterior using the SMC algorithm and default parameters outlined in Section \ref{sec:smcsamplers}.

\subsection{Quantile ``g-and-k'' distribution \label{sec:gandk}}

We first consider an example where the
likelihood can be approximated to high precision, which allows comparisons
between the standard posterior and WABC approximations. We observe 
that the WABC posterior converges to the true posterior in the
univariate g-and-k example, as suggested by Proposition
\ref{prop:as_distn_fixedn}. We also compare WABC to a method
developed by \citet{fearnhead:prangle:2012} that uses a semi-automatic
construction of summary statistics. Lastly, we compare the
Wasserstein distance with the other distances described in Section \ref{sec:distancecalculations} on the
bivariate g-and-k distribution.

\subsubsection{Univariate ``g-and-k''} \label{sec:gandk_univariate}
A classical example in the ABC literature \citep[see e.g.][]{fearnhead:prangle:2012,mengersen2013bayesian}, the univariate g-and-k distribution is defined in terms of its quantile function:
\begin{equation}
r\in (0,1) \mapsto a + b \left(1 + 0.8\frac{1-\exp(-gz(r)}{1+\exp(-gz(r)}\right) \left(1+z(r)^2\right)^k z(r),
    \label{eq:gandk}
\end{equation}
where $z(r)$ refers to the $r$-th quantile of the standard Normal distribution.

Sampling from the g-and-k distribution 
can be done by plugging standard Normal variables into  \eqref{eq:gandk}
in place of $z(r)$. The probability density function is intractable, but
can be numerically calculated with high precision since 
it only involves one-dimensional inversions and differentiations of the quantile function in  \eqref{eq:gandk},
as described in \citet{rayner2002numerical}. Therefore,
Bayesian inference can be carried out with e.g. Markov chain Monte Carlo.

We generate $n=250$ observations from the model using $a = 3, b = 1, g = 2, k = 0.5$, and the parameters are assigned a uniform prior on $[0,10]^4$. We estimate the posterior distribution by running $5$ Metropolis--Hastings chains for $75,000$ iterations, and discard the first $50,000$ as burn-in. For the WABC approximation, we use the SMC sampler outlined in Section \ref{sec:smcsamplers} with $N=2,048$ particles, for a total of $2.4\times 10^6$ simulations from the model. The resulting marginal WABC posteriors are also compared to the marginal posteriors obtained with the semi-automatic ABC approach of \citet{fearnhead:prangle:2012}. We used the rejection sampler in the \texttt{abctools} package \citep{nunes2015abctools}, also for a total of $2.4\times 10^6$ model simulations, of which $N=2,048$ draws from the prior are accepted. We observed no benefit to accepting fewer draws. The semi-automatic approach requires the user to specify a set of initial summary statistics, for which we used every 25th order statistic as well as the minimum (that is, $y_{(1)},y_{(25)},y_{(50)},\dots,y_{(250)}$) and their powers up to fourth order, following guidance in \citet{fearnhead:prangle:2012}.

\begin{figure}[hp]
        \centering
        \begin{subfigure}[b]{0.42\textwidth}
            \centering
            \includegraphics[width=\textwidth]{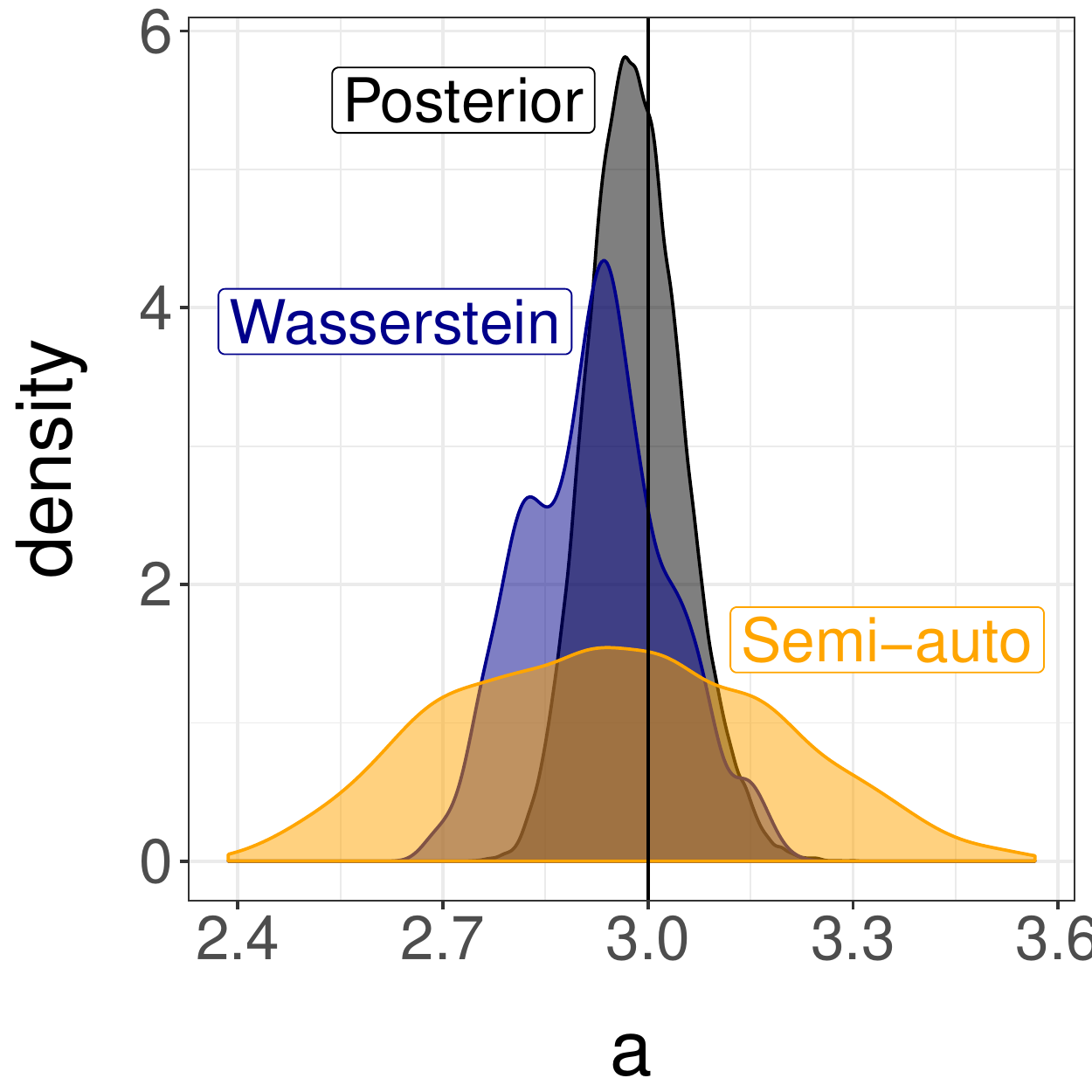}
            \caption{{\small Posteriors of $a$.}}    
        \end{subfigure}
         \hskip 0.8cm
        \begin{subfigure}[b]{0.42\textwidth}
            \centering
            \includegraphics[width=\textwidth]{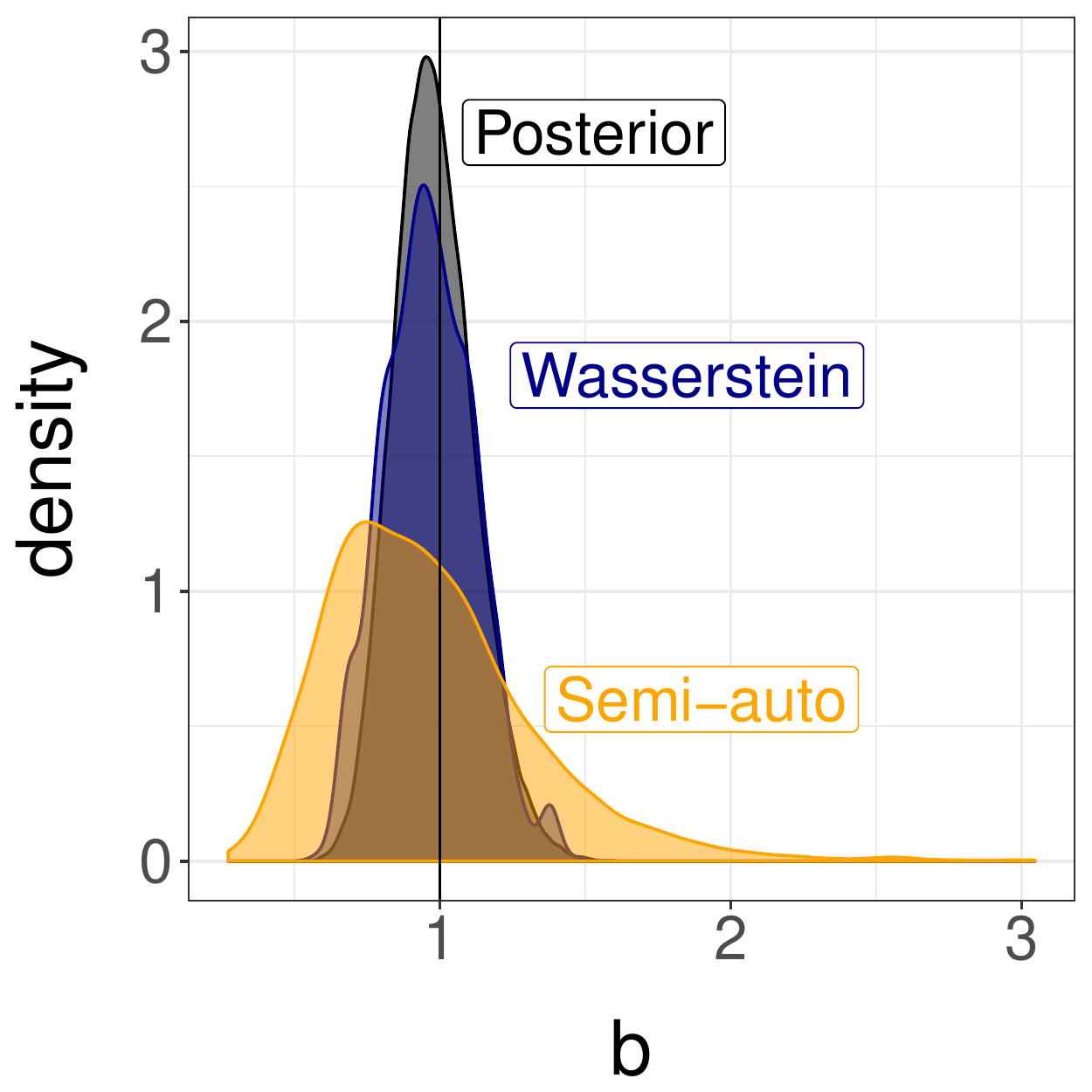}
            \caption{{\small Posteriors of $b$. }}    
        \end{subfigure}
        
        \begin{subfigure}[b]{0.42\textwidth}
            \centering
            \includegraphics[width=\textwidth]{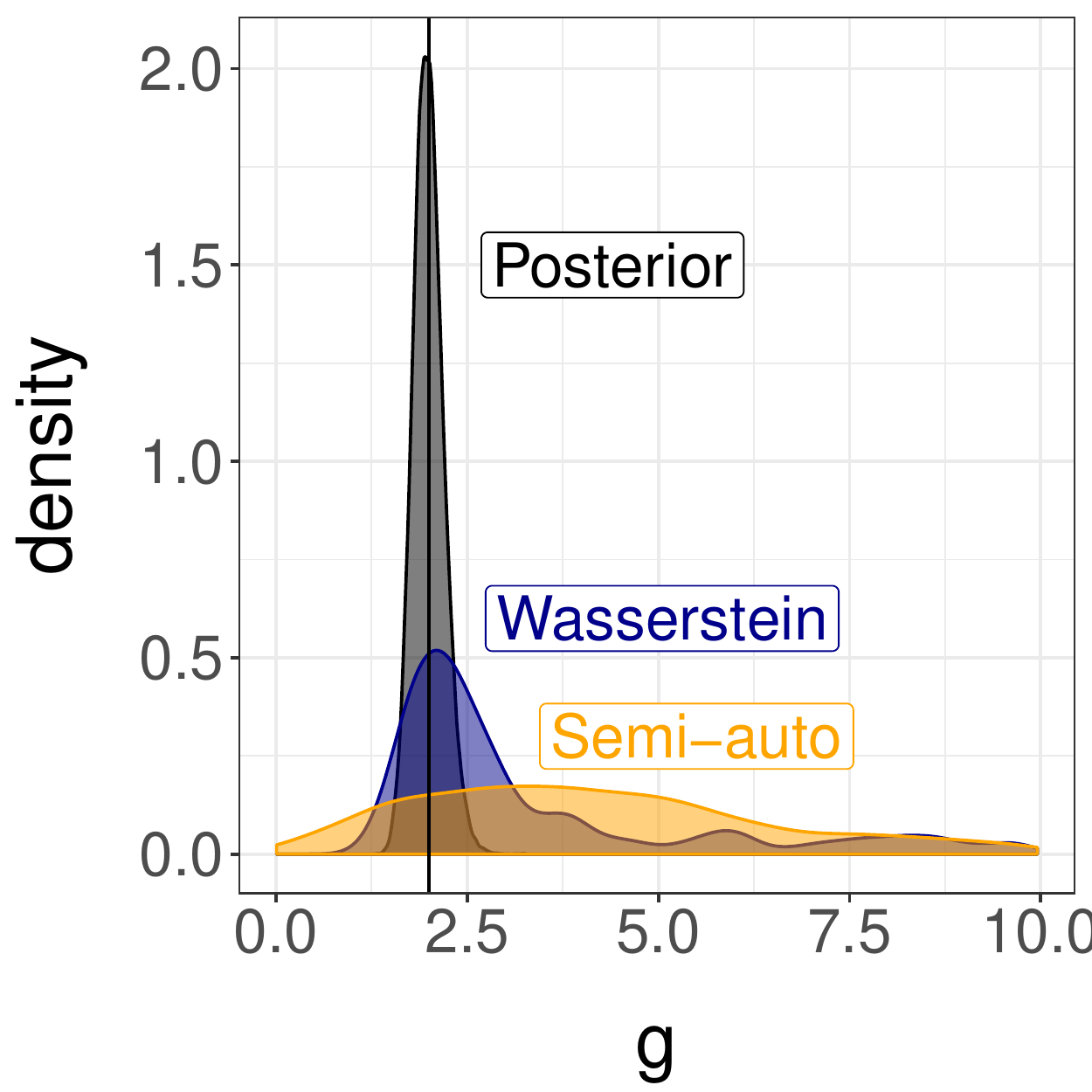}
            \caption{{\small Posteriors of $g$.}}    
        \end{subfigure}
        \hskip 0.8cm
        \begin{subfigure}[b]{0.42\textwidth}
            \centering
            \includegraphics[width=\textwidth]{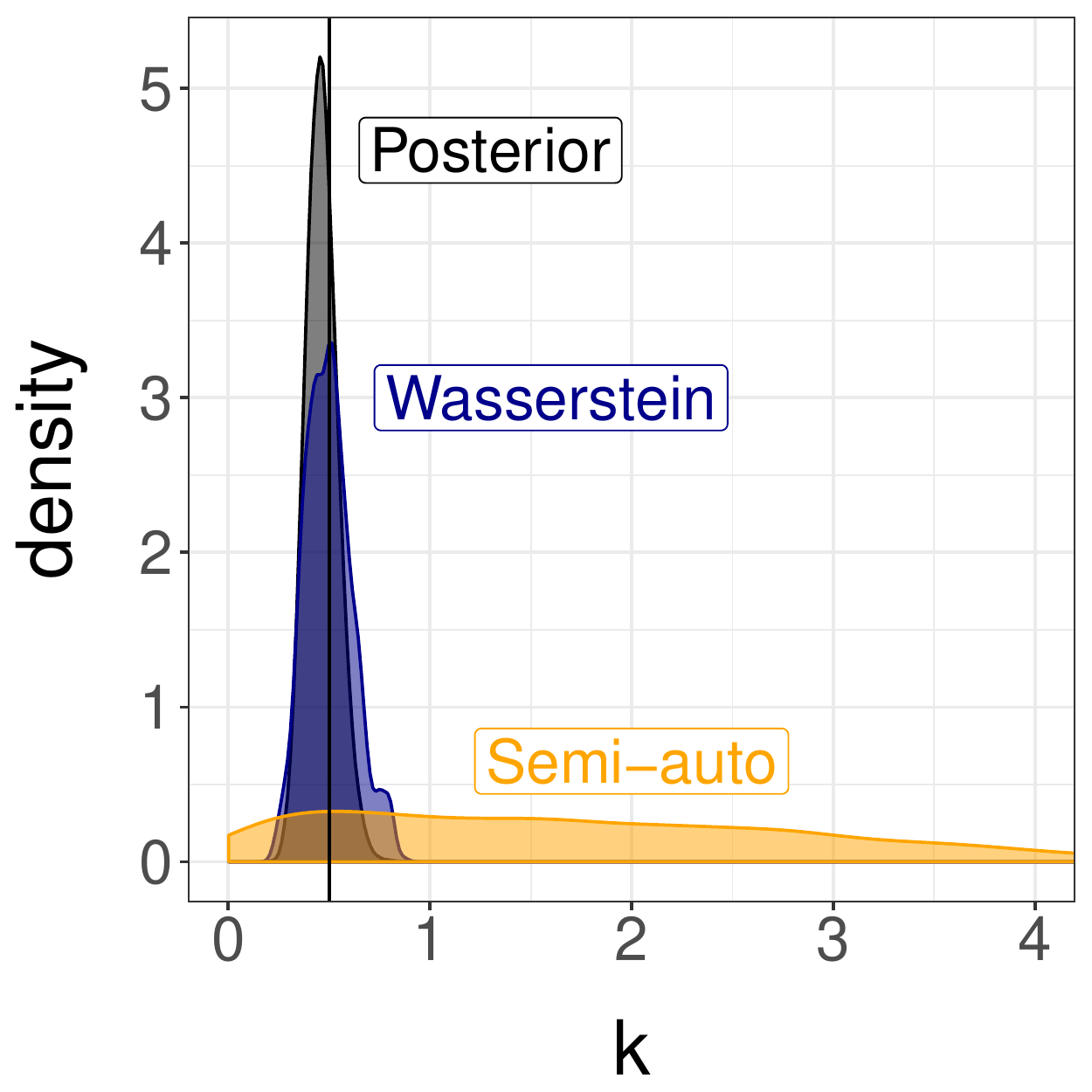}
            \caption{{\small Posteriors of $k$.}}    
        \end{subfigure}
         \caption{\small Posterior marginals in the univariate g-and-k example of Section \ref{sec:gandk_univariate} (obtained via MCMC), approximations by Wasserstein ABC and semi-automatic ABC, each with a budget of $2.4\times 10^6$ model simulations. Data-generating values are indicated by vertical lines.}
        \label{fig:gandk:abc}
\end{figure}

Figure \ref{fig:gandk:abc} shows the marginal posterior distributions and their approximations obtained with WABC and semi-automatic ABC. The plots show that the WABC posteriors appear to be closer to the target distributions, especially on the $a,b$ and $k$ parameters. Neither method captures the marginal posterior of $g$ well, though the WABC posterior appears more concentrated in the region of interest on that parameter as well.

In both of the ABC approaches, the main computational costs stem from simulating from the model and sorting the resulting data sets. Over 1,000 repetitions, the average wall-clock time to simulate a data set was $7.7\times 10^{-5}s$ on an Intel Core i5 (2.5GHz). The average time to sort the resulting data sets was $8.9 \times 10^{-5}s$, and computing the Wasserstein distance to the observed data set was negligibly different from this. In semi-automatic ABC, one additionally has to perform a regression step. The rejection sampler in semi-automatic ABC is
easier to parallelize than our SMC approach, but on the other hand requires more memory due to the
regression used in constructing the summary statistic. This makes the method hard to scale up beyond the number of
model simulations considered here, without using specialized tools for large scale 
regression.

\begin{sidewaysfigure}[hp]
        \centering
        \begin{subfigure}[b]{0.21\textwidth}
            \centering
            \includegraphics[width=\textwidth]{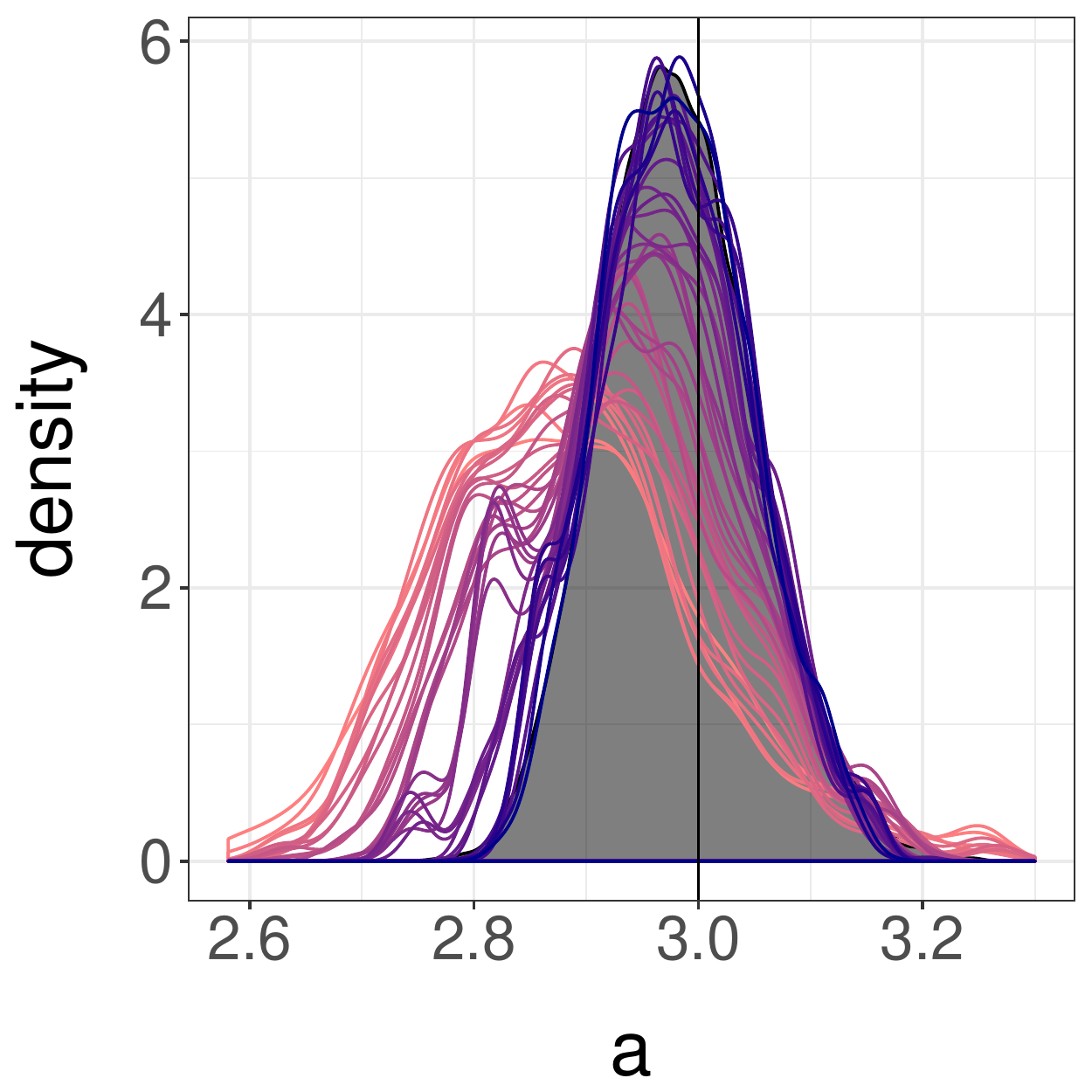}
            \caption{{\small Posteriors of $a$.}} 
            \label{fig:gandk_marginal1}
        \end{subfigure}
                                     \hskip 0.3cm
        \begin{subfigure}[b]{0.21\textwidth}
            \centering
            \includegraphics[width=\textwidth]{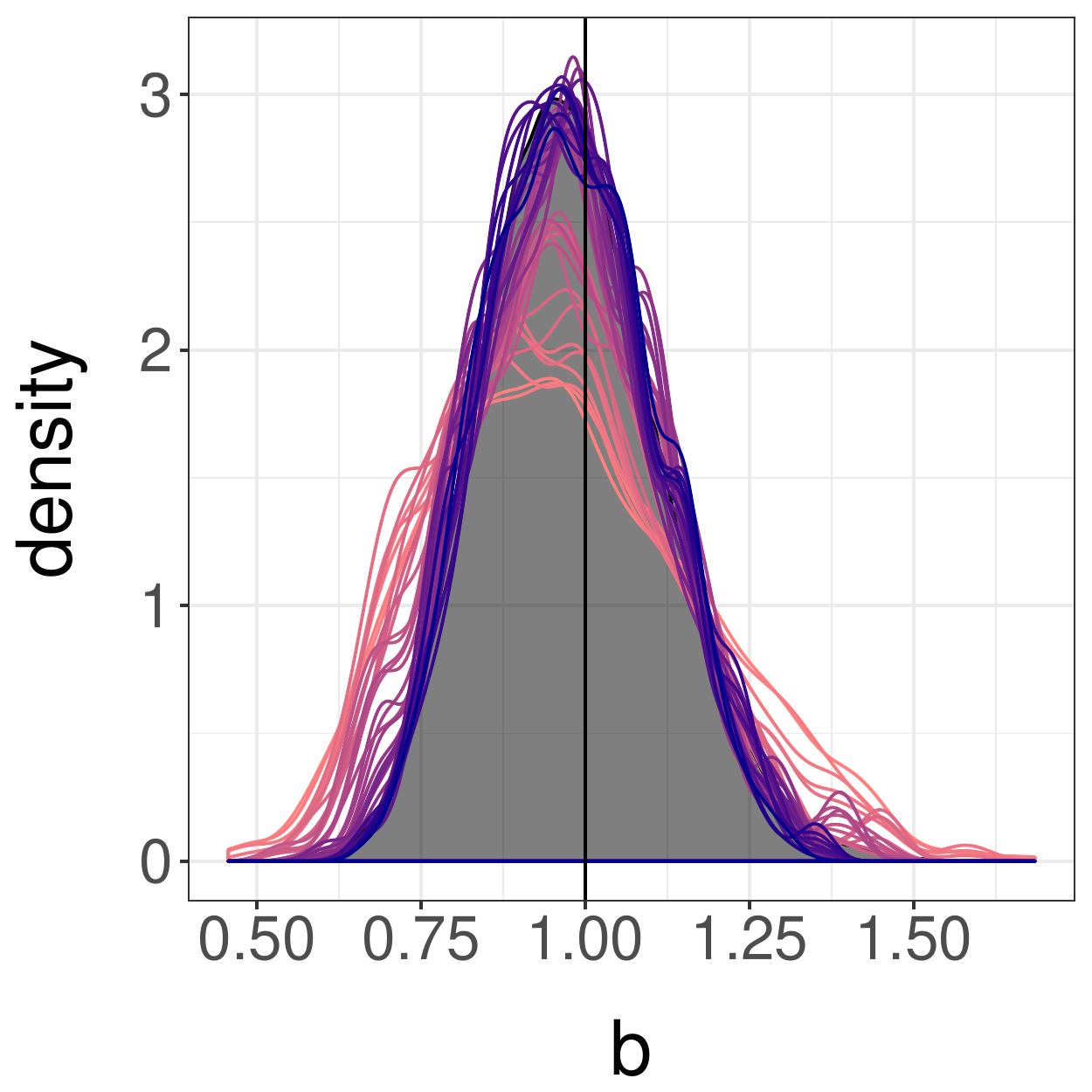}
            \caption{{\small Posteriors of $b$. }}    
            \label{fig:gandk_marginal2}
        \end{subfigure}
                                     \hskip 0.3cm
        \begin{subfigure}[b]{0.21\textwidth}
            \centering
            \includegraphics[width=\textwidth]{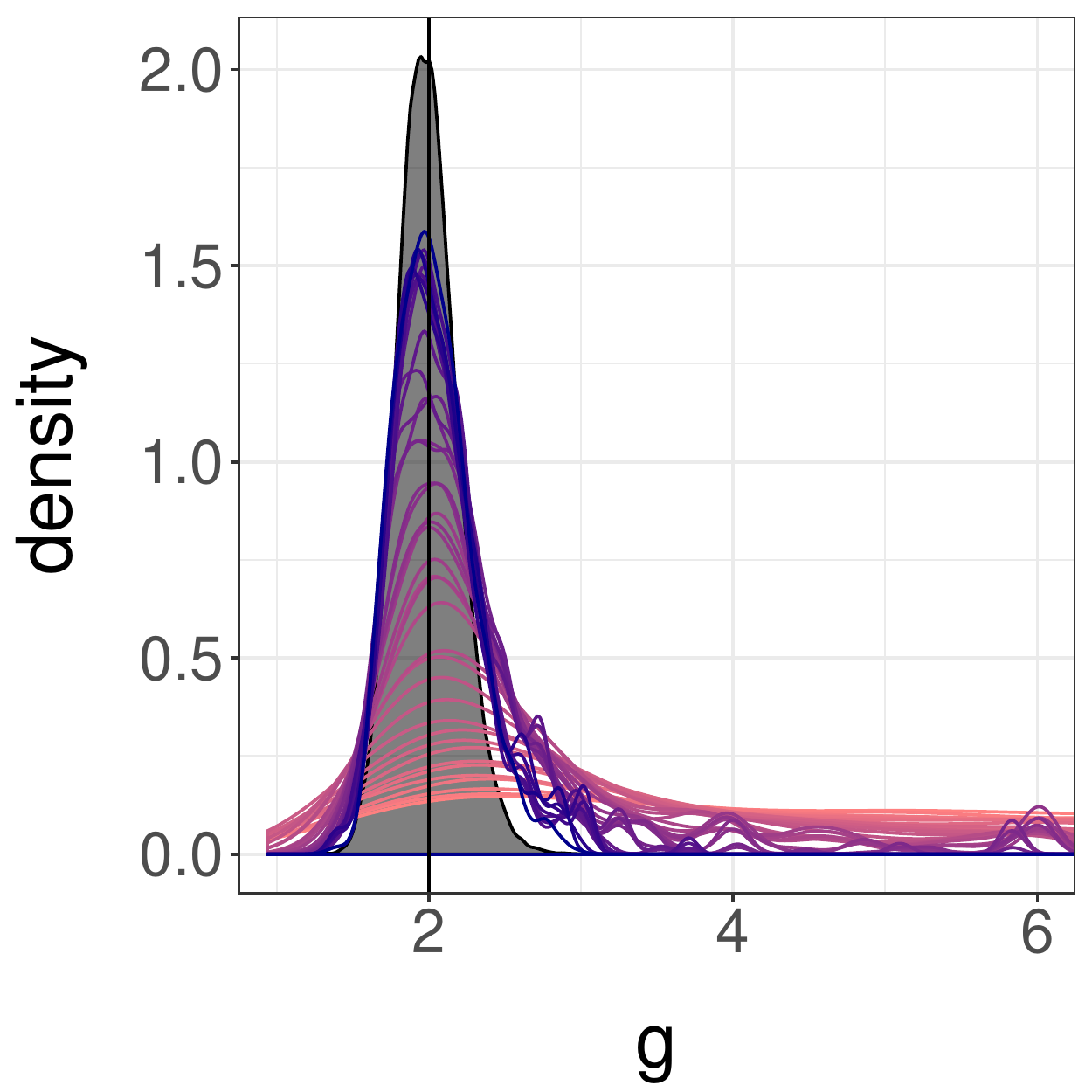}
            \caption{{\small Posteriors of $g$.}}
            \label{fig:gandk_marginal3}   
        \end{subfigure}
                                     \hskip 0.3cm
        \begin{subfigure}[b]{0.21\textwidth}
            \centering
            \includegraphics[width=\textwidth]{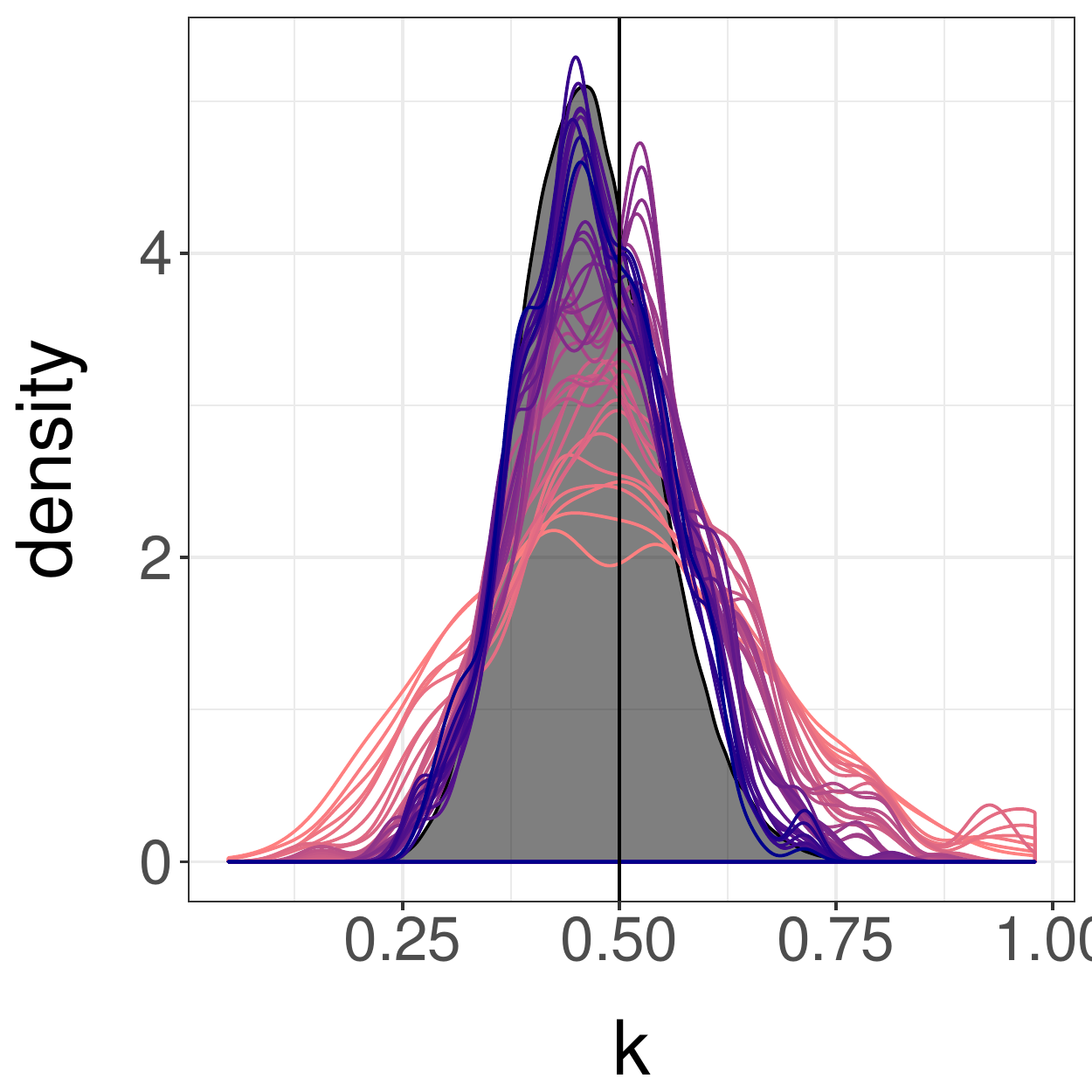}
            \caption{{\small Posteriors of $k$.}}
            \label{fig:gandk_marginal4}    
        \end{subfigure}
        
                \begin{subfigure}[b]{0.21\textwidth}
            \centering
            \includegraphics[width=\textwidth]{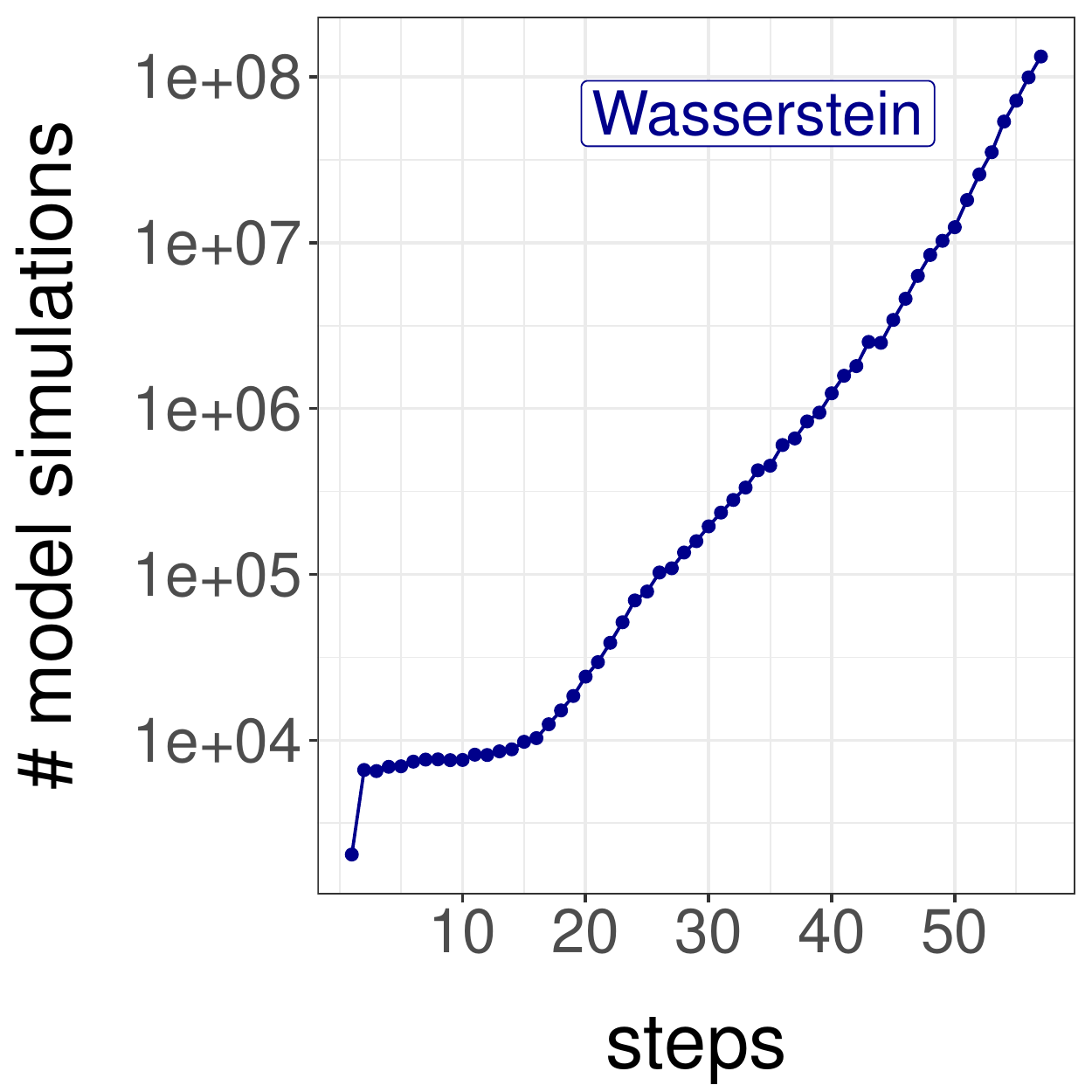}
            \caption{{\small Model simulations vs. SMC steps.}}
            \label{fig:gandk_ncomp_vs_steps}    
        \end{subfigure}
                                     \hskip 0.3cm
        \begin{subfigure}[b]{0.21\textwidth}
            \centering
            \includegraphics[width=\textwidth]{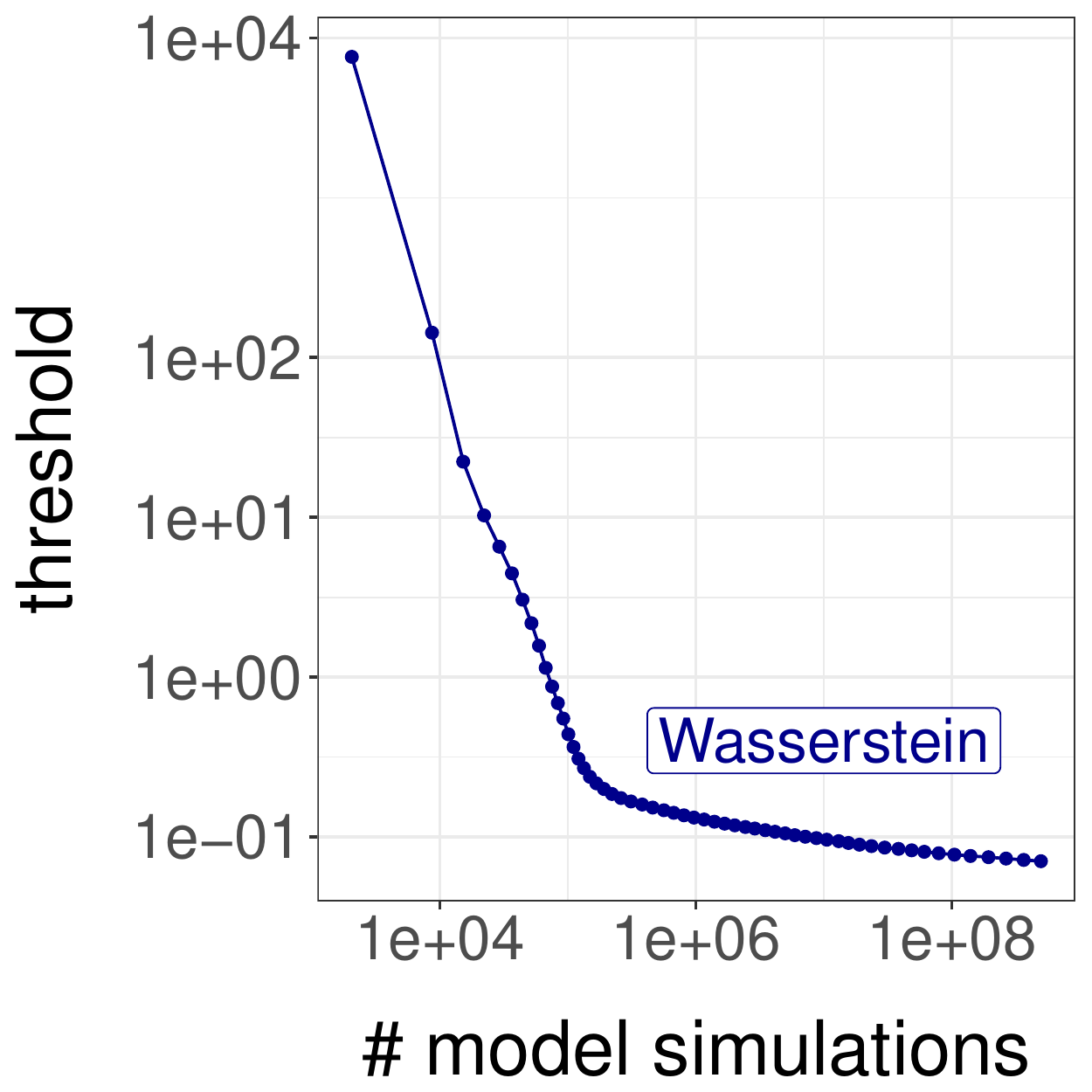}
            \caption{{\small Threshold $\varepsilon$ vs. model simulations}}    
            \label{fig:gandk_threshold_vs_ncomp}   
        \end{subfigure}
                                     \hskip 0.3cm
        \begin{subfigure}[b]{0.42\textwidth}
            \centering
            \includegraphics[width=\textwidth]{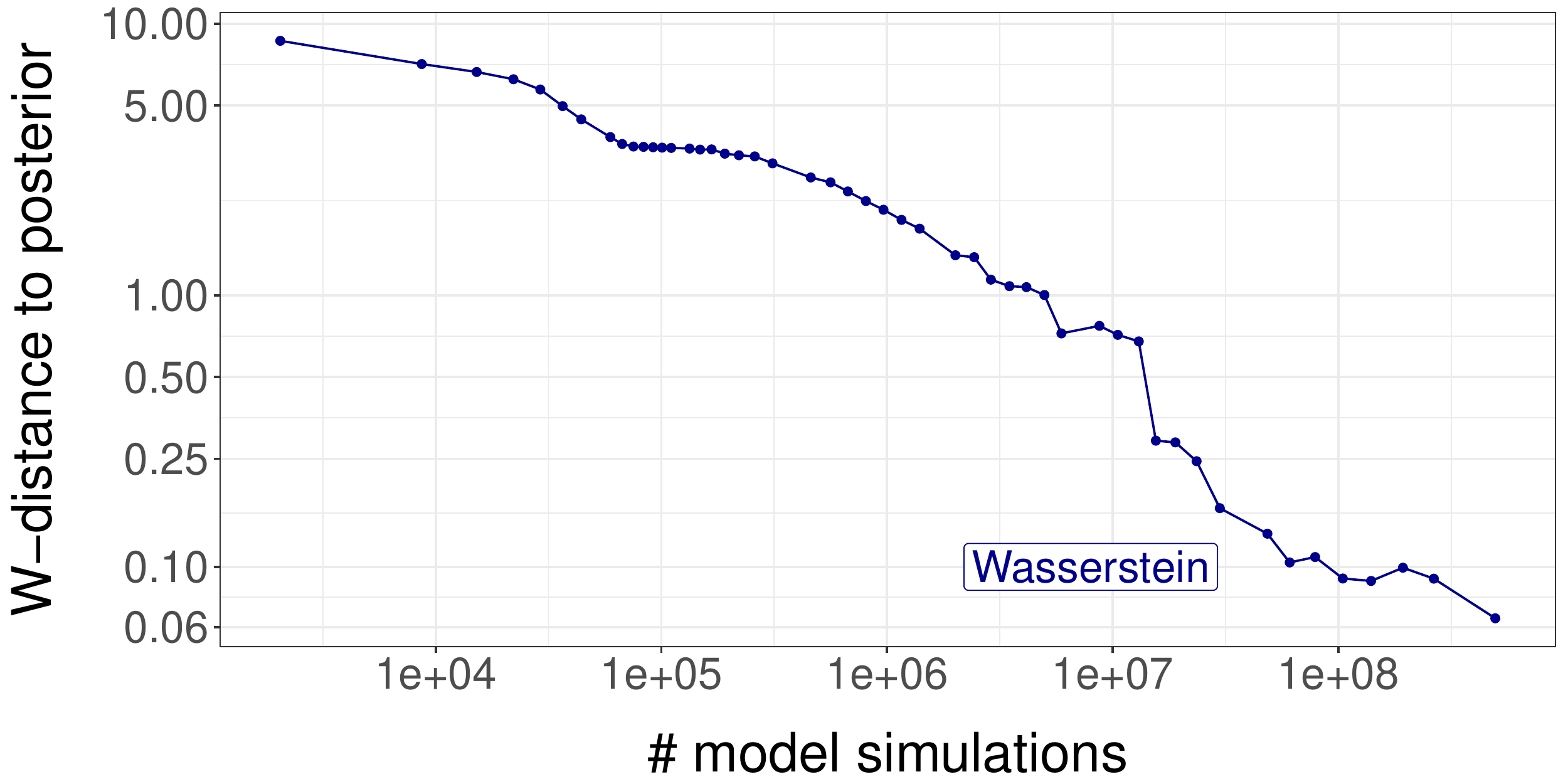}
            \caption{{\small $\was_1$-distance to posterior vs. number of model simulations.}}
            \label{fig:gandk_wdist_vs_ncomp}       
        \end{subfigure}
         \caption{\small \ref{fig:gandk_marginal1}-\ref{fig:gandk_marginal4}: Posterior marginals in the univariate g-and-k example of Section \ref{sec:gandk_univariate} (grey, obtained via MCMC) and approximations by Wasserstein ABC from step 20 to step 57 of the SMC algorithm.  The color of the WABC approximation changes from red to blue as more steps of the SMC sampler are performed, decreasing the threshold $\varepsilon$. For the densities plotted here, the threshold reduces from $\varepsilon =  0.20$ to $\varepsilon =  0.07$. The range of the plot for $g$ has been truncated to $(1,6)$ to highlight the region of interest, despite the densities from the earlier steps of the SMC having support outside this region. Data-generating values are indicated by vertical lines. Figure \ref{fig:gandk_ncomp_vs_steps} shows the number of simulations from the model used in each step of the SMC algorithm ($y$-axis in log scale). This number is increasing due to use of the $r$-hit kernel within the SMC.  Figures \ref{fig:gandk_threshold_vs_ncomp} and \ref{fig:gandk_wdist_vs_ncomp} show the threshold $\varepsilon$ and the $\was_1$-distance to the posterior respectively, against the number of model simulations (both plots in log-log scale).}
        \label{fig:gandk:abc_conv}
\end{sidewaysfigure}

This problem does not arise in the WABC sequential Monte Carlo sampler, and
Figure \ref{fig:gandk:abc_conv} illustrates the behavior of the marginal WABC
posteriors as more steps of the SMC sampler are performed. In particular, we
can see that the WABC approximations for $a,b$ and $k$ converge to the
corresponding posteriors (up to some noise). The approximations for $g$ also
shows convergence towards the posterior, but have not yet reached the target
distribution at the stage when the sampler was terminated. The convergence is
further illustrated in Figure \ref{fig:gandk_wdist_vs_ncomp}, where the
$\was_1$- distance between the joint WABC posterior and joint posterior is
plotted as a function of the number of simulations from the model. The plot
shows that the Wasserstein distance between the distributions decreases from
around $10$ to around $0.06$ over the course of the SMC run. The distances are
approximated by thinning the MCMC samples left after burn-in down to $2,048$
samples, and computing the Wasserstein between the corresponding empirical
distribution and the empirical distributions supported at the $N = 2,048$ SMC
particles at each step. 

Figure \ref{fig:gandk_threshold_vs_ncomp} shows the development of the
threshold as a function of the number of model simulations, showing that
$\varepsilon$ decreases to $0.07$ over the course of the
SMC. The threshold decreases at a slower rate as it approaches zero, suggesting
that the underlying sampling problem becomes harder as $\varepsilon$ becomes
smaller. This is also illustrated by Figure \ref{fig:gandk_ncomp_vs_steps},
which shows the number of model simulations performed at each step of the SMC
algorithm. This number is increasing throughout the run of the algorithm, as
the $r$-hit kernel requires more and more attempts before it reaches the
desired number of hits.

\subsubsection{Bivariate ``g-and-k''} \label{sec:gandk_multivariate}
We also consider the bivariate extension of the g-and-k distribution \citep{drovandi2011likelihood},
where one generates bivariate Normals with mean zero, variance one, and correlation $\rho$,
and substitutes $z(r)$ with them in  \eqref{eq:gandk}, with parameters $(a_i,b_i,g_i,k_i)$ for each 
component $i\in \{1,2\}$.  Since the model generates bivariate data, we can no longer rely on simple sorting to 
calculate the Wasserstein distance. We compare the exact Wasserstein distance to
the approximations discussed in Section \ref{sec:distancecalculations}, as well as the maximum mean discrepancy, whose use within ABC was proposed by \citet[][]{park2015k2} (see Section \ref{subsect:related_works}).

We generate $n=500$ observations from the model using 
$a_1 = 3, b_1 = 1, g_1 = 1, k_1 = 0.5, a_2 = 4, b_2 = 0.5, g_2 = 2, k_2 = 0.4, \rho = 0.6$,
as in Section 5.2 of \citet{drovandi2011likelihood}.
The parameters $(a_i,b_i,g_i,k_i)$ are assigned a uniform prior on $[0,10]^4$, independently for $i\in\{1,2\}$,
and $\rho$ a uniform prior on $[-1,1]$. We estimate the posterior distribution by running $8$ Metropolis--Hastings chains for $150,000$ iterations, and discard the first $50,000$ as burn-in. For each of the ABC approximations, we run the SMC sampler outlined in Section \ref{sec:smcsamplers} for a total of $2\times 10^6$ simulations from the model. For the MMD, we use the estimator 
\begin{equation}
\text{MMD}^2(y_{1:n},z_{1:n}) = \frac{1}{n^2} \sum_{i,j=1}^n k(y_i,y_j) +  \frac{1}{n^2} \sum_{i,j=1}^n k(z_i,z_j) - \frac{2}{n^2} \sum_{i,j=1}^n k(y_i,z_j),
\end{equation}
with the kernel $k(x,x^\prime) = \exp\left(-\|x-x'\|^2/ 2h^2\right)$. The bandwidth $h$ was fixed to be the median of the set $\{\|y_i - y_j\|_1: i,j = 1, \dots, n\}$, following guidance in \citet{park2015k2}.

\begin{figure}[hp]
        \centering
        \begin{subfigure}[t]{0.42\textwidth}
            \centering
            \includegraphics[width=\textwidth]{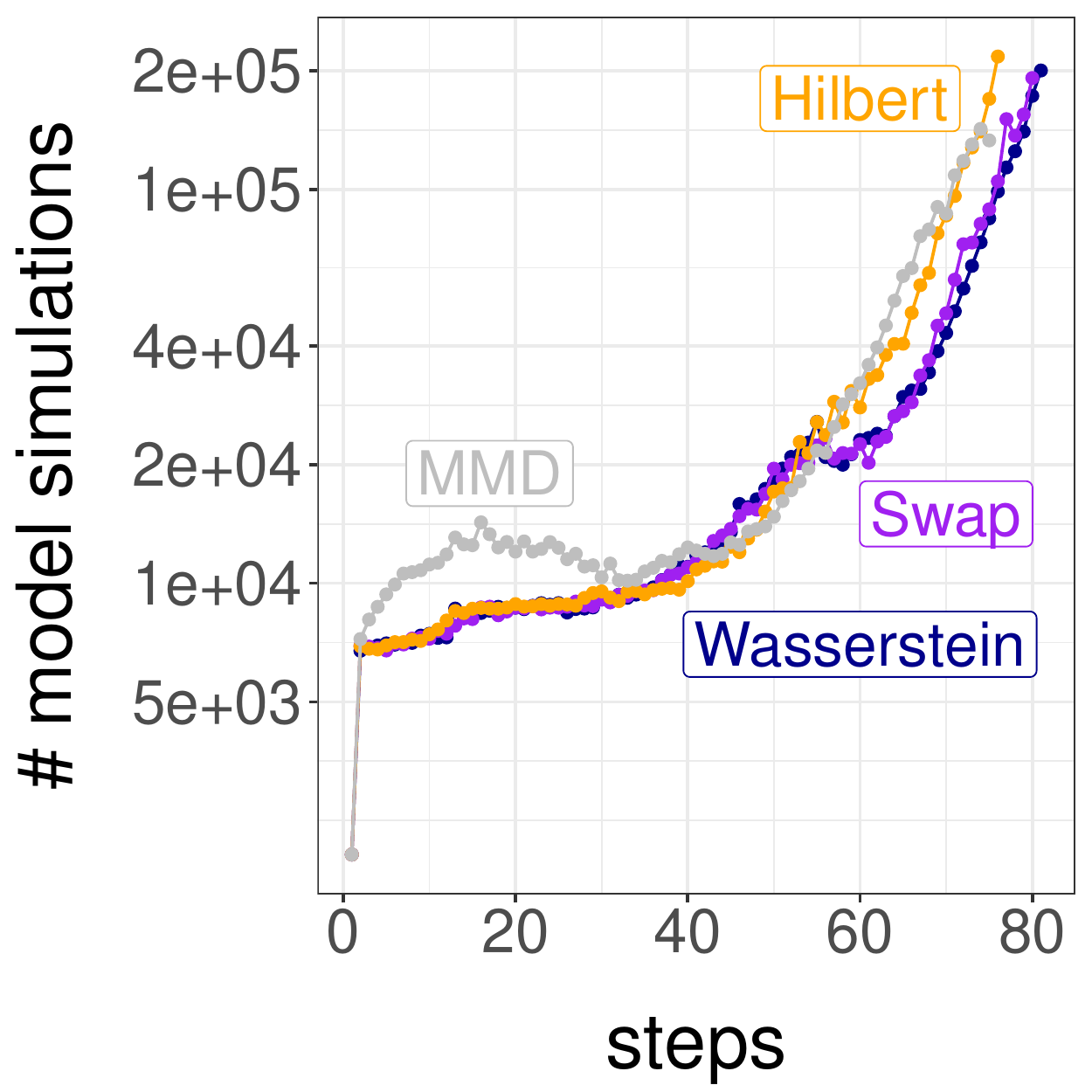}
            \caption{{\small Number of model simulations vs. SMC steps.}}    
            \label{fig:mgandk_ncomp_vs_steps}
        \end{subfigure}
        \hskip0.5cm
        \begin{subfigure}[t]{0.42\textwidth}
            \centering
            \includegraphics[width=\textwidth]{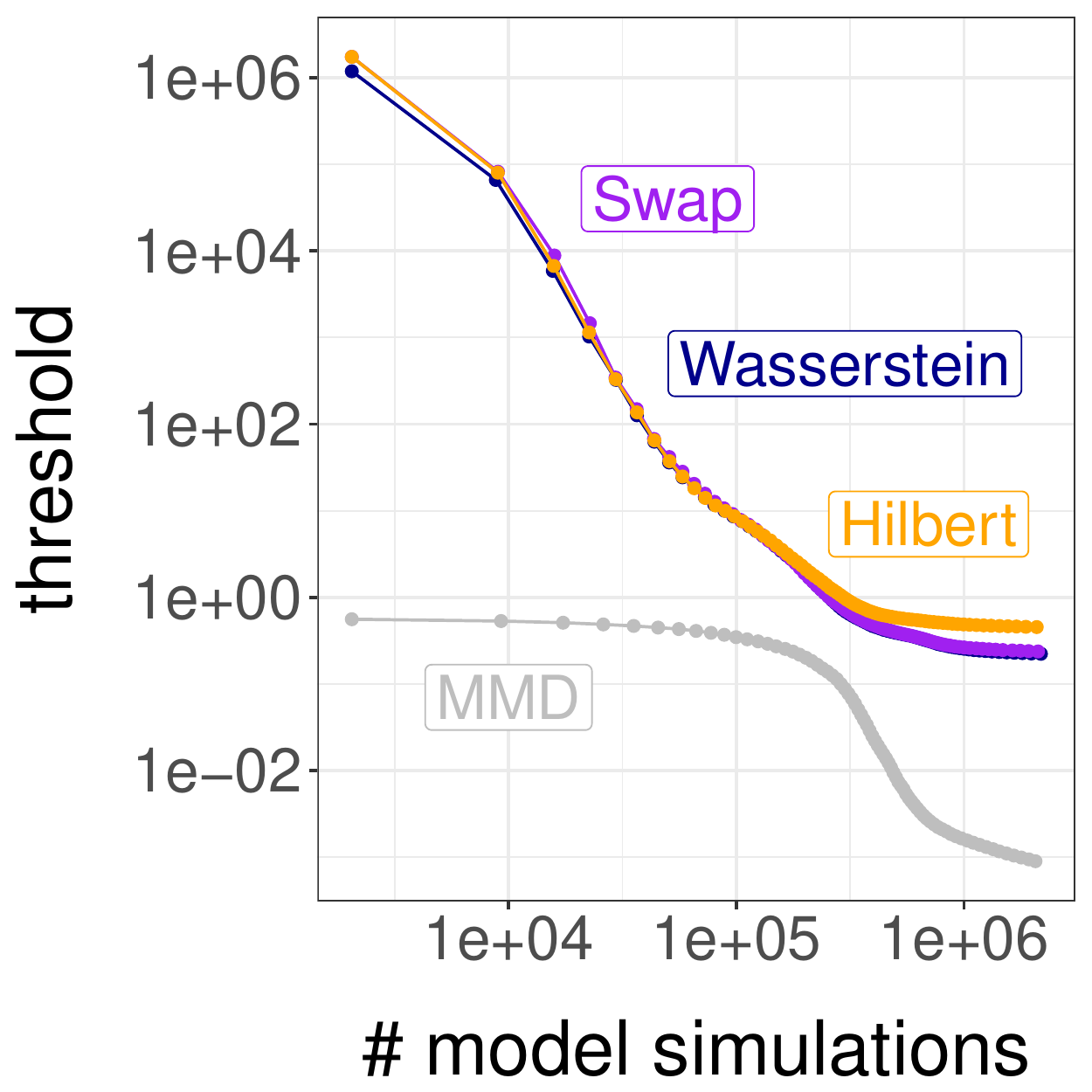}
            \caption{{\small Threshold $\varepsilon$ vs. number of model simulations}}    
            \label{fig:mgandk_threshold_vs_ncomp}
        \end{subfigure}
        \vskip1cm
        \begin{subfigure}[t]{0.86\textwidth}
            \centering
            \includegraphics[width=\textwidth]{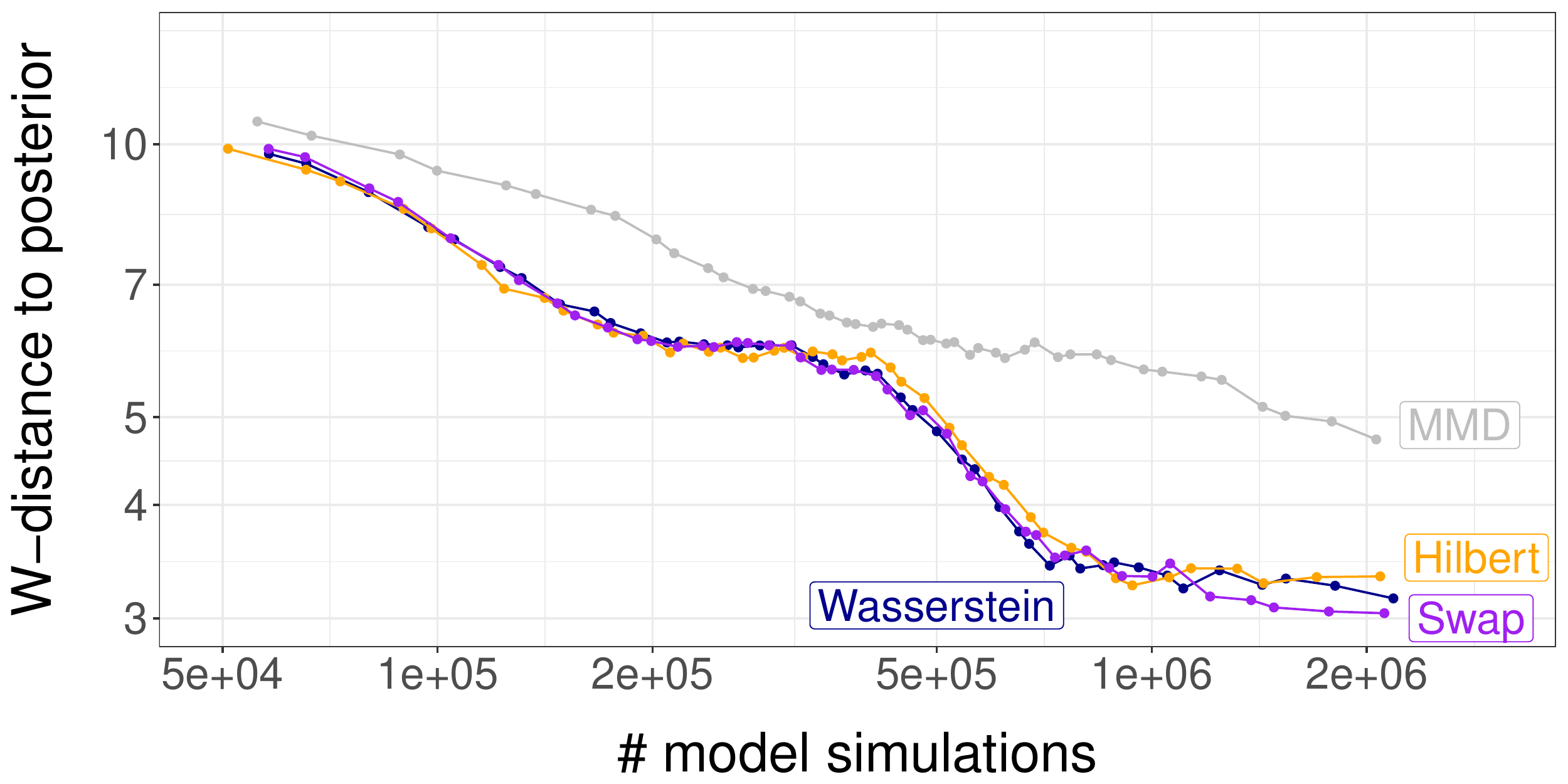}
            \caption{{\small $\was_1$-distance to posterior vs. number of model simulations.}}    
            \label{fig:mgandk_wdist_vs_ncomp}
        \end{subfigure}
        \caption{\small \ref{fig:mgandk_ncomp_vs_steps} shows the number of simulations from the model used in each step of the SMC algorithm for the four distances applied to the bivariate g-and-k model of Section \ref{sec:gandk_multivariate} ($y$-axis in log scale). This number is increasing due to use of the $r$-hit kernel within the SMC. Figure \ref{fig:mgandk_threshold_vs_ncomp} shows the thresholds $\varepsilon$ against the number of model simulations (in log-log scale). Note that the MMD is not on the same scale as the Wasserstein distance and its approximations, and therefore the MMD thresholds are not directly comparable to the those of the other distances. Figure \ref{fig:mgandk_wdist_vs_ncomp} shows the $\was_1$-distance between the joint ABC posteriors based on the different distances to the joint true posterior, against the number of model simulations (in log-log scale).} 
        \label{fig:mgandk:abc}
\end{figure}

Figure \ref{fig:mgandk_wdist_vs_ncomp} shows the $\was_1$-distance between the $N = 2,048$ ABC posterior samples, obtained
with the different distances, and a sample of $2,048$ points thinned out from the Markov chains targeting the posterior. 
This distance is plotted against the number of model simulations. 
It shows that all distances yield ABC posteriors that get closer to the actual posterior. On the other hand,
for this number of model simulations, all of the ABC posteriors are significantly different from the actual posterior.
For comparison, the $\was_1$-distance between two samples of size $2,048$ thinned out from the Markov chains
is on average about $0.07$. In this example, it appears that the MMD leads to ABC posteriors
that are not as close to the posterior as the other distances given the same budget of model simulations. The Hilbert distance
provides a particularly cheap and efficient alternative to the Wasserstein distance in this bivariate case, providing a very similar approximation to
the posterior as both the exact Wasserstein and swapping distances.

Figure \ref{fig:mgandk_threshold_vs_ncomp} shows the development of the threshold as a function of the number of model simulations for the different distances. Note that the MMD is not on the same scale as the Wasserstein distance and its approximations, and therefore the MMD thresholds are not directly comparable to the those of the other distances. As for the univariate g-and-k distribution, the thresholds decrease at a slower rate as they become smaller for each of the distances, suggesting that the underlying sampling problem becomes harder as $\varepsilon$ becomes smaller. The behaviors of the thresholds based on the exact Wasserstein, swapping, and Hilbert distances appear negligibly different. Figure \ref{fig:mgandk_ncomp_vs_steps} shows the number of model simulations performed at each step of the SMC algorithm for each of the distances. As before, these numbers are increasing throughout the run of the algorithm, as the $r$-hit kernel requires more and more attempts before it reaches the desired number of hits.

An important distinction between the distances is the time they take to compute. For data sets of size $n = 500$ simulated using the data-generating parameter, the average
wall-clock times to compute distances between simulated and observed data, on an Intel Core i7-5820K (3.30GHz), are as follows: 
$0.002s$ for the Hilbert distance, $0.01s$ for the MMD, $0.03s$ for the swapping distance, and $0.22s$
for the exact Wasserstein distance; these average times were computed on $1,000$ 
independent data sets. In this example, simulating from the model takes a negligible amount of time, even compared to the Hilbert distance.
Calculating the likelihood over $1,000$ parameters drawn from the prior, we find an average computing time of $0.05s$. In combination with the information
conveyed by Figure \ref{fig:mgandk:abc}, the Hilbert and swapping-based ABC posteriors provide good approximations of the exact Wasserstein-based ABC posteriors in only a fraction of the time the latter takes to compute.

\subsection{Toggle switch model \label{sec:toggleswitch}}

\begin{sidewaysfigure}[hp]
        \centering
        \begin{subfigure}[b]{0.23\textwidth}
            \centering
            \includegraphics[width=\textwidth]{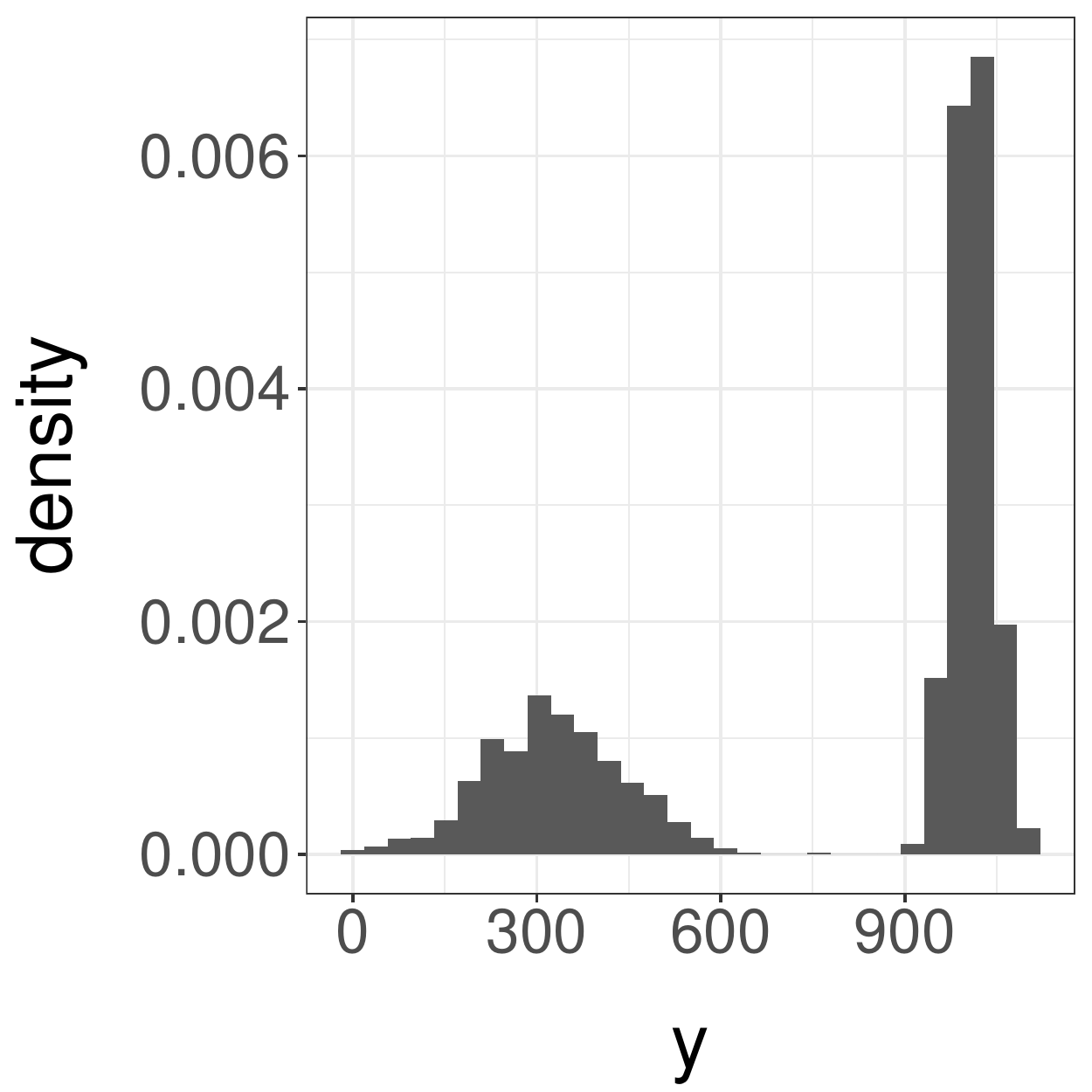}
            \caption{{\small Observations. }}    
            \label{fig:toggleswitch:data}
        \end{subfigure}
        \begin{subfigure}[b]{0.23\textwidth}
            \centering
            \includegraphics[width=\textwidth]{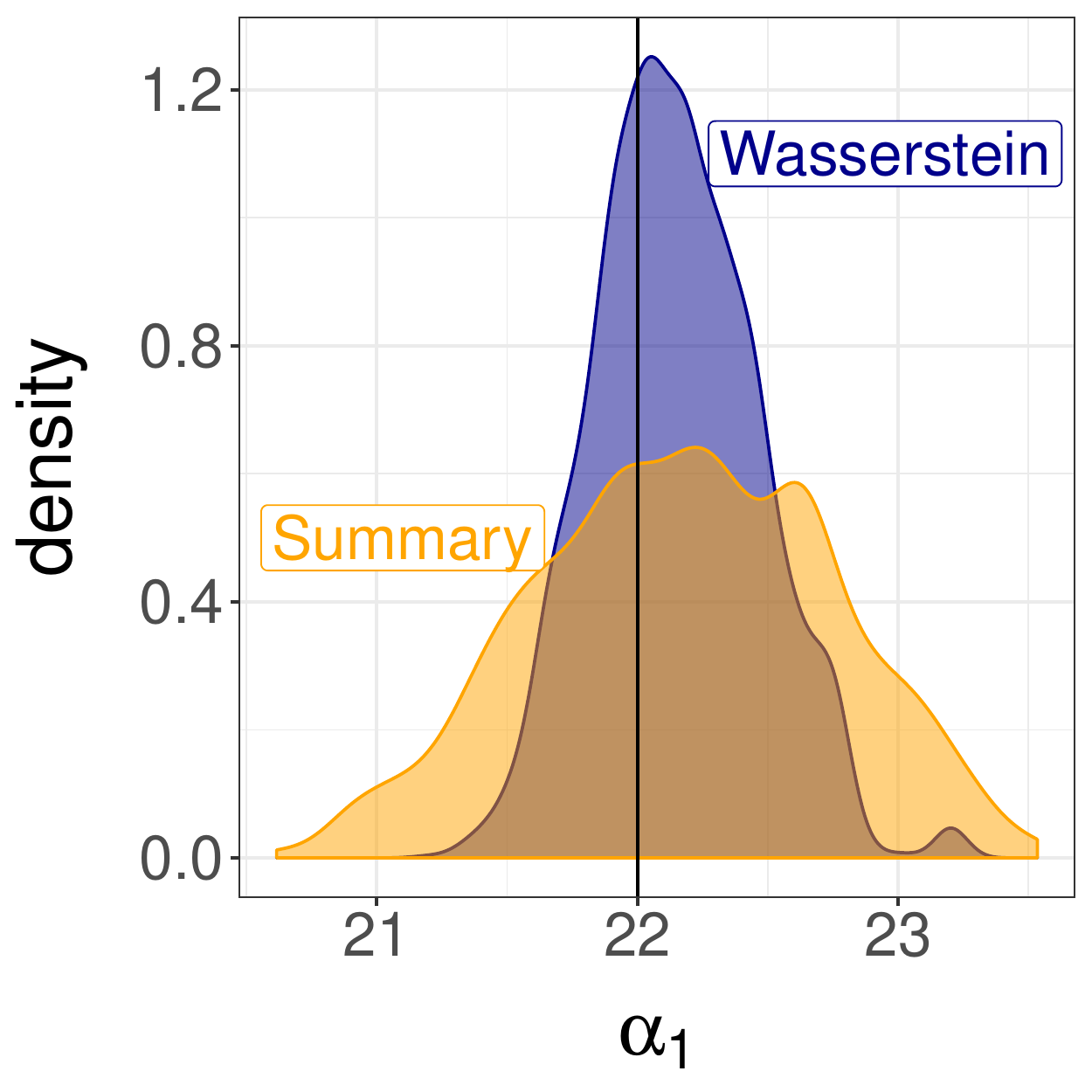}
            \caption{{\small Posteriors of $\alpha_1$. }}    
        \end{subfigure}
        \begin{subfigure}[b]{0.23\textwidth}
            \centering
            \includegraphics[width=\textwidth]{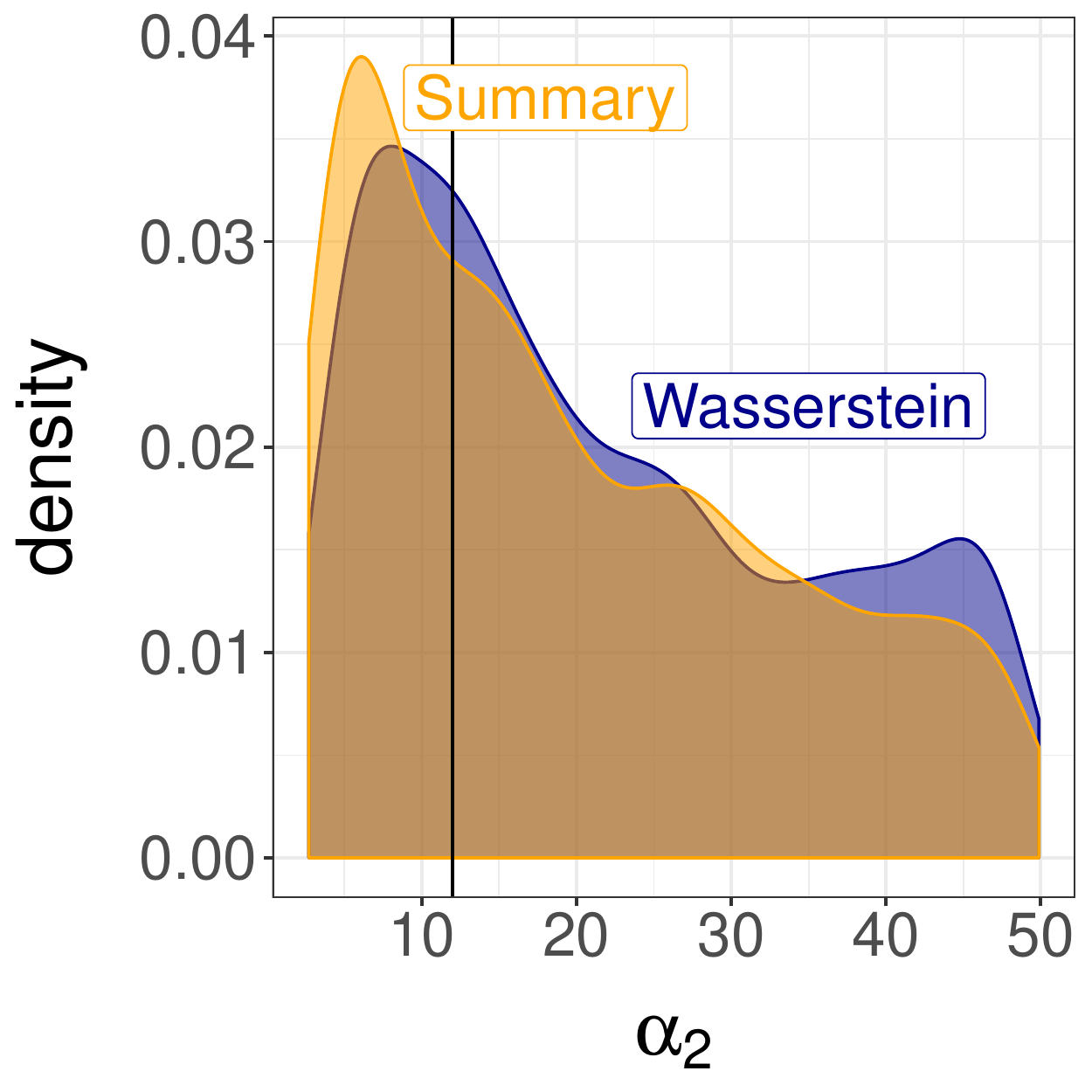}
            \caption{{\small Posteriors of $\alpha_2$.}}    
        \end{subfigure}
        \begin{subfigure}[b]{0.23\textwidth}
            \centering
            \includegraphics[width=\textwidth]{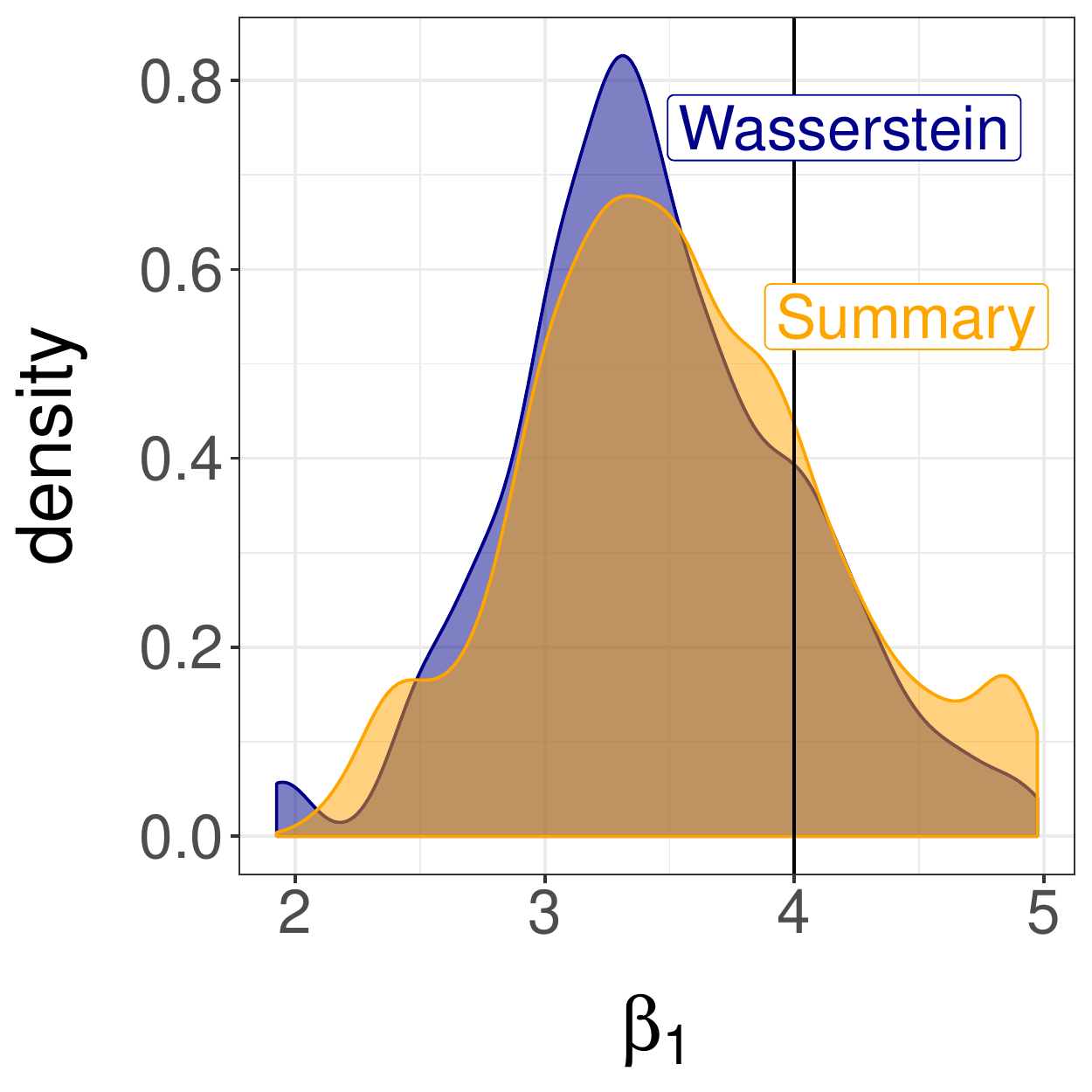}
            \caption{{\small Posteriors of $\beta_1$.}}    
        \end{subfigure}

        \begin{subfigure}[b]{0.23\textwidth}
            \centering
            \includegraphics[width=\textwidth]{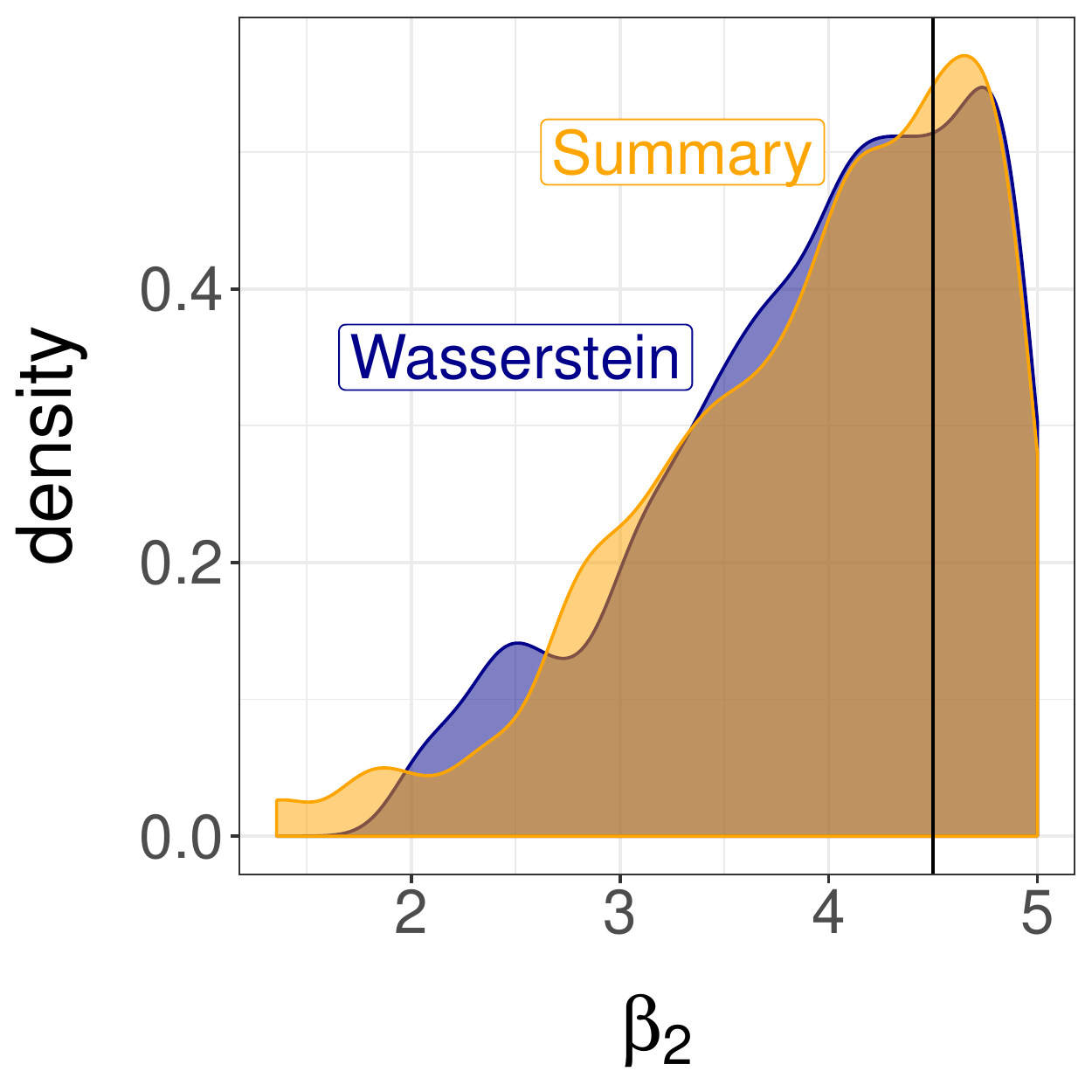}
            \caption{{\small Posteriors of $\beta_2$.}}    
        \end{subfigure}
        \begin{subfigure}[b]{0.23\textwidth}
            \centering
            \includegraphics[width=\textwidth]{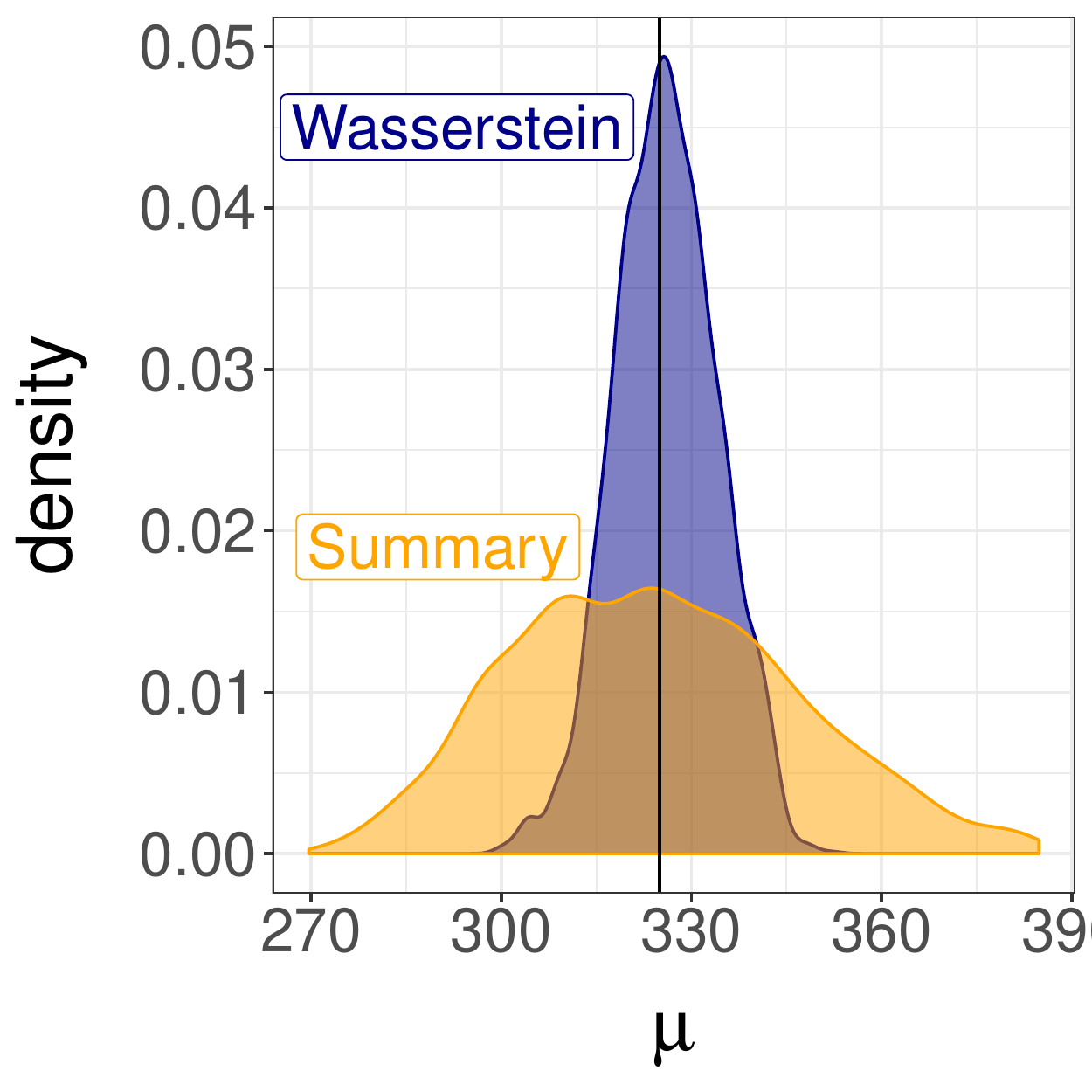}
            \caption{{\small Posteriors of $\mu$.}}    
        \end{subfigure}
        \begin{subfigure}[b]{0.23\textwidth}
            \centering
            \includegraphics[width=\textwidth]{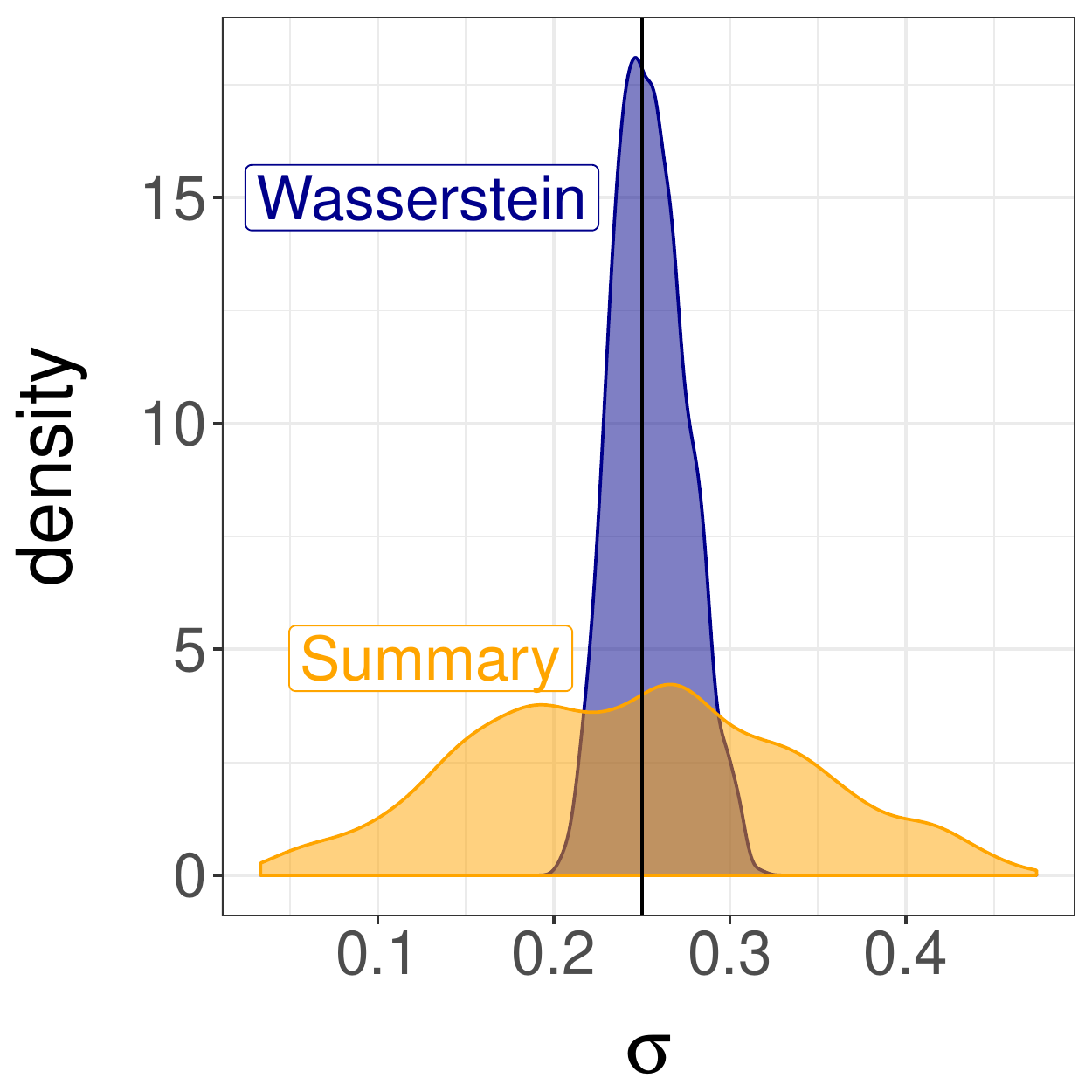}
            \caption{{\small Posteriors of $\sigma$.}}    
        \end{subfigure}
        \begin{subfigure}[b]{0.23\textwidth}
            \centering
            \includegraphics[width=\textwidth]{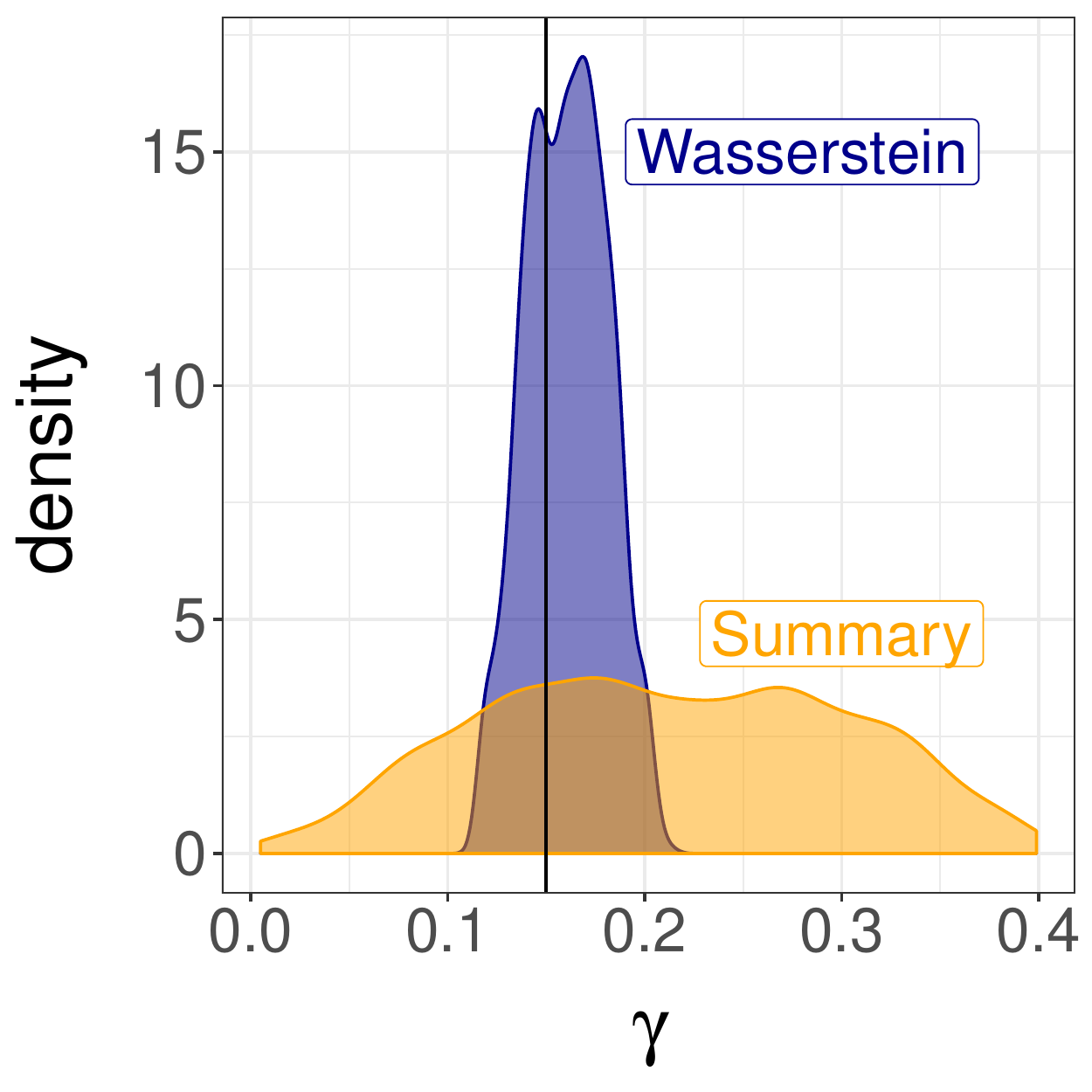}
            \caption{{\small Posteriors of $\gamma$.}}    
        \end{subfigure}
        %\hfill
        \caption{\small Histogram of observations (\ref{fig:toggleswitch:data}), and marginal posteriors based on WABC and the summary statistic from \citet{bonassi2011bayesian,bonassi2015sequential} in the toggle switch model. The ABC posteriors are computed using the SMC sampler from Section \ref{sec:smcsamplers}, for a total number
of $10^6$ model simulations. Data-generating values are indicated by vertical lines.}
        \label{fig:toggleswitch:marginal}
\end{sidewaysfigure}

We borrow the system biology ``toggle switch'' model used in \citet{bonassi2011bayesian,bonassi2015sequential},
inspired by studies of dynamic cellular networks. This provides an example where the design of specialized summaries can be replaced by the Wasserstein distance between empirical distributions. 
For $i\in \{1,\dots,n\}$ and $t\in \{1,\dots,T\}$, let $(u_{i,t}, v_{i,t})$ denote the expression levels of two genes in cell $i$ at time $t$.
Starting from $(u_{i,0},v_{i,0}) = (10,10)$, the evolution of $(u_{i,t}, v_{i,t})$ is given by 
\begin{align*}
    u_{i,t+1} &= u_{i,t} + \alpha_1/(1+v_{i,t}^{\beta_1}) - (1+0.03 u_{i,t}) + 0.5 \xi_{i,1,t},\\
    v_{i,t+1} &= v_{i,t} + \alpha_2/(1+u_{i,t}^{\beta_2}) - (1+0.03 v_{i,t}) + 0.5 \xi_{i,2,t},
\end{align*}
where $\alpha_1, \alpha_2, \beta_1, \beta_2$ are parameters, and $\xi$'s are standard Normal variables, truncated 
so that $(u_{i,t}, v_{i,t})$ only takes non-negative values. For each cell $i$, we only observe a noisy measurement of the terminal expression level $u_{i,T}$. Specifically, the observations $y_i$ are assumed to be independently 
distributed as $\mathcal{N} (\mu + u_{i,T}, \mu^2 \sigma^2 / u_{i,T}^{2\gamma})$ random variables truncated to be non-negative, where $\mu,\sigma,\gamma$ are parameters. 
We generate $n=2,000$ observations using
$\alpha_1 = 22$, $\alpha_2 = 12$, $\beta_1 = 4$, $\beta_2 = 4.5$, $\mu = 325$, $\sigma = 0.25$, $\gamma = 0.15$.
A histogram of the data is shown in Figure \ref{fig:toggleswitch:data}.

We consider the task of estimating the data-generating values, using uniform prior distributions 
on $[0,50]$ for $\alpha_1,\alpha_2$, on $[0,5]$ for $\beta_1,\beta_2$, on $[250,450]$ for $\mu$, 
$[0,0.5]$ for $\sigma$ and on $[0,0.4]$ for $\gamma$. These ranges are derived from Figure 5 in \citet{bonassi2015sequential}.
We compare our method using $p=1$ with a summary-based approach using the 11-dimensional tailor-made summary statistic from \citet{bonassi2011bayesian,bonassi2015sequential}. Since the data are one-dimensional, the Wasserstein distance between data sets can be computed quickly via sorting.
For both methods, we use the SMC sampler outlined in Section \ref{sec:smcsamplers}, for a total number
of $10^6$ model simulations. 

The seven marginal ABC posterior distributions obtained in the final step of the SMC sampler are shown in
Figure \ref{fig:toggleswitch:marginal}.  We find that the marginal WABC and summary-based posteriors concentrate to the same distributions for the $\alpha_2, \beta_1$ and $\beta_2$ parameters. On the remaining parameters, the marginal WABC posteriors show stronger concentration around the data-generating parameters than the summary-based approach. Comparing the results,
we see that the design of a custom summary can be bypassed using the Wasserstein distance
between empirical distributions: the resulting posterior approximations appear to be more concentrated around
the data-generating parameters,
and our proposed approach is fully black-box. The time to simulate data from the model does not seem to depend noticeably on the parameter, and the average wall-clock time to simulate a data set over 1,000 repetitions was $0.523s$ on an Intel Core i5 (2.5GHz). The average time compute the Wasserstein distance to the observed data set was $0.0002s$, whereas the average time to compute the summary statistic was $0.176s$. 

\subsection{Queueing model \label{sec:queue}}

We turn to the M/G/1 queueing model, which has appeared frequently as a test case in the
ABC literature, see e.g. \citet{fearnhead:prangle:2012}. It
provides an example where the observations are dependent, but where the parameters can
be identified from the marginal distribution of the data. In the model,
customers arrive at a server with independent interarrival times $w_i$,
exponentially distributed with rate $\theta_3$. Each customer is served with
independent service times $u_i$, taken to be uniformly distributed on
$[\theta_1,\theta_2]$. We observe only the interdeparture times $y_i$, given by
the process
$y_i = u_i + \max\{ 0, \sum_{j=1}^i w_j - \sum_{j=1}^{i-1} y_j\}$. The prior on
$(\theta_1, \theta_2-\theta_1, \theta_3)$ is Uniform on
$[0,10]^2\times[0,1/3]$.

We use the data set given in \citet{shestopaloff2014bayesian}, which was
generated using the parameters $(\theta_1,\theta_2-\theta_1,\theta_3) = (4,3,0.15)$ and
$n=50$. The WABC posterior based on the empirical distribution of
$y_{1:n}$, ignoring dependencies, is approximated using the SMC algorithm of Section
\ref{sec:smcsamplers}, with a budget of $10^7$ model simulations. We compare with the semi-automatic ABC approach of
\citet{fearnhead:prangle:2012} with the same budget of model simulations, using a subset of 20 evenly spaced order statistics as the initial summary statistics in that method. The semi-automatic ABC posteriors are computed using the rejection sampler in the \texttt{abctools} package \citep{nunes2015abctools}, accepting the $100$ best samples.
The actual posterior distribution is approximated with a particle marginal Metropolis--Hastings (PMMH) run \citep{andrieu:doucet:holenstein:2010},
using $4,096$ particles and $10^5$ iterations. The
use of PMMH was suggested in \citet{shestopaloff2014bayesian} as an alternative to the model-specific Markov chain Monte Carlo algorithm they propose. 

\begin{figure}[h]
        \centering
        \begin{subfigure}[b]{0.32\textwidth}
            \centering
            \includegraphics[width=\textwidth]{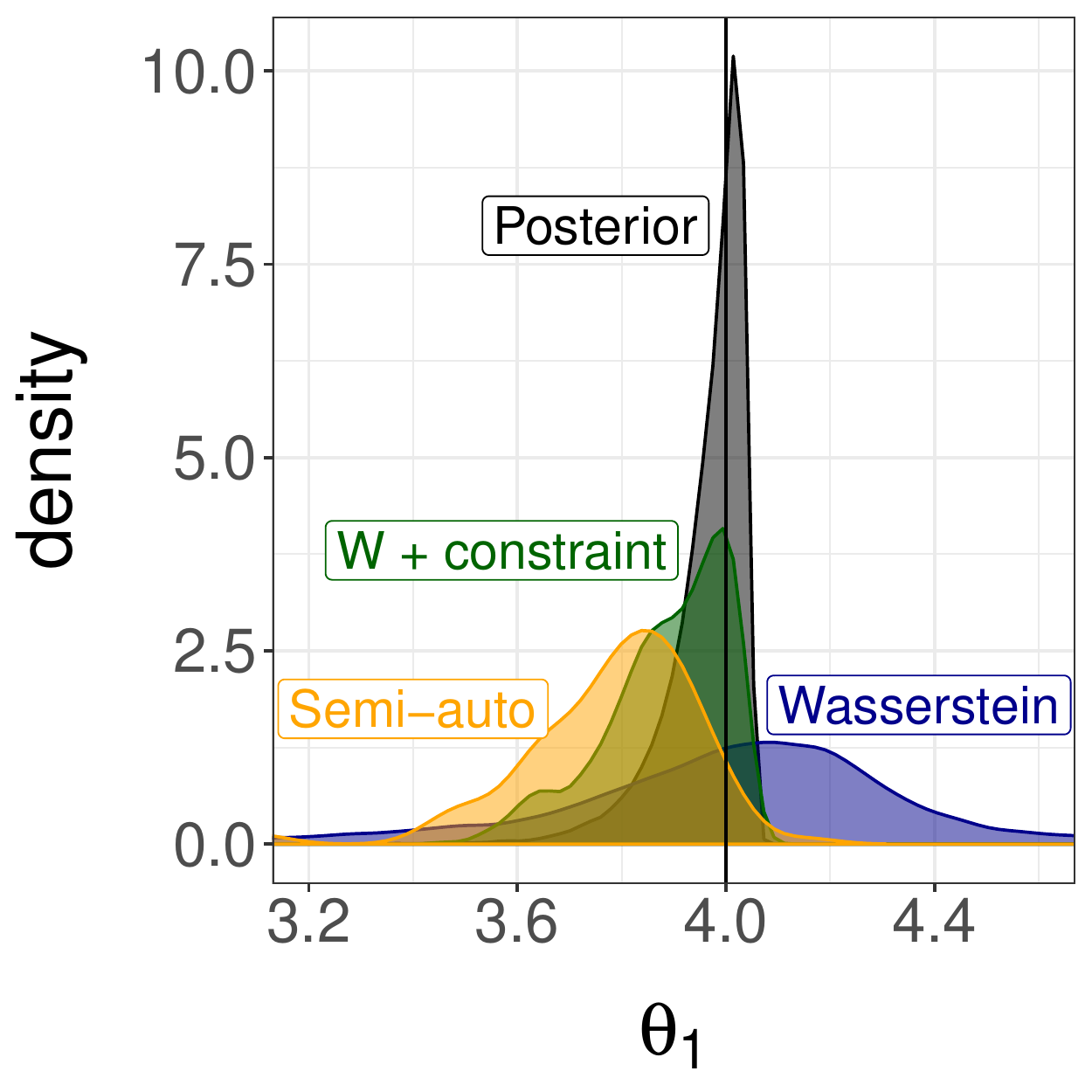}
            \caption{{\small Posteriors of $\theta_1$.}}    
        \end{subfigure}
        %\hspace*{1cm}
        \begin{subfigure}[b]{0.32\textwidth}
            \centering
            \includegraphics[width=\textwidth]{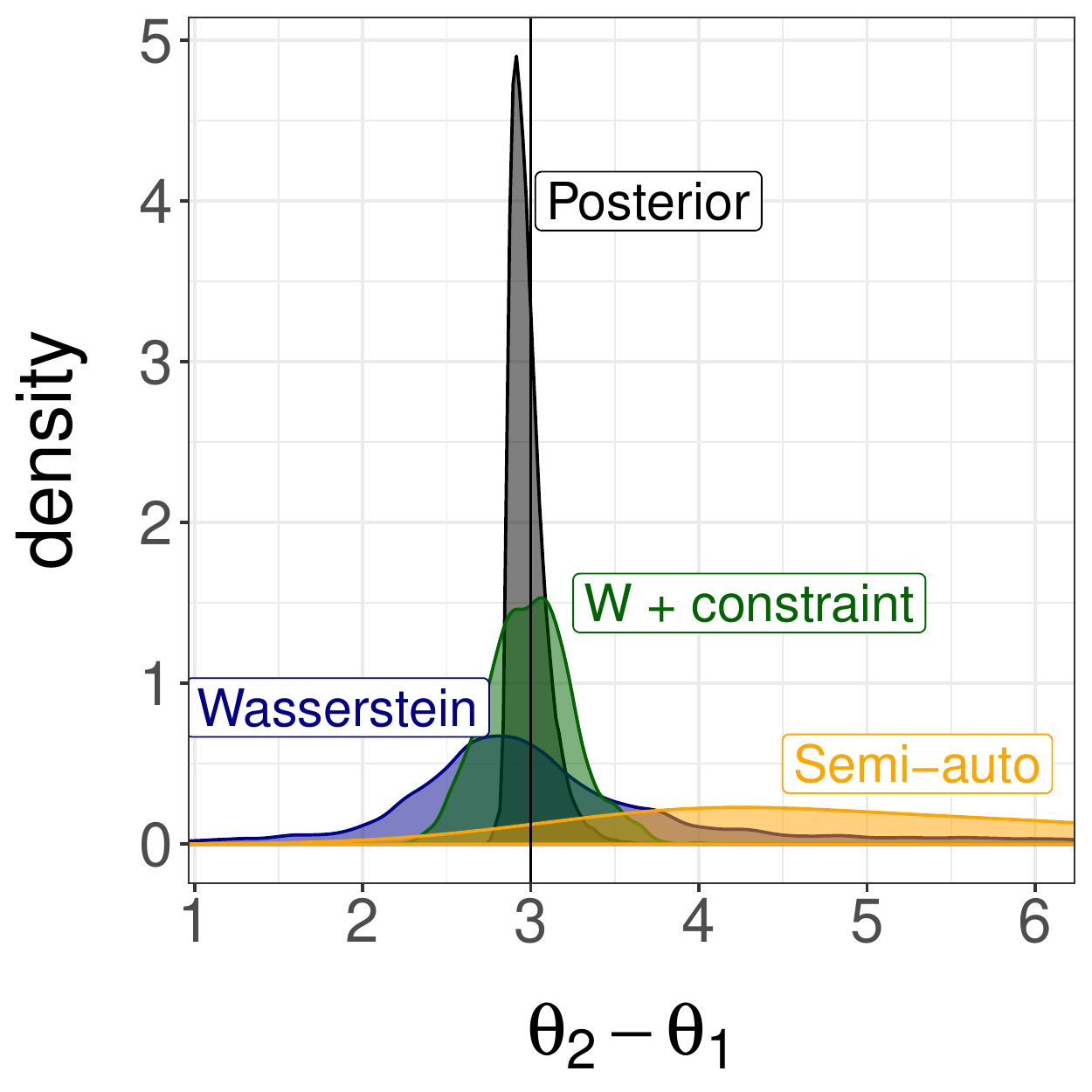}
            \caption{{\small Posteriors of $\theta_2-\theta_1$.}}    
        \end{subfigure}
        \begin{subfigure}[b]{0.32\textwidth}
            \centering
            \includegraphics[width=\textwidth]{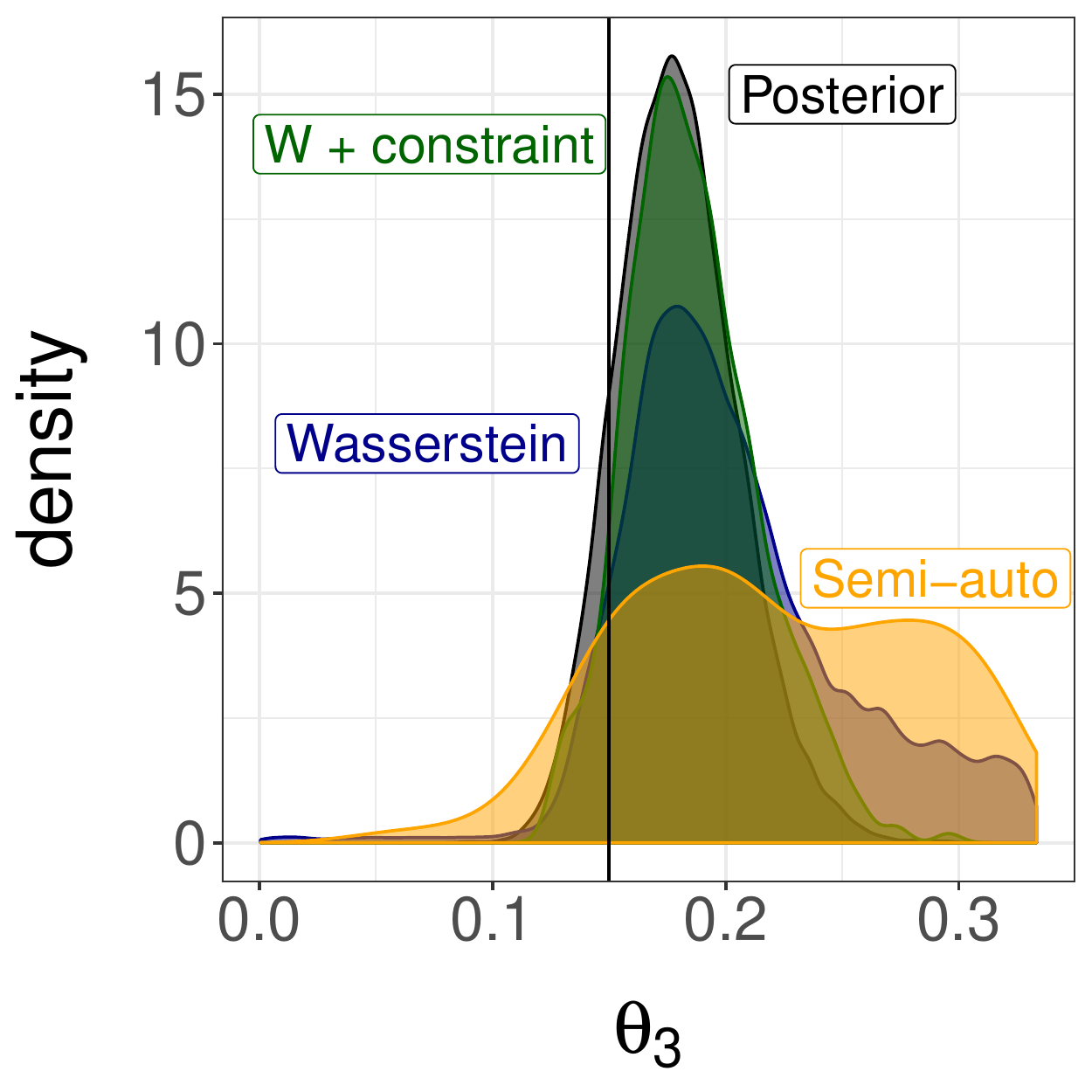}
            \caption{{\small Posteriors of $\theta_3$.}}    
        \end{subfigure}
        \caption{\small Posterior marginals in the M/G/1 queueing model of Section \ref{sec:queue} (obtained via particle marginal Metropolis--Hastings), approximations by Wasserstein ABC and semi-automatic ABC, and Wasserstein ABC accounting for the constraint that $\theta_1$ has to be less than $\min_{i\in 1:n} y_i$, each with a budget of $10^7$ model simulations. Data-generating values are indicated by vertical lines.}
        \label{fig:queue:intermediate}
\end{figure}

Upon observing $y_{1:n}$, $\theta_1$ has to be less than
$\min_{i\in 1:n} y_i$, which is implicitly encoded in the likelihood, but
not in an ABC procedure.  One can add this constraint explicitly, rejecting
parameters that violate it, which is equivalent to redefining the prior on $\theta_1$
to be uniform on $[0,\min_{i\in 1:n} y_i]$.  Figure
\ref{fig:queue:intermediate} shows the marginal distributions
of the parameters obtained with PMMH, semi-automatic ABC, and WABC,
with or without the additional constraint. 

Overall, the WABC approximations are close to the posterior,
in comparison to the relatively vague prior distribution on $(\theta_1,\theta_2-\theta_1)$. Furthermore, 
we see that incorporating the constraint leads to marginal WABC approximations that are closer to the marginal posteriors.
Both variations of WABC appear to perform better than semi-automatic ABC, except on $\theta_1$, where the semi-automatic ABC approximation is closer to the posterior than the unconstrained WABC approximation.
We observed no significant difference in the semi-automatic ABC posterior when incorporating the constraint on $\theta_1$, 
and hence only show the approximated posterior for the unconstrained approach.

As in the univariate g-and-k model of Section \ref{sec:gandk_univariate}, the computation costs for the WABC and semi-automatic approaches are similar, as they both rely on simulating from the model and sorting the resulting data. Over 1,000 repetitions, the average wall-clock time to simulate a data set was $7.5\times 10^{-5} s$ on an Intel Core i5 (2.5GHz). Sorting a data set took on average $7.7\times 10^{-5}s$, and computing the Wasserstein distance was negligibly different from this. For the semi-automatic ABC approach, one additionally has to perform the regression step.   The model simulations in semi-automatic ABC are easier to parallelize, but the method is hard to scale up without specialized tools for large-scale regression,
due to memory requirements of the regression used to construct the summary statistics.

\subsection{L\'evy-driven stochastic volatility model\label{sec:levydriven}}

We consider a L\'evy-driven stochastic volatility model \citep[e.g.][]{bns:real},
used in \citet{chopin2012smc2} as a challenging example of parameter inference in state space models.
We demonstrate how ABC with transport distances can identify some of the parameters in a black-box fashion,
and can be combined with summaries to identify the remaining parameters.
The observation $y_{t}$ at time $t$ is the log-return of a financial asset, assumed Normal with mean $\mu+\beta v_t$ and variance $v_t$,
where $v_t$ is the actual volatility. Together with the spot volatility $z_t$,
the pair $(v_t,z_t)$ constitutes a latent Markov chain, assumed to follow a L\'evy process.
Starting with $z_0 \sim \Gamma(\xi^2/\omega^2,\xi/\omega^2)$ (where the second parameter is the rate), and an arbitrary $v_0$,
the evolution of the process goes as follows:
\begin{equation}
  \label{eq:ssf}
  \begin{split}
      k  & \sim  \mathcal{P}\text{oisson}\left( \lambda \xi^2/\omega^2 \right)\,, \quad 
      c_{1:k}  \overset{\text{i.i.d.}}{\sim} \mathcal{U}(t,t+1)\,, \quad 
 e_{1:k}  \overset{\text{i.i.d.}}{\sim}  \mathcal{E}\text{xp}\left(\xi/\omega^2 \right),  \\ 
  z_{t+1}  & =  e^{-\lambda} z_t + \sum_{j=1}^k  e^{-\lambda(t+1-c_j)}
  e_j\,, \quad
v_{t+1} =  {\frac{1}{\lambda}} [ z_t - z_{t+1} + \sum_{j=1}^k e_j
].
  \end{split}
\end{equation}
The random variables $(k,c_{1:k},e_{1:k})$ are generated
independently for each time period, and $1:k$ is the
empty set when  $k=0$. The parameters are $(\mu,\beta,\xi,\omega^2,\lambda)$. We specify the prior as Normal with mean zero and 
variance $2$ for $\mu$ and $\beta$, Exponential with rate $0.2$
for $\xi$ and $\omega^2$, and Exponential with rate $1$ for $\lambda$.

\begin{figure}[hp]
        \centering
        \begin{subfigure}[t]{0.33\textwidth}
            \centering
            \includegraphics[width=\textwidth]{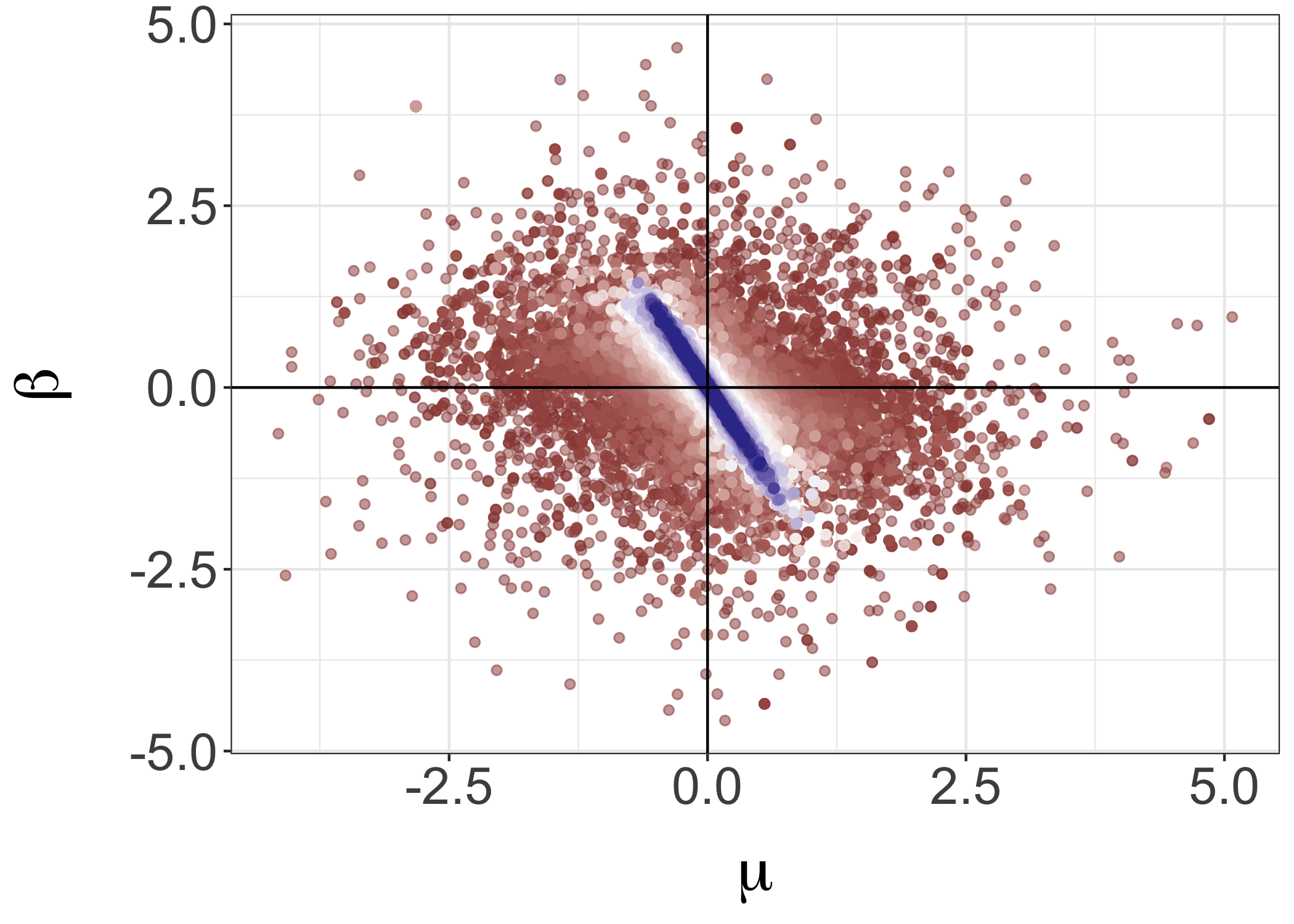}
            \caption{{\small Posteriors of $(\mu,\beta)$.}}    
        \end{subfigure}
        %\hspace*{1cm}
        \begin{subfigure}[t]{0.33\textwidth}
            \centering
            \includegraphics[width=\textwidth]{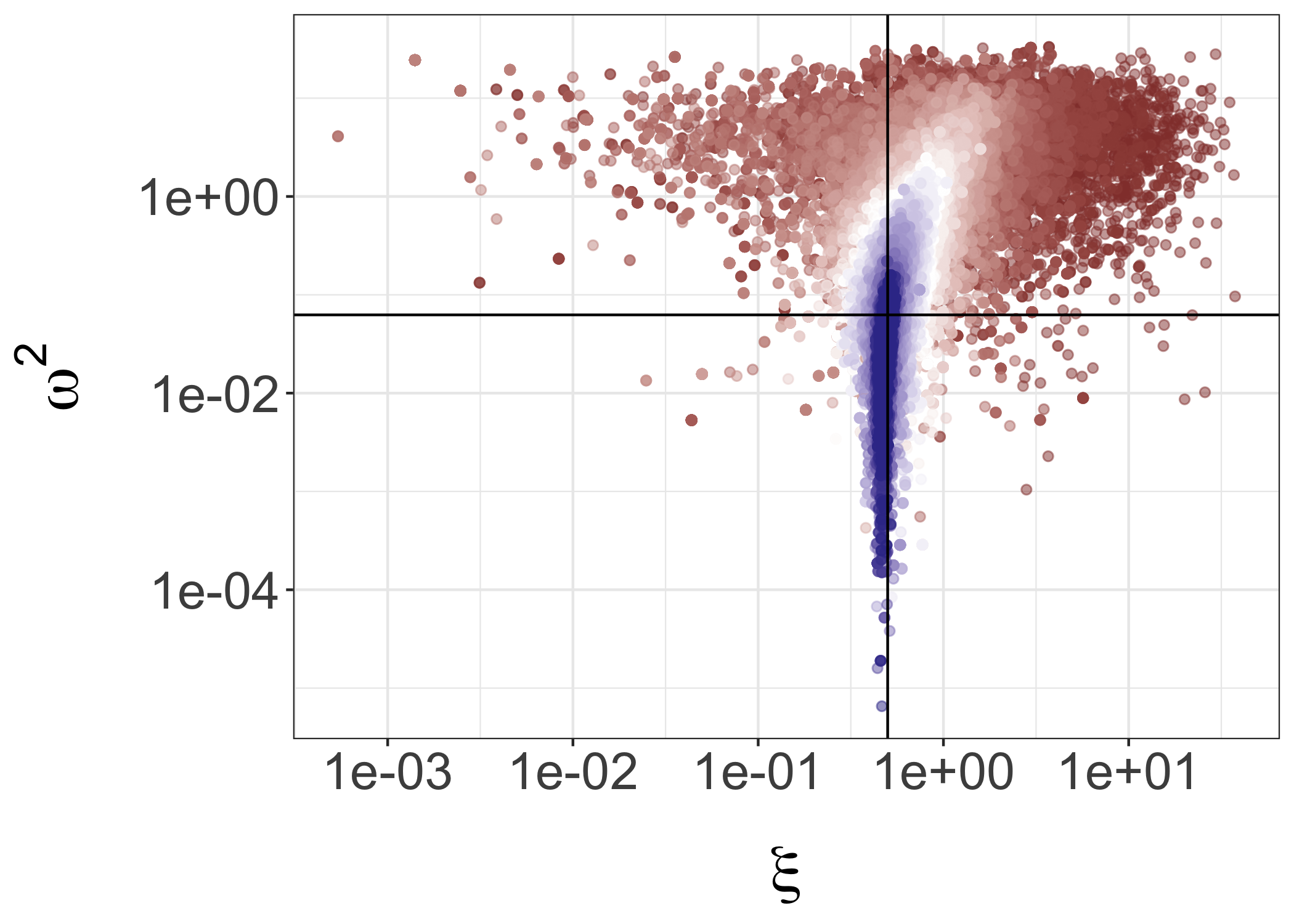}
            \caption{{\small Posteriors of $(\xi,\omega^2)$ (log-log).}}    
        \end{subfigure}
        \begin{subfigure}[t]{0.32\textwidth}
            \centering
            \includegraphics[width=\textwidth]{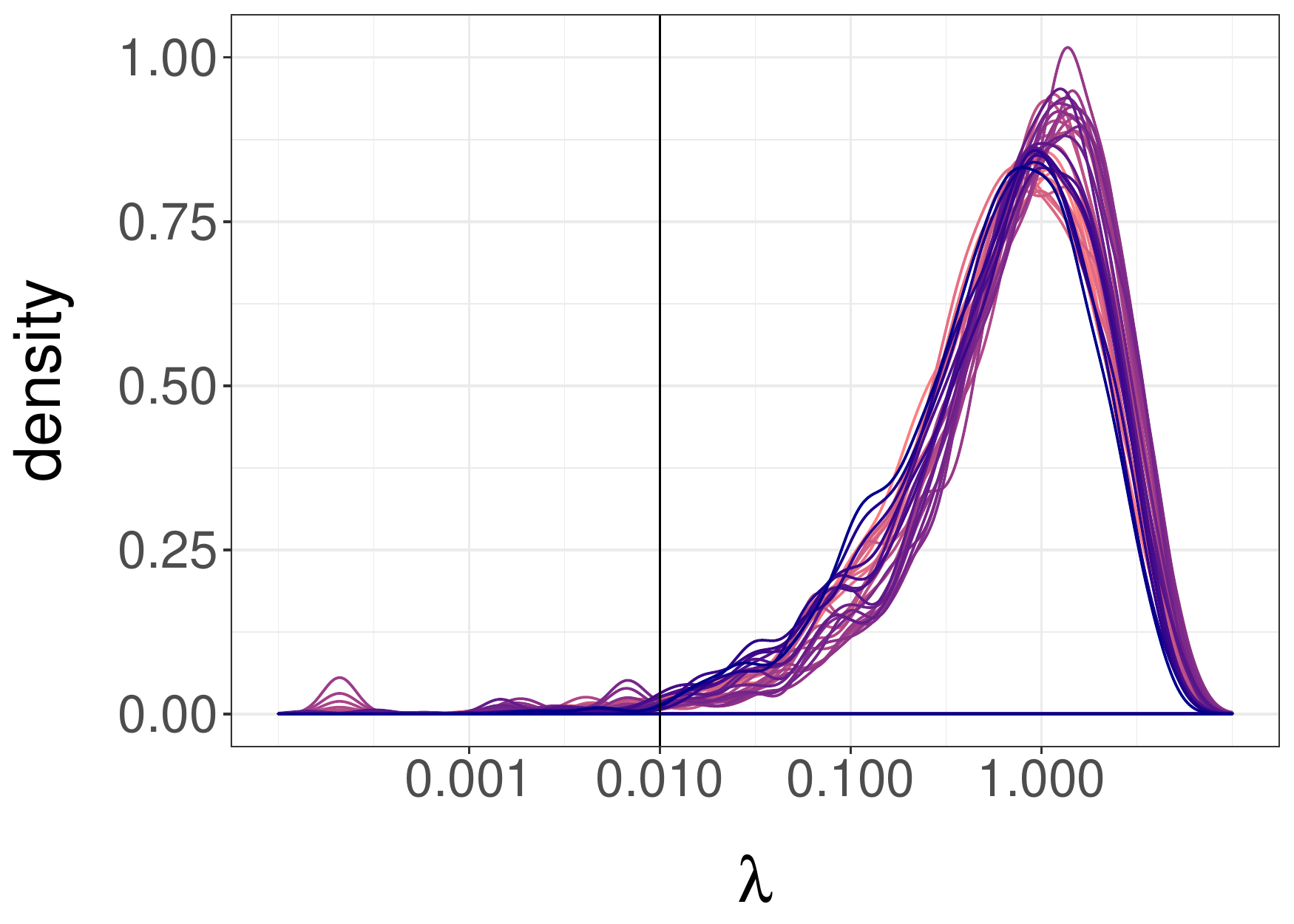}
            \caption{{\small Posteriors of $\lambda$ (in log scale).}}    
        \end{subfigure}
        \caption{\small ABC approximations in the L\'evy-driven stochastic
            volatility model, using the Hilbert distance between delay
            reconstructions with lag $k=1$.  The plots show samples from the bivariate
            marginals of $(\mu, \beta)$ (left), $(\xi,\omega^2)$ (middle), and
            the marginal distributions of $\lambda$ (right), as the threshold $\varepsilon$ decreases during the steps of the
            SMC sampler (colors from red to blue).
             The total budget was $4.2 \times 10^5$ model simulations.  Data-generating parameters
    are indicated by full lines.}
        \label{fig:levydriven:quasiposterior}
\end{figure}

We generate synthetic data with $\mu = 0$, $\beta = 0$, $\xi =
0.5$, $\omega^2 = 0.0625$, $\lambda = 0.01$, which were used also in the
simulation study of \cite{bns:real,chopin2012smc2}, 
of length $n=10,000$. We use delay
reconstruction with a lag $k=1$, and the Hilbert distance $\mathfrak{H}_p$ of Section
\ref{sec:hilbert} with $p=1$. Given the length of the time series,
the cost of computing the Hilbert distance is much smaller than that of the other 
distances discussed in Section \ref{sec:distancecalculations}.
We ran the SMC algorithm outlined in Section \ref{sec:smcsamplers} until a total of $4.2\times 10^5$ data sets had been simulated.
Figure \ref{fig:levydriven:quasiposterior} shows the resulting quasi-posterior
marginals for $(\mu,\beta)$, $(\xi,\omega^2)$, and $\lambda$. 
The parameters $(\mu,\beta,\xi,\omega^2)$ are accurately identified,
from a vague prior to a region close to the data-generating values.
On the other hand, the approximation of $\lambda$ is barely different from the prior 
distribution. Indeed, the parameter $\lambda$ represents a discount rate
which impacts the long-range dependencies
of the process, and is thus not captured by the bivariate marginal distribution
of $(y_t,y_{t-1})$.

\begin{figure}[hp]
        \centering
        \begin{subfigure}[t]{0.32\textwidth}
            \centering
            \includegraphics[width=\textwidth]{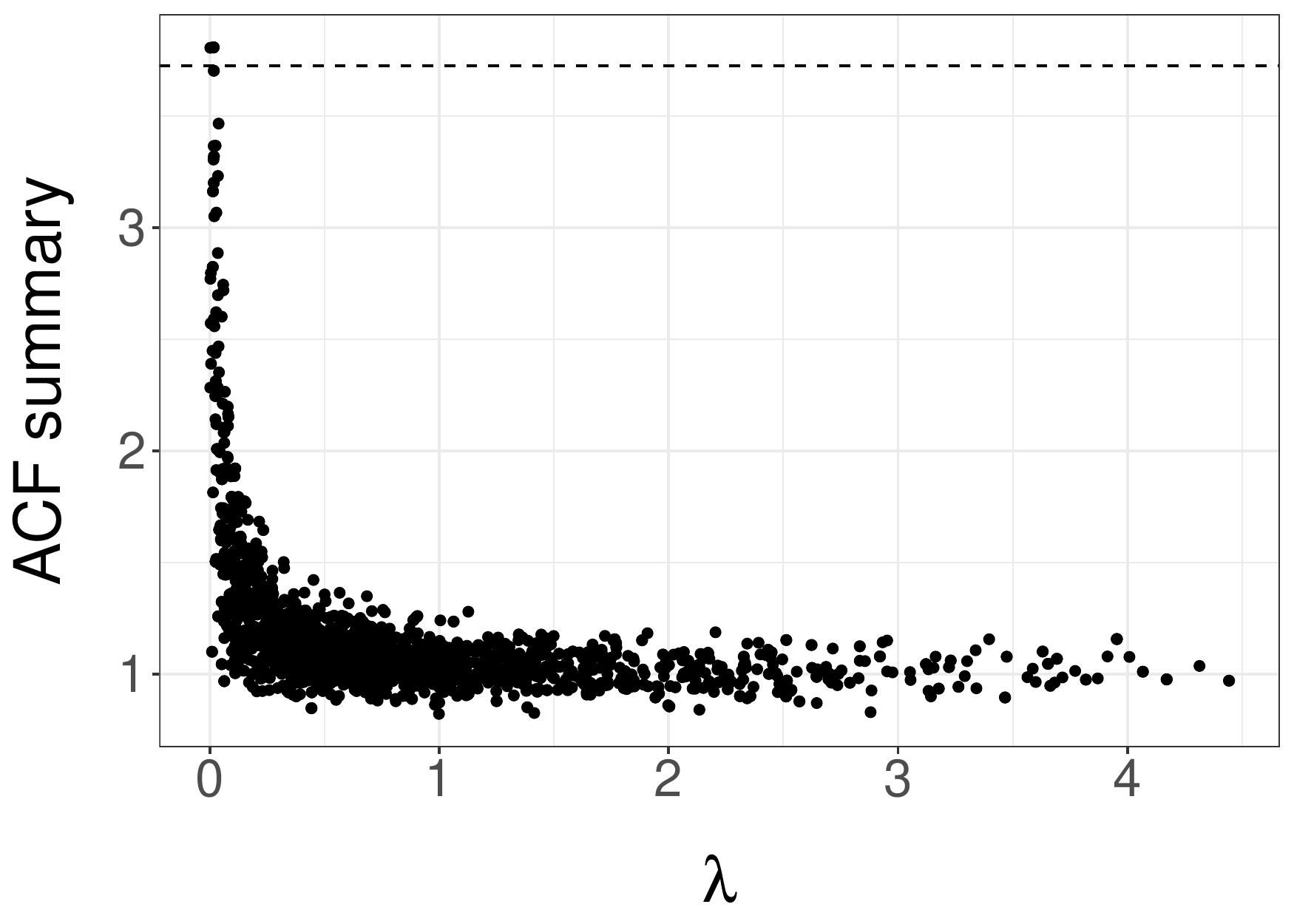}
            \caption{{\small Summary  against $\lambda$.} \label{fig:levydriven:acfsummary}}    
        \end{subfigure}
        %\hspace*{1cm}
        \begin{subfigure}[t]{0.32\textwidth}
            \centering
            \includegraphics[width=\textwidth]{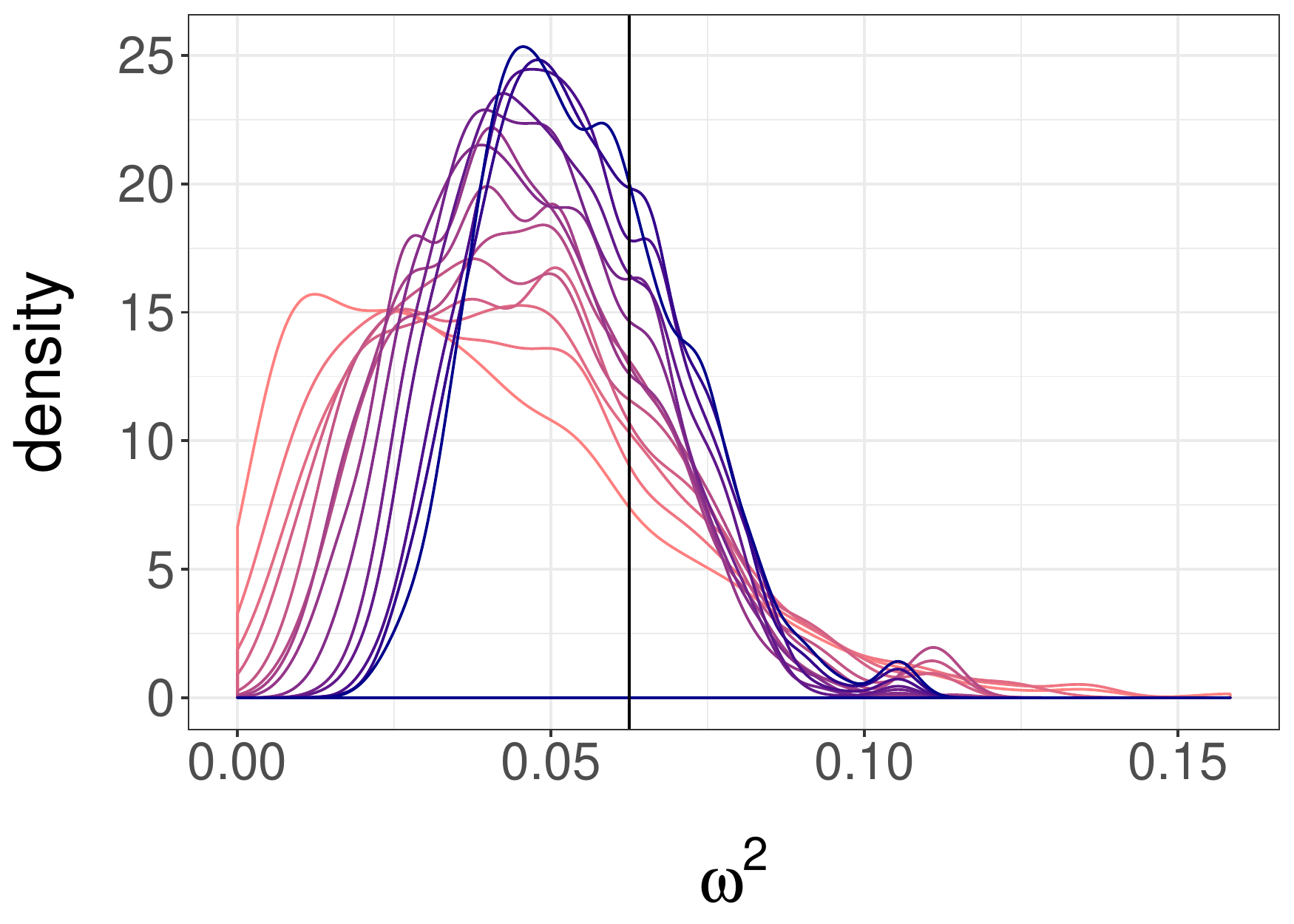}
            \caption{{\small Distributions of $\omega^2$.}\label{fig:levydriven:omega2summary}}    
        \end{subfigure}
        \begin{subfigure}[t]{0.32\textwidth}
            \centering
            \includegraphics[width=\textwidth]{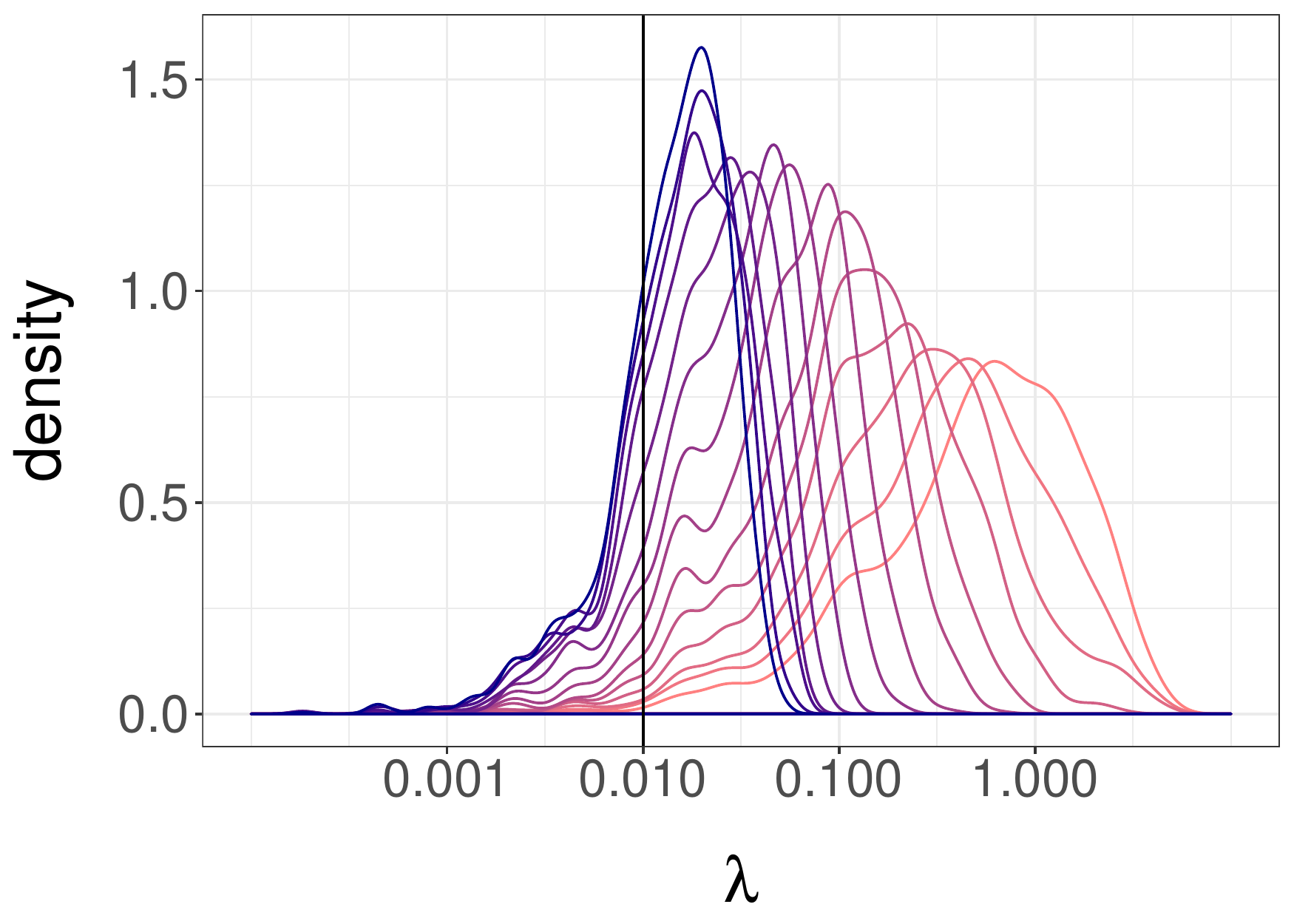}
            \caption{{\small Distributions of $\lambda$ (log-scale).}\label{fig:levydriven:lambdasummary}}    
        \end{subfigure}
        \caption{\small Left: summary, defined as the sum of the first $50$ sample autocorrelations
            of the squared series, against $\lambda$, computed for the output of the WABC algorithm using the Hilbert
            distance between delay reconstructions, applied to the L\'evy-driven stochastic volatility model of Section \ref{sec:levydriven}.
            Middle and right: approximations of $\omega^2$ and $\lambda$, from the second run of WABC 
            using the Hilbert distance between delay reconstructions combined with
            the summary for another $6.6 \times 10^5$ model simulations. The colors change from red to blue as more steps of the SMC sampler are performed. The horizontal axis in the right plot is in log-scale, and
            illustrates the concentration of the ABC posterior towards the data-generating value
            of $\lambda$.}
        \label{fig:levydriven:withsummary}
\end{figure}

Hoping to capture long-range dependencies in the series, we define a summary
$\eta(y_{1:n})$ as the sum of the first $50$ sample autocorrelations
among the squared observations.  
For each of the parameters obtained with the first run of WABC described above,
we compute the summary of the associated synthetic data set.
We plot the summaries against $\lambda$ in Figure
\ref{fig:levydriven:acfsummary}.  The dashed line indicates the value of the
summary calculated on the observed data. The plot shows that
the summaries closest to the observed summary are those obtained with
the smallest values of $\lambda$. Therefore,
we might be able to learn more about $\lambda$ by 
combining the previous Hilbert distance with a distance between summaries.

Denote by $\mathfrak{H}_1(\tilde{y}_{1:n},\tilde{z}_{1:n})$ the Hilbert
distance between delay reconstructions, and by $\varepsilon_h$ the threshold
obtained after the first run of the algorithm.  A new distance between data
sets is defined as $|\eta(y_{1:n}) - \eta(z_{1:n})|$ if
$\mathfrak{H}_1(\tilde{y}_{1:n},\tilde{z}_{1:n}) < \varepsilon_h$, and
$+\infty$ otherwise. We then run the SMC sampler of Section
\ref{sec:smcsamplers}, initializing with the results of the first run, but using the
new distance. In this second run, a new threshold is introduced and adaptively
decreased, keeping the first threshold $\varepsilon_h$ fixed. One could also
decrease both thresholds together or alternate between decreasing either.  Note
that the Hilbert distance and the summaries could have been combined in other
ways, for instance in a weighted average.

We ran the algorithm with the new distance 
for an extra $6.6 \times 10^5$ model simulations.
Figures \ref{fig:levydriven:omega2summary} and
\ref{fig:levydriven:lambdasummary} show the evolution 
of the WABC posterior distributions of $\omega^2$ and $\lambda$ during the second run. The WABC posteriors concentrate
closer to the data-generating values, particularly for $\lambda$;
for $(\mu,\beta,\xi)$, the effect is minimal and not shown.
In terms of computing time, it took on average $1.3\times 10^{-1}s$ to generate time series given the data-generating parameter, $2.4\times 10^{-2}s$ to compute the Hilbert distance, and $1.5\times 10^{-3}s$ to compute the summary statistic, on an Intel Core i5 (2.5GHz). Thus, most of the  time consumed by the algorithm was spent generating data.

The WABC posterior could then be used to initialize a particle MCMC algorithm
\citep{andrieu:doucet:holenstein:2010} targeting the posterior. 
The computational budget of roughly $1.1 \times 10^6$
model simulations, as performed in total by the WABC procedure in this section,
would be equivalent to relatively few iterations of particle MCMC in terms of number of model transitions. 
Therefore, the cost of initializing a particle MCMC algorithm
with the proposed ABC approach is likely to be negligible. The approach
could be valuable in settings where it is difficult to initialize particle MCMC algorithms, 
for instance due to the large variance of the likelihood estimator 
for parameters located away from the posterior mode,
as illustrated in Figure 2 (c) of \citet{murray2013disturbance}.

\section{Discussion}

Using the Wasserstein distance in approximate Bayesian computation leads to a
new way of inferring parameters in generative models, bypassing the choice
of summaries. The approach can also be readily used for deterministic models. We have demonstrated how the proposed
approach can identify high posterior density regions, in settings of both
i.i.d. (Section \ref{sec:gandk}) and dependent data (Section \ref{sec:queue}). 
In some examples the proposed approximations appear to be at least as close to the posterior distribution 
as those produced by state-of-the-art summary-based ABC.
For instance, in the toggle switch model of Section \ref{sec:toggleswitch}, our black-box method
obtained posterior approximations that are more concentrated on the data-generating parameters than those obtained with sophisticated, case-specific 
summaries, while being computationally cheaper.  Furthermore, we have shown
how summaries and transport distances can be fruitfully combined in Section \ref{sec:levydriven}.
There are various ways of combining distances in the ABC approach, which could 
deserve more research.

% time series
We have proposed multiple ways of defining empirical distributions
of time series data, in order to identify model parameters.  The proposed
approaches have tuning parameters, such as $\lambda$ in the
curve matching approach of Section \ref{sec:curvematching} or the lags in
delay reconstruction in Section \ref{sec:reconstructions}.  The choice of these
parameters has been commented on by \citet{thorpe2017transportation} in the case of curve matching,
and by \citet{muskulus2011,stark2003,kantz2004nonlinear} in the case of delay reconstructions.
Making efficient choices of these parameters might be easier than choosing
summary statistics. Further research could leverage e.g. the literature on Skorokhod distances for $\lambda$ \citep{majumdar2015computing}. 
The investigation of similar methods for the setting of spatial data would also be interesting.

% theory
We have established some theoretical properties of the 
WABC distribution, adding to the existing literature on asymptotic properties
of ABC posteriors \citep{frazier2016,li2015asymptotic}. In particular, we have considered settings where the threshold $\varepsilon$ goes to zero for a fixed set of observations, 
and where the number of observations $n$ goes
to infinity with a slowly decreasing threshold sequence $\varepsilon_n$. 
In the first case, we establish conditions under which the WABC posterior converges to the posterior, as illustrated empirically
in Section \ref{sec:gandk}. In the second case, our results show that under certain conditions, the WABC posterior can
concentrate in different regions of the parameter space compared to the posterior. We also derive upper bounds on the concentration rates, which 
highlight the potential impact of the order $p$ of the Wasserstein
distance, of the dimension of the observation
space, and of model misspecification. The dependence on dimension of the observation space could be a particularly interesting avenue of future research

In comparison with the asymptotic regime, less is known about the properties of ABC posteriors for fixed $\varepsilon$.
Viewing the WABC posterior as a coarsened posterior \citep{miller2015robust},
one can justify its use in terms of
robustness to model misspecification.
On the other hand,  ABC posteriors in general do not yield conservative
statements about the posterior for a fixed threshold $\varepsilon$ and data set $y_{1:n}$. For instance, Figure \ref{fig:cosine:post3}
shows that ABC posteriors can have little overlap with the posterior,
despite having shown signs of concentration away from the prior distribution.

% distance calculations 
As Wasserstein distance calculations scale super-quadratically with the number
of observations $n$, we have introduced a new distance based on the Hilbert
space-filling curve, computable in order $n \log n$, which can be used to
initialize a swapping distance with a cost of order $n^2$.  We have derived some
posterior concentration results for the ABC posterior distributions using the
Hilbert and swapping distances, similarly to Proposition
\ref{theorem:concentration} obtained for the Wasserstein distance. 
Many other distances related to optimal transport could be used;
we mentioned \citet{park2015k2} who used the maximum mean discrepancy,
and recently \citet{genevay2017learning} consider Sinkhorn divergences,
and \citet{jiang2018approximate} consider the Kullback--Leibler divergence.
A thorough comparison between these different distances, none of which involve summary statistics,
could be of interest to ABC practitioners.

\textbf{Acknowledgements} We are grateful to Marco Cuturi, Jeremy Heng, Guillaume Pouliot, Neil Shephard for helpful comments. Pierre E. Jacob gratefully acknowledges support by the National Science Foundation through grant DMS-1712872.

\bibliography{biblio}
\bibliographystyle{apalike}

\begin{appendix}

    \section{Proofs}
    \begin{proof}[of Proposition \ref{prop:as_distn_fixedn}]
        We follow a similar approach to that in Proposition 1 of \citet{rubio2012maximum}. Fix $y_{1:n}$ and let $\bar{\varepsilon}$ be as in the statement of our proposition. For any $0<\varepsilon <\bar{\varepsilon}$, let $q^\varepsilon(\theta)$ denote the normalized quasi-likelihood induced by the ABC procedure, i.e.
$$q^\varepsilon(\theta)= \frac{\int_{\mathcal{Y}^n}\mathds{1}\left(\pdist(y_{1:n},z_{1:n})\leq\varepsilon\right)f_\theta^{(n)}(z_{1:n})dz_{1:n}}{\int_{\mathcal{Y}^n}\mathds{1}\left(\pdist(y_{1:n},z'_{1:n})\leq\varepsilon\right) dz'_{1:n}} = \int_{\mathcal{Y}^n}K^\varepsilon(y_{1:n},z_{1:n})f_\theta^{(n)}(z_{1:n})dz_{1:n},$$
where $K^\varepsilon(y_{1:n},z_{1:n})$ denotes the density of the uniform distribution on  $\mathcal{A}^\varepsilon = \{z_{1:n} : \pdist(y_{1:n},z_{1:n})\leq\varepsilon\}$, evaluated at some $z_{1:n}$. Note that the sets $\mathcal{A}^\varepsilon$ are compact, due to the continuity of $\pdist$. Now, for any $\theta\in\mathcal{H}\setminus\mathcal{N}_{\mathcal{H}}$ we have
\begin{align*} \left\lvert q^\varepsilon(\theta) -  f_\theta^{(n)}(y_{1:n})\right\rvert 
&\leq \int_{\mathcal{Y}^n}K^\varepsilon(y_{1:n},z_{1:n})\left\lvert f_\theta^{(n)}(z_{1:n}) - f_\theta^{(n)}(y_{1:n}) \right\rvert dz_{1:n}\\
&\leq \sup_{z_{1:n}\in \mathcal{A}^\varepsilon} \left\lvert f_\theta^{(n)}(z_{1:n}) - f_\theta^{(n)}(y_{1:n}) \right\rvert \\
& = \left\lvert f_\theta^{(n)}(z^{\varepsilon}_{1:n}) - f_\theta^{(n)}(y_{1:n}) \right\rvert
\end{align*}
for some $z^\varepsilon_{1:n}\in \mathcal{A}^\varepsilon$, where the second inequality
holds since $\int_{\mathcal{Y}^n}K^\varepsilon(y_{1:n},z_{1:n})dz_{1:n} = 1$,
and the last equality holds by compactness of $\mathcal{A}^\varepsilon$ and continuity of
$f_{\theta}^{(n)}$. Since for each $\varepsilon >0$, $z^{\varepsilon}_{1:n} \in \mathcal{A}^\varepsilon$, we know
$\lim_{\varepsilon \to 0} z^{\varepsilon}_{1:n} \in
\cap_{\varepsilon\in\mathbb{Q}^+}\mathcal{A}^\varepsilon$. Under condition
\ref{cond:exch}, $\cap_{\varepsilon\in\mathbb{Q}^+}\mathcal{A}^\varepsilon =
\{y_{\sigma(1:n)} : \sigma\in\mathcal{S}_n\}$, by continuity of $\pdist$. Similarly,
under condition \ref{cond:exact},
$\cap_{\varepsilon\in\mathbb{Q}^+}\mathcal{A}^\varepsilon = \{y_{1:n}\}$. In both cases,
taking the limit $\varepsilon \to 0$ yields $\lvert q^\varepsilon(\theta) -
f_\theta^{(n)}(y_{1:n})\rvert \to 0$, due to the continuity of $f_\theta^{(n)}$
(and $n$-exchangeability under condition \ref{cond:exch}).

Let $\varepsilon \leq \bar{\varepsilon}$, so that 
\begin{align*}\sup_{\theta\in\mathcal{H}\setminus \mathcal{N}_{\mathcal{H}}} q^\varepsilon(\theta) &= \sup_{\theta\in\mathcal{H}\setminus \mathcal{N}_{\mathcal{H}}}  \int_{\mathcal{Y}^n}K^\varepsilon(y_{1:n},z_{1:n})f_\theta^{(n)}(z_{1:n}) dz_{1:n}\\
& \leq \sup_{\theta\in\mathcal{H}\setminus \mathcal{N}_{\mathcal{H}}}\sup_{z_{1:n}\in \mathcal{A}^{\bar{\varepsilon}}} f_\theta^{(n)}(z_{1:n}) < M,
\end{align*}
for some $0<M<\infty$. By the bounded convergence theorem, for any measurable $\mathcal{B}\subset\mathcal{H}$ we have that
$\int_\mathcal{B}\pi(d\theta) q^\varepsilon(\theta) \to \int_\mathcal{B}\pi(d\theta) f_\theta^{(n)}(y_{1:n})$ as $\varepsilon \to 0$. Hence,
$$\lim_{\varepsilon \to 0}\int_\mathcal{B}\pi^{\varepsilon}_{y_{1:n}}\left(d\theta \right) = \frac{\lim_{\varepsilon \to 0}\int_\mathcal{B}\pi\left(d\theta\right) q^\varepsilon(\theta)}{\lim_{\varepsilon \to 0}\int_{\mathcal{H}} \pi\left(d\vartheta\right) q^\varepsilon(\vartheta)} = \frac{\int_\mathcal{B}\pi\left(d\theta\right) f_\theta^{(n)}(y_{1:n})}{\int_{\mathcal{H}} \pi\left(d\vartheta\right) f_\theta^{(n)}(y_{1:n})} = \int_\mathcal{B}\pi\left(d\theta | y_{1:n}\right).$$
\end{proof}

\begin{proof}[of Proposition \ref{theorem:concentration}]
We first look at the WABC posterior probability of the sets 
$\{\theta\in\mathcal{H}: \mathfrak{W}_p(\mu_\star, \mu_{\theta}) > \delta \}$. Note that, using Bayes' formula, for all $\varepsilon,\delta >0$, 
\begin{align*}
\pi^{\varepsilon + \varepsilon_\star}_{y_{1:n}} \left(\was_p(\mu_\star, \mu_{\theta}) > \delta \right) & = \frac{\mathbb{P}_\theta(\was_p(\mu_\star, \mu_{\theta}) > \delta, \; \was_p(\hat{\mu}_n,\hat{\mu}_{\theta,n}) \leq \varepsilon +\varepsilon_\star)}{\mathbb{P}_\theta(\was_p(\hat{\mu}_n,\hat{\mu}_{\theta,n}) \leq \varepsilon + \varepsilon_\star)},
\end{align*}
where $\mathbb{P}_\theta$ denotes the distribution of $\theta\sim \pi$ and of the synthetic data $z_{1:n} \sim \mu_\theta^{(n)}$, keeping the observed data $y_{1:n}$ and hence $\hat{\mu}_n$ fixed. We aim to upper bound this expression, and proceed by upper bounding the numerator and lower bounding the denominator.

 By the triangle inequality, 
$$\was_p(\mu_\star,\mu_\theta) \leq  \was_p(\mu_\star, \hat\mu_{n}) + \was_p(\hat\mu_n, \hat\mu_{\theta,n})  + \was_p(\hat\mu_{\theta,n}, \mu_{\theta}).$$
On the events $\{\was_p(\mu_\star, \mu_{\theta}) > \delta, \; \was_p(\hat{\mu}_n,\hat{\mu}_{\theta,n}) \leq \varepsilon + \varepsilon_\star\}$,
we have 
$$\delta < \mathfrak{W}_p(\mu_\star, \mu_{\theta}) \leq \mathfrak{W}_p(\mu_\star, \hat\mu_{n}) + \mathfrak{W}_p(\hat\mu_{\theta,n}, \mu_{\theta}) + \varepsilon + \varepsilon_\star. $$
Let $A(n,\varepsilon) = \{y_{1:n} : \was_p(\hat{\mu}_n,\mu_\star)\leq \varepsilon/3\}$. Assuming $y_{1:n} \in A(n,\varepsilon)$ implies that
$$\delta <  \mathfrak{W}_p(\hat\mu_{\theta,n}, \mu_{\theta}) + \frac{4\varepsilon}{3} + \varepsilon_\star.$$
Using this to bound the numerator, we get by a simple reparametrization that for any $\zeta > 0$,
\begin{align*}
\pi^{\varepsilon + \varepsilon_\star}_{y_{1:n}}  \left(\was_p(\mu_\star, \mu_{\theta}) > 4\varepsilon/3 + \varepsilon_\star +\zeta \right) & \leq \frac{\mathbb{P}_\theta(\mathfrak{W}_p(\hat\mu_{\theta,n}, \mu_{\theta}) > \zeta)}{\mathbb{P}_\theta(\was_p(\hat{\mu}_n,\hat{\mu}_{\theta,n}) \leq \varepsilon + \varepsilon_\star)}.
\end{align*}

The remainder of the proof follows from further bounding this fraction using the assumptions we made on the convergence rate of empirical measures in the Wasserstein distance. Focusing first on the numerator, for any $\zeta>0$ we have by Assumption \ref{as:concentration} that
\begin{align*}
\mathbb{P}_\theta(\mathfrak{W}_p(\hat\mu_{\theta,n}, \mu_{\theta}) > \zeta) & =
\int_\mathcal{H} \mu_\theta^{(n)}(\was_p(\mu_\theta,\hat{\mu}_{\theta,n}) > \zeta ) \pi(d\theta)  \\
& \leq \int_\mathcal{H} c(\theta)f_n(\zeta) \pi(d\theta) \leq c_1 f_n(\zeta),
\end{align*}
for some constant $c_1<+\infty$. For the denominator,
\begin{align*}
&\mathbb{P}_\theta(\was_p(\hat{\mu}_n,\hat{\mu}_{\theta,n}) \leq \varepsilon + \varepsilon_\star)  = \int_\mathcal{H} \mu_{\theta}^{(n)}(\was_p(\hat{\mu}_n,\hat{\mu}_{\theta,n}) \leq \varepsilon +\varepsilon_\star) \pi(d\theta) \\
& \geq \int_{\was_p(\mu_\star, \mu_{\theta}) \leq \varepsilon/3 +\varepsilon_\star} \mu_{\theta}^{(n)}(\was_p(\hat{\mu}_n,\hat{\mu}_{\theta,n}) \leq \varepsilon +\varepsilon_\star) \pi(d\theta) \quad \text{(by non-negativity of integrand)}\\
& \geq \int_{\was_p(\mu_\star, \mu_{\theta}) \leq \varepsilon/3+\varepsilon_\star} \mu_{\theta}^{(n)}(\was_p(\mu_\star, \mu_{\theta}) + \was_p(\hat{\mu}_n,\mu_\star) + \was_p(\mu_\theta,\hat{\mu}_{\theta,n}) \leq  \varepsilon +\varepsilon_\star) \pi(d\theta) \\ 
&\quad\quad\quad\quad\quad\quad\quad\quad\quad\quad\quad\quad\quad\quad\quad\quad\quad\quad\quad\quad\quad\quad\quad\quad\quad\quad\quad \text{(by the triangle inequality)}\\
	& \geq \int_{\was_p(\mu_\star, \mu_{\theta}) \leq \varepsilon/3+\varepsilon_\star} \mu_{\theta}^{(n)}(\was_p(\mu_\theta,\hat{\mu}_{\theta,n}) \leq \varepsilon/3) \pi(d\theta) \\ & \quad\quad\quad\quad\quad\quad\quad\quad\quad\quad\quad\quad\quad\quad
    \quad \text{(since $\was_p(\mu_\star, \mu_{\theta}) \leq \varepsilon/3+\varepsilon_\star$ and $\was_p(\hat{\mu}_n,\mu_\star)\leq \varepsilon/3$)}\\
	& = \pi(\was_p(\mu_\star, \mu_{\theta})\leq \varepsilon/3+\varepsilon_\star) - \int_{\was_p(\mu_\star, \mu_{\theta}) \leq \varepsilon/3+\varepsilon_\star} \mu_{\theta}^{(n)}(\was_p(\mu_\theta,\hat{\mu}_{\theta,n}) > \varepsilon/3) \pi(d\theta) \\
& \geq \pi(\was_p(\mu_\star, \mu_{\theta})\leq \varepsilon/3+\varepsilon_\star) - \int_{\was_p(\mu_\star, \mu_{\theta}) \leq \varepsilon/3+\varepsilon_\star} c(\theta)f_n(\varepsilon/3) \pi(d\theta) \quad \text{(by Assumption \ref{as:concentration})}.
\end{align*}

We now make more specific choices for $\varepsilon$ and $\zeta$, starting with assuming that $\varepsilon/3 \leq \delta_0$, such that $c(\theta) \leq c_0$ for some constant $c_0>0$ in the last integrand above, by Assumption \ref{as:concentration}. The last line above is then greater than or equal to $\pi(\was_p(\mu_\star, \mu_{\theta})\leq \varepsilon/3+\varepsilon_\star)\left(1-c_0 f_n(\varepsilon/3)\right).$ Replacing $\varepsilon$ with $\varepsilon_n$ such that $f_n(\varepsilon_n/3) \to 0$ implies that $c_0 f_n(\varepsilon_n/3) \leq 1/2$ for sufficiently large $n$. Hence, 
$$\pi(\was_p(\mu_\star, \mu_{\theta})\leq \varepsilon_n/3+\varepsilon_\star)\left(1-c_0 f_n(\varepsilon_n/3)\right) \geq \frac{1}{2}\pi(\was_p(\mu_\star, \mu_{\theta})\leq \varepsilon_n/3+\varepsilon_\star) \geq c_\pi \varepsilon_n^L,$$
 for sufficiently large $n$, by Assumption \ref{as:prior}. We can summarize the bounds derived above as follows,
$$\pi^{\varepsilon_n + \varepsilon_\star}_{y_{1:n}}  \left(\was_p(\mu_\star, \mu_{\theta}) > 4\varepsilon_n/3 + \varepsilon_\star +\zeta \right) \leq Cf_n(\zeta)\varepsilon_n^{-L},$$
where $C = c_1/c_\pi$. 

Set some $R>0$ and note that for any $n\geq 1$, because the function $f_n$ is strictly decreasing under Assumption \ref{as:concentration}, $f_n^{-1}(\varepsilon_n^{L}/R)$ is well-defined  in the sense that $f_n^{-1}$ is defined at $\varepsilon_n^{L}/R$.
Choosing $\zeta_n = f_n^{-1}(\varepsilon_n^{L}/R)$ leads to
$$\pi^{\varepsilon_n + \varepsilon_\star}_{y_{1:n}}  \left(\was_p(\mu_\star, \mu_{\theta}) > 4\varepsilon_n/3 + \varepsilon_\star + f_n^{-1}(\varepsilon_n^{L}/R)\right) \leq \frac{C}{R}.$$
Since we assumed that $\mathbb{P}\left(\{\omega: y_{1:n}(\omega) \in A(n,\varepsilon_n)\right) \to 1$ as $n\to\infty$, the statement above holds with probability going to one.
\end{proof}

\begin{proof}[of Corollary \ref{cor:concentration}]
Let $\delta>0$ be such that $\{\theta\in\mathcal{H} : \rho_\mathcal{H}(\theta,\theta_\star) \leq \delta\} \subset U$, where $U$ is the set in Assumption \ref{as:functional}. By Assumption \ref{as:wellsep}, there exists $\delta'>0$ such that $\rho_\mathcal{H}(\theta,\theta_\star) > \delta$ implies $\was_p(\mu_\theta,\mu_\star) - \varepsilon_\star > \delta'$. Let $n$ be large enough such that $4\varepsilon_n/3  + f_n^{-1}(\varepsilon_n^{L}/R) < \delta'$, which implies $\{\theta\in\mathcal{H}:\was_p(\mu_\star, \mu_{\theta}) - \varepsilon_\star \leq  4\varepsilon_n/3  + f_n^{-1}(\varepsilon_n^{L}/R)\} \subset U$. 

From Proposition \ref{theorem:concentration}, we know that
$$\pi^{\varepsilon_n + \varepsilon_\star}_{y_{1:n}}  \left(\was_p(\mu_\star, \mu_{\theta}) - \varepsilon_\star \leq  4\varepsilon_n/3  + f_n^{-1}(\varepsilon_n^{L}/R) \right) \geq 1- \frac{C}{R},$$
with probability going to one. Applying the inequality in Assumption \ref{as:functional} gives 
$$\pi^{\varepsilon_n + \varepsilon_\star}_{y_{1:n}}  \left(\rho_\mathcal{H}(\theta,\theta_\star) \leq K[4\varepsilon_n/3 + f_n^{-1}(\varepsilon_n^L/R)]^{\alpha} \right) \geq 1- \frac{C}{R},$$
with probability going to one.
\end{proof}

\begin{proof}[of Proposition \ref{prop:hilbert}]
Let $x_{1:n}$, $y_{1:n}$  and $z_{1:n}$ be three vectors in $\mathcal{Y}^n$ and denote by $\hat{\mu}_n^x$, $\hat{\mu}_n^y$ and $\hat{\mu}_n^z$ the corresponding empirical distributions of size $n$.
Since $\rho$ is a metric on $\mathcal{Y}$,
$$
\mathfrak{H}_p(\hat{\mu}_n^x,\hat{\mu}_n^z)\geq 0,\quad \mathfrak{H}_p(\hat{\mu}_n^x,\hat{\mu}_n^z)=\mathfrak{H}_p(\hat{\mu}_n^z,\hat{\mu}_n^x)
$$
and $\mathfrak{H}_p(\hat{\mu}_n^x,\hat{\mu}_n^z)=0$ if and only if $\hat{\mu}_n^x=\hat{\mu}_n^z$. To conclude the  proof it therefore remains to show that
$$
\mathfrak{H}_p(\hat{\mu}_n^x,\hat{\mu}_n^z)\leq \mathfrak{H}_p(\hat{\mu}_n^x,\hat{\mu}_n^y)+\mathfrak{H}_p(\hat{\mu}_n^y,\hat{\mu}_n^z).
$$ 
To this end, we define
\begin{align*}
&\rho_{xy}=\big(\rho\big(x_{\sigma_x(1)}, y_{\sigma_y(1)} \big),\dots,\rho\big(x_{\sigma_x(n)}, y_{\sigma_y(n)} \big)\big),\quad \rho_{xz}=\big(\rho\big(x_{\sigma_x(1)}, z_{\sigma_z(1)} \big),\dots,\rho\big(x_{\sigma_x(n)}, z_{\sigma_z(n)} \big)\big)\\
&\rho_{yz}=\big(\rho\big(y_{\sigma_y(1)}, z_{\sigma_z(1)} \big),\dots,\rho\big(y_{\sigma_y(n)}, z_{\sigma_z(n)} \big)\big)
\end{align*}
and denote by $\|\cdot\|_p$ the $L_p$-norm on $\mathbb{R}^n$. Then,
\begin{align*}
 \mathfrak{H}_p(\hat{\mu}_n^x,\hat{\mu}_n^z)&= n^{-1/p}\|\rho_{xz}\|_p\\
 &\leq n^{-1/p}\|\rho_{xy}\|_p+n^{-1/p}\|\rho_{xz}-\rho_{xy}\|_p\\
 &=\mathfrak{H}_p(\hat{\mu}_n^x,\hat{\mu}_n^y)+n^{-1/p} \left(\sum_{i=1}^n \big|\rho\big(x_{\sigma_x(i)}, z_{\sigma_z(i)} \big)-\rho\big(x_{\sigma_x(i)}, y_{\sigma_y(i)} \big)\big|^p \right)^{1/p}\\
 &\leq \mathfrak{H}_p(\hat{\mu}_n^x,\hat{\mu}_n^y)+n^{-1/p} \left(\sum_{i=1}^n \rho\big(y_{\sigma_y(i)}, z_{\sigma_z(i)} \big)^p \right)^{1/p}\\
 &=\mathfrak{H}_p(\hat{\mu}_n^x,\hat{\mu}_n^y)+\mathfrak{H}_p(\hat{\mu}_n^y,\hat{\mu}_n^z),
\end{align*}
where the first inequality uses the triangle inequality, and the last uses the reverse triangle inequality. 
\end{proof}

\end{appendix}

\end{document}